\documentclass[preprint2]{aastex_mod}
\usepackage{ulem}
\usepackage[dvips]{color}
\usepackage{graphicx}
\usepackage{txfonts}
\usepackage{rotating}
\usepackage[USenglish]{babel}

\usepackage{longtable}

\begin{document}

\title{The intriguing stellar populations in the globular clusters
  NGC~6388 and NGC~6441$^{\ast}$}

\author{A.\ Bellini\altaffilmark{1,2},
G.\ Piotto\altaffilmark{2},
A.\ P. Milone\altaffilmark{3,4},
I.\ R.\ King\altaffilmark{5},
A.\ Renzini\altaffilmark{6},
S.\ Cassisi\altaffilmark{7},
J.\ Anderson\altaffilmark{1},
L.\ R.\ Bedin\altaffilmark{6},
D.\ Nardiello\altaffilmark{2},
A.\ Pietrinferni\altaffilmark{7}
and A.\ Sarajedini\altaffilmark{8} 
}

\altaffiltext{1}{Space Telescope Science Institute, 3700 San Martin
  Dr., Baltimore, 21218, MD, USA}

\altaffiltext{2}{Dipartimento di Fisica e Astronomia ``Galileo
  Galilei'', Universit\`a di Padova, v.co dell'Osservatorio 3,
  I-35122, Padova, Italy, EU}

\altaffiltext{3}{Instituto de Astrofisica de Canarias, E-38200 La
  Laguna, Tenerife, Canary Islands, Spain, EU}

\altaffiltext{4}{Department of Astrophysics, University of La Laguna,
  E-38200 La Laguna, Tenerife, Canary Islands, Spain, EU}

\altaffiltext{5}{Department of Astronomy, University of Washington,
  Box 351580, Seattle, 98195, WA, USA}

\altaffiltext{6}{Istituto Nazionale di Astrofisica, Osservatorio
  Astronomico di Padova, v.co dell'Osservatorio 5, I-35122, Padova,
  Italy, EU}

\altaffiltext{7}{Istituto Nazionale di Astrofisica, Osservatorio
  Astronomico di Collurania, via Mentore Maggini, I-64100 Teramo,
  Italy, EU}

\altaffiltext{8}{University of Florida, Department of Astronomy, 211
  Bryant Space Science Center, Gainesville, 32611, FL, USA}

\altaffiltext{$^\ast$}{Based on proprietary and
    archival observations with the NASA/ESA Hubble Space Telescope,
    obtained at the Space Telescope Science Institute, which is
    operated by AURA, Inc., under NASA contract NAS 5-26555.}

\email{bellini@stsci.edu}

\date{\today}

\begin{abstract} { NGC~6388 and NGC~6441 are two massive Galactic
    bulge globular clusters which share many properties, including the
    presence of an extended horizontal branch (HB), quite unexpected
    because of their high metal content. In this paper we use
    \textit{HST}'s WFPC2, ACS, and WFC3 images and present a broad
    multicolor study of their stellar content, covering all main
    evolutionary branches. The color-magnitude diagrams (CMDs) give
    compelling evidence that both clusters host at least two stellar
    populations, which manifest themselves in different ways. NGC~6388
    has a broadened main sequence (MS), a split sub-giant branch
    (SGB), and a split red giant branch (RGB) that becomes evident
    above the HB in our data set; its red HB is also split into two
    branches. NGC~6441 has a split MS, but only an indication of two
    SGB populations, while the RGB clearly splits in two from the SGB
    level upward, and no red HB structure.  The multicolor analysis of
    the CMDs confirms that the He difference between the two main
    stellar populations in the two clusters must be similar. This is
    observationally supported by the HB morphology, but also confirmed
    by the color distribution of the stars in the MS optical band
    CMDs. However, a MS split becomes evident in NGC~6441 using UV
    colors, but not in NGC~6388, indicating that the chemical patterns
    of the different populations are different in the two clusters,
    with C, N, O abundance differences likely playing a major role.
    We also analyze the radial distribution of the two populations.}
\end{abstract}

\keywords{globular clusters: individual (NGC~6388, NGC~6441) --
  photometry -- Stars: Population II -- C-M diagrams -- proper
  motions}

\maketitle

\begin{table*}[th!]
\label{tab:1}
\begin{center}
\begin{tabular}{ccccccc}
\multicolumn{7}{c}{\textsc{Table 1}}\\
\multicolumn{7}{c}{\textsc{Log of \textit{HST} Observations for NGC~6388}}\\
\hline\hline
\textbf{GO}&\textbf{PI}&\textbf{Instrument}&\textbf{Filter}&\textbf{Exposures}&\textbf{$\Delta$r$^\dagger$}&\textbf{Epoch}\\
(1)&(2)&(3)&(4)&(5)&(6)&(7)\\
\hline
9835 & G.\ Drukier & ACS/HRC & F555W & 51$\times$155$\,$s &$0\farcm05$& 30 Oct 2003\\
     &             &         & F814W & 5$\times$25$\,$s, 2$\times$469$\,$s, 10$\times$505$\,$s &$0\farcm05$& 30 Oct 2003\\
10350 & H.\ Cohn & ACS/HRC & F330W & 2$\times$1266$\,$s 4$\times$1314$\,$s &$0\farcm05$& 7 Apr 2006\\
      &          &         & F555W & 3$\times$155$\,$s &$0\farcm05$& 7 Apr 2006\\
      &          & ACS/WFC & F660N & 2$\times$1123$\,$s, 4$\times$1171$\,$s &$4\farcm05$& 7 Apr 2006\\
10474 & G.\ Drukier & ACS/HRC & F814W & 5$\times$25$\,$s, 8$\times$501$\,$s, 4$\times$508$\,$s &$0\farcm05$& 4--8 Apr 2006\\
9821$^{\rm a}$ & B.\ Protz & ACS/WFC & F435W & 6$\times$11$\,$s & $0\farcm08$& 2003--2004\\
     &           &         & F555W & 6$\times$7$\,$s  & $0\farcm08$& 2003--2004\\
10775 & A. Sarajedini & ACS/WFC & F606W & 1$\times$40$\,$s, 5$\times$340$\,$s & $0\farcm27$& 6 Apr 2006\\
      &               &         & F814W &  1$\times$40$\,$s, 5$\times$350$\,$s   & $0\farcm27$& 7 Apr 2006\\
11739 & G.\ Piotto & ACS/WFC   & F475W & 4$\times$788$\,$s & $3\farcm86$& 3 Jul 2010\\
      &            &           & F814W & 1$\times$67$\,$s, 4$\times$155$\,$s & $4\farcm02$& 30 Jun 2010\\
      &            & WFC3/UVIS & F390W & 6$\times$880$\,$s & $2\farcm35$& 3 Jul 2010\\
      &            & WFC3/IR  & F160W & 6$\times$199$\,$s, 4$\times$249$\,$s & $2\farcm18$& 30 Jun 2010\\
\hline
\multicolumn{7}{l}{\small{$\dagger$ Average distance of data set from
    cluster center.}}\\
\multicolumn{6}{l}{$^{\rm a}$SNAP program.}\\
\end{tabular}
\end{center}
\end{table*}

\begin{table*}[th!]
\begin{center}
\begin{tabular}{ccccccc}
\multicolumn{7}{c}{\textsc{Table 2}}\\
\multicolumn{7}{c}{\textsc{Log of \textit{HST} Observations for NGC~6441}}\\
\hline\hline
\textbf{GO}&\textbf{PI}&\textbf{Instrument}&\textbf{Filter}&\textbf{Exposures}&\textbf{$\Delta$r$^\dagger$}&\textbf{Epoch}\\
(1)&(2)&(3)&(4)&(5)&(6)&(7)\\
\hline
5667 & B.\ H.\ Margon     & WFPC2  & F336W & 2$\times$500$\,$s & $0\farcm05$& 8 Aug 1994\\
8251 & H.\ A.\ Smith      & WFPC2  & F439W & 3$\times$50$\,$s,  15$\times$80$\,$s & $0\farcm10$& 1999--2000\\
     &                    &        & F555W & 18$\times$10$\,$s & $0\farcm01$& 1999--2000\\
8718 & G.\ Piotto     & WFPC2  & F336W & 1$\times$260$\,$s, 2$\times$400$\,$s & $0\farcm02$& 16 Sep 2001\\
9835 & G.\ Drukier & ACS/HRC & F555W & 36$\times$240$\,$s & $0\farcm07$& 1 Sep 2003\\
     &             &         & F814W & 5$\times$40$\,$s, 1$\times$413$\,$s, 10$\times$440$\,$s & $0\farcm07$& 1 Sep 2003\\
10775 & A. Sarajedini & ACS/WFC & F606W & 1$\times$45$\,$s, 5$\times$340$\,$s   & $0\farcm27$&25 May 2006\\
      &               &         & F814W & 1$\times$45$\,$s, 5$\times$350$\,$s   & $0\farcm27$&25 May 2006\\
11739 & G.\ Piotto & ACS/WFC   & F475W & 3$\times$768$\,$s, 3$\times$803$\,$s & $3\farcm86$&4--8 May 2010\\
      &            &           &       & 1$\times$765$\,$s, 1$\times$800$\,$s & $3\farcm86$&30 May 2011\\
      &            &           & F814W & 2$\times$67$\,$s, 6$\times$155$\,$s & $4\farcm02$&4--8 May 2010\\
      &            &           &       & 2$\times$155$\,$s & $4\farcm02$&30 May 2011\\
      &            & WFC3/UVIS & F390W & 9$\times$884$\,$s & $2\farcm35$&4--8 May 2010\\
      &            &           &       & 3$\times$884$\,$s & $2\farcm35$&30 May 2011\\
      &            & WFC3/IR  & F160W & 8$\times$199$\,$s, 8$\times$249$\,$s & $2\farcm18$&7 May 2010\\
\hline
\multicolumn{7}{l}{\small{$\dagger$ Average distance of data set from
    cluster center.}}\\
\end{tabular}
\label{tab2}
\end{center}
\end{table*}

%F1
\begin{figure*}[t!]
\centering
\includegraphics[height=8.2cm]{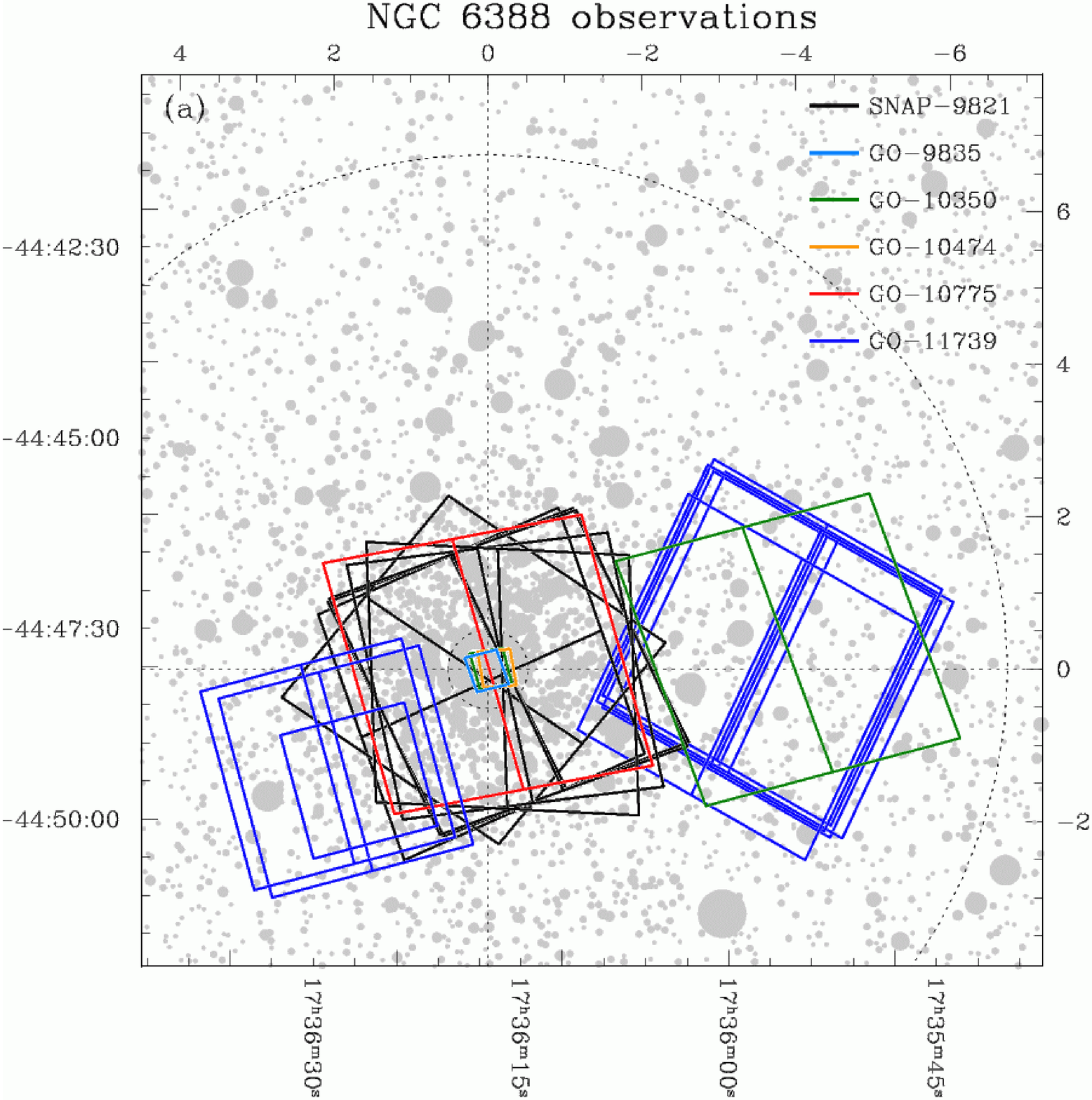}
\includegraphics[height=8.2cm]{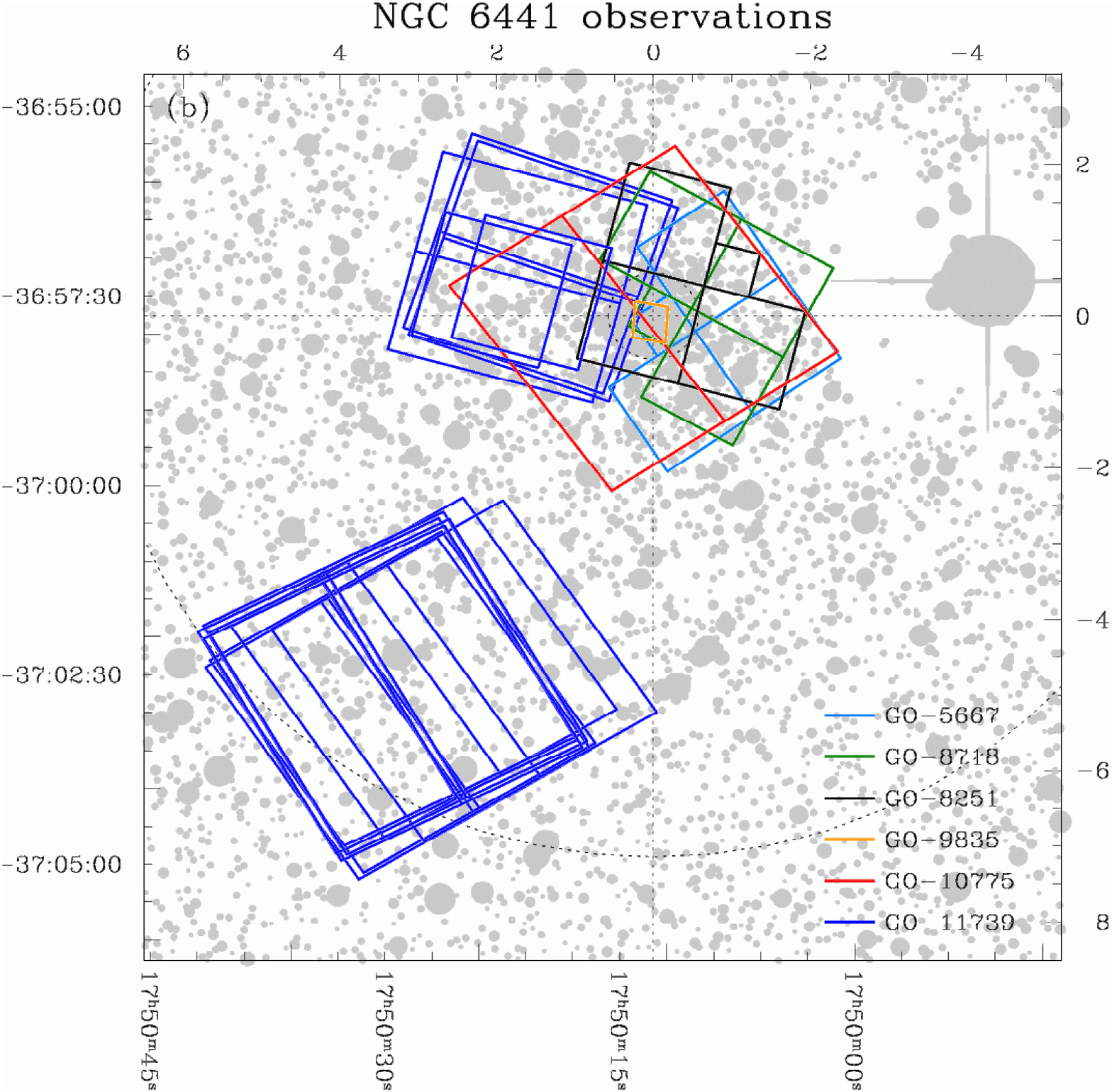}
\includegraphics[height=8.7cm]{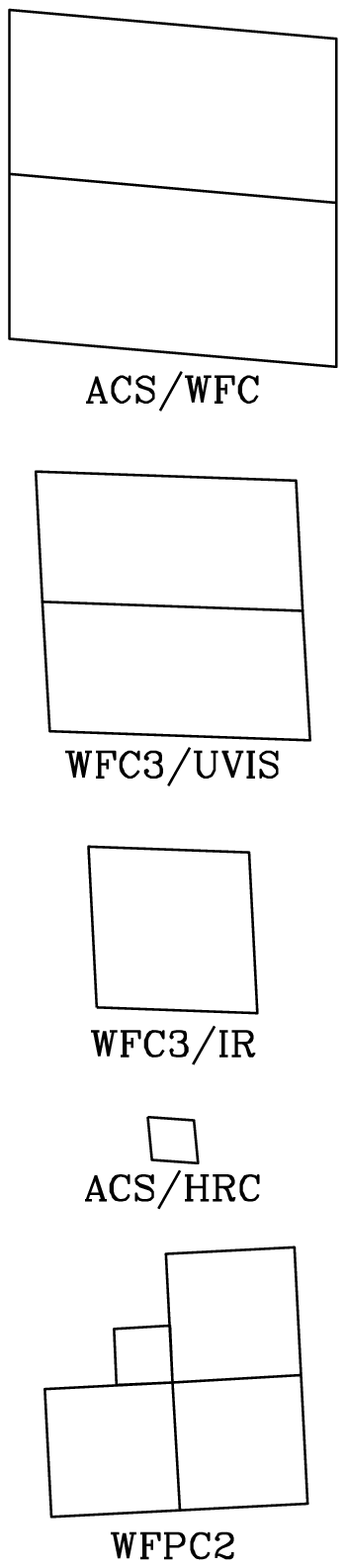}\\
\caption{Finding charts for NGC~6388 (a) and NGC~6441 (b) superimposed
  on star maps from the $J$-band 2MASS catalog.  Data sets are
  color-coded according to the GO program.  There is no ambiguity in
  using the same color for different cameras (e.g., for GO-11739
  WFC3/UVIS, WFC3/IR, and ACS/WFC) because of the different size of
  each camera's FoV (which are shown, to scale, on the right edge of
  the figure).  ACS/WFC is the larger 2-CCD footprint, and WFC3/UVIS
  is the smaller one; the larger 1-CCD footprint is WFC3/IR, and the
  smaller one is ACS/HRC (always placed at the cluster center).
  Finally, WFPC2 has its own distinctive L-shaped footprint, and is
  used here only for NGC~6441.  We give both right ascension and
  declination (lower abscissa and left ordinate) and components
  relative to the cluster center in arcmin (upper abscissa and right
  ordinate).  The inner dotted circle marks the half-light radius
  $r_{\rm h}$, while the outer one marks the tidal radius ($r_{\rm
    t}$, from Harris 1996, 2010 edition). The core radius (not marked)
  falls within the ACS/HRC pointings.}
\label{fig:1}
\end{figure*}

\section{Introduction}
\label{sec:1}

Since the 1990s, NGC~6388 ($l=345^{\circ}\!\! .56$, $b=-6^{\circ}\!\!
.74$) and NGC~6441 ($l=353^{\circ}\!\! .53$, $b=-5^{\circ}\!\! .01$)
have been considered as ``twin'' clusters because of many common
features.  They are both located in the bulge of our Galaxy, they are
two of the most luminous and massive ($\sim$1.6$\times$10$^6$
M$_\odot$, Pryor \& Meylan 1993) globular clusters (GCs) of the Milky
Way, and they are both metal rich ([Fe/H]=$-$0.55 for NGC~6388 and
[Fe/H]=$-$0.46 for NGC~6441, Harris 1996, 2010 edition).  Despite
their high metal content, their most intriguing common property is an
extended blue horizontal branch (HB, Rich et al.\ 1997), which makes
them among the most extreme examples of the so-called second-parameter
problem.  Moreover, the red clump (RC) of the HB is noticeably tilted,
with bluer stars being brighter (in the optical bands) than redder
ones (Raimondo et al.\ 2002).

Since the work of Rich et al.\ (1997), attempts to understand what is
special about these two clusters have focused on these two
peculiarities. Sweigart \& Catelan (1998) first proposed high helium
as the origin of the blue HB extension, in a scenario in which the
envelope of fast-rotating stars, while ascending the red giant branch
(RGB), would be enriched in helium by meridional currents.  Others,
focusing on the HB tilt, argued for differential reddening (DR) as the
cause of the tilt itself (e.g., Raimondo et al.\ 2002), but even
though these clusters are certainly affected by DR, the size of the
tilt is too large to be explained by DR alone, and finally, DR has
been excluded by Busso et al.\ (2007), who found that the HB tilt
persisted in all parts of their field, in each of the two clusters.

It became evident that traditional evolutionary scenarios could not
account for the properties of these two clusters following the
discovery that their RR Lyrae variables have exceptionally long
periods (something that could be explained by high helium but
certainly not by DR). With an average fundamental-mode period of
$\langle P_{ab}\rangle=0\fd71$ for NGC~6388, and $0\fd76$ for
NGC~6441, these two clusters were legitimately considered to be the
prototype of a third Oosterhoff type (the average periods of
Oosterhoff I and Oosterhoff II clusters being $0\fd55$ and $0\fd65$,
respectively, Layden et al.\ 1999; Pritzl et al.\ 2000, 2001, 2002,
2003).

The discovery of multiple stellar populations with enhanced helium in
several GCs (see Piotto 2009 for a recent review) offers a new
perspective for understanding the many apparent anomalies that these
two clusters share.  After it was shown that high helium could be the
cause of both the blue HB extension and the unusual period of the RR
Lyrae (D'Antona \& Caloi 2004; Caloi \& D'Antona 2007; Busso et al.\
2007), the question became whether high helium was the result of
deep-mixing in a minority of RGB stars (as originally suggested by
Sweigart \& Catelan 1998), or was due to the presence of a second,
helium-enriched stellar population. In the former case no splitting of
the main sequence (MS) was expected, whereas a sizable split should
exist if the blue HB and RR Lyrae stars belong to a distinct,
helium-rich sub-population.

Evidence for different stellar generations has already been found in
both clusters with the detection of light-element star-to-star
variations and the Na--O anticorrelation (Gratton et al.\ 2007;
Carretta et al.\ 2007, 2009a). Moreover, NGC~6388 also exhibits a
split sub-giant branch (SGB, Piotto 2008; Moretti et al.\ 2009; Piotto
et al.\ 2012).  Thus, if the blue HB and RR Lyrae stars belong to
helium-enriched populations in these clusters, we should be able to
see the effect of enhanced He on the MSs (a spread or a split), to
quantify the enhancement, and to check whether it is consistent with
the amount of helium enhancement required to account for the HB
properties.

To this end, we submitted a \textit{Hubble Space Telescope}
(\textit{HST}) proposal (GO-11739, PI G. Piotto) to perform accurate
photometry of NGC~6388 and NGC~6441 with the Wide Field Camera 3
(WFC3). In the present paper, we combine GO-11739 data with a large
set of archival images from the Advanced Camera for Surveys (ACS) and
the Wide-Field Planetary Camera 2 (WFPC2), in order to thoroughly map
the multiple stellar populations of these two clusters from their MS
all the way to the SGB, RGB and HB.

This paper is organized as follows:\ in Section~2 the vast set of
\textit{HST} data used for the project is presented, along with the
reduction procedures. In Section~3 distinct stellar populations are
identified in all the color-magnitude-diagram (CMD) sequences of the
two clusters, and in Section~4 their radial distributions within the
clusters are presented.  In Section~5 we estimate the the amount of
helium enhancement $\Delta Y$ required to account for the size of the
MS split, and discuss its implications at the light of our new
observational findings.

\section{Data sets and reductions}
\label{sec:2}

This work is based on proprietary and on archival \textit{HST} images.
We made use of exposures taken with three different cameras:\ (1) ACS,
with both the Wide-Field Channel and the High-Resolution Channel (WFC
and HRC); (2) WFC3, with both the Ultraviolet-Visible (UVIS) and the
Infrared (IR) channels;\ and (3) WFPC2.  A list of the observations is
given in Table~1 for NGC~6388 and in Table~2 for NGC~6441.

To summarize briefly:\ ACS/WFC observations come from programs
SNAP-9821, GO-10775 and GO-11739 for NGC~6388; from GO-10775 and
GO-11739 for NGC~6441.  ACS/HRC observations are from programs
GO-9835, GO-10350, and GO-10474.  WFPC2 exposures come from GO-5667,
GO-8251, and GO-8718.  Much of our material is archival, but as an
essential part of this study we specifically designed the new program
GO-11739 (PI:\ Piotto) to detect multiple stellar populations along
the MSs of NGC~6388 and NGC~6441 by exploiting the power of the new
WFC3 cameras.  We used both WFC3/UVIS and WFC3/IR, in each case taking
parallel exposures with ACS/WFC.

Figure~1 shows the footprints of all exposures that are listed in
Tables~1 and 2, superimposed on stars from the 2MASS $J$-band catalog.
A distinctive color is used for each program and a distinctive shape
and size for each camera.

Figure~1 and Tables 1 and 2 are crucial for this paper, as they
exhibit the overlaps of the different data sets, which govern much of
what follows. Careful study of the figure, along with the tables, will
enable the reader to see possible filter combinations that can be used
to build CMDs at different radial locations in the clusters.
Moreover, since some overlapping data sets encompass different epochs,
one can see where proper motions are measurable, information that will
prove invaluable for selection of cluster members.  The reader should
particularly note, for each cluster, the overlap between GO-11739 and
GO-10775 (much smaller for NGC~6388 than for NGC~6441).

In this paper we will focus our attention on the more central fields,
leaving the analysis of the outer fields to a later paper.
This choice will not affect our results, since the outer field
  has been imaged only with filters not designed to highlight the
  multi-population phenomenon along the MS of these two
  clusters. Moreover, the outer field does not have enough RGB and SGB
  stars for a study of their radial distribution.

\subsection{ACS/WFC and HRC}

All ACS/WFC \texttt{\_flt} images (i.e., images that were dark- and
bias-subtracted and flat-fielded, but not resampled) were corrected
for charge-transfer deficiencies by using the procedure and the
software developed by Anderson \& Bedin (2010).

Photometry and astrometry were carried out with the software tools
described by Anderson et al.\ (2008).  Briefly, all exposures
belonging to a specific program were analyzed simultaneously to
generate an astrophotometric catalog of stars in the field of view
(FoV).  Stars were measured independently in each image by using the
spatially varied point-spread function (PSF) library models from
Anderson \& King (2006b), along with a spatially constant perturbation
specifically derived for each exposure, to compensate for small focus
changes due to spacecraft ``breathing''. Star positions were corrected
for geometric distortion using Anderson \& King (2006b) and Anderson
(2007).

Photometry was calibrated into the Vega-mag flight system following
recipes in Bedin et al.\ (2005), and using the zero points given in
Sirianni et al.\ (2005).

The measurement of stellar fluxes and positions in each ACS/HRC image
was performed by using the publicly available routine img2xym\_HRC,
library PSFs, and the distortion correction described in Anderson \&
King (2006a). We corrected the photometric zero points in the same way
as for the WFC photometry.

\subsection{WFPC2}

The WFPC2 images are not \texttt{\_flt}-type (introduced only for
later instruments), but rather the earlier equivalent \texttt{\_c0f}.
We reduced them with the algorithms described in Anderson \& King
(2000). The field was calibrated to the photometric Vega-mag flight
system of WFPC2 according to the prescriptions of Holtzman et
al.\ (1995).

\subsection{WFC3/UVIS and IR}

For the WFC3 exposures we used the \texttt{\_flt} images.  Star
positions and fluxes on WFC3/UVIS images were measured with software
based mostly on \allowbreak img2xym\_WFI (Anderson et al.\ 2006),
which implements a set of spatially variable empirical PSFs for each
individual exposure. We corrected positions for geometric distortion
using the recipes given in Bellini \& Bedin (2009) and Bellini,
Anderson \& Bedin (2011). Photometry was calibrated as in Bedin et
al.\ (2005).

Infrared images were reduced with software based on that of Anderson
\& King (2006a) for ACS/HRC. We constructed an empirical F160W
``effective PSF'' using the 3-step procedures devised by Anderson \&
King (2000).  A detailed description of the software, PSF derivation,
and geometric distortion solution will be presented
elsewhere. Photometry was calibrated as in Bedin et al.\ (2005).

%F2
\begin{figure*}[t!]
\centering
\includegraphics[width=18cm]{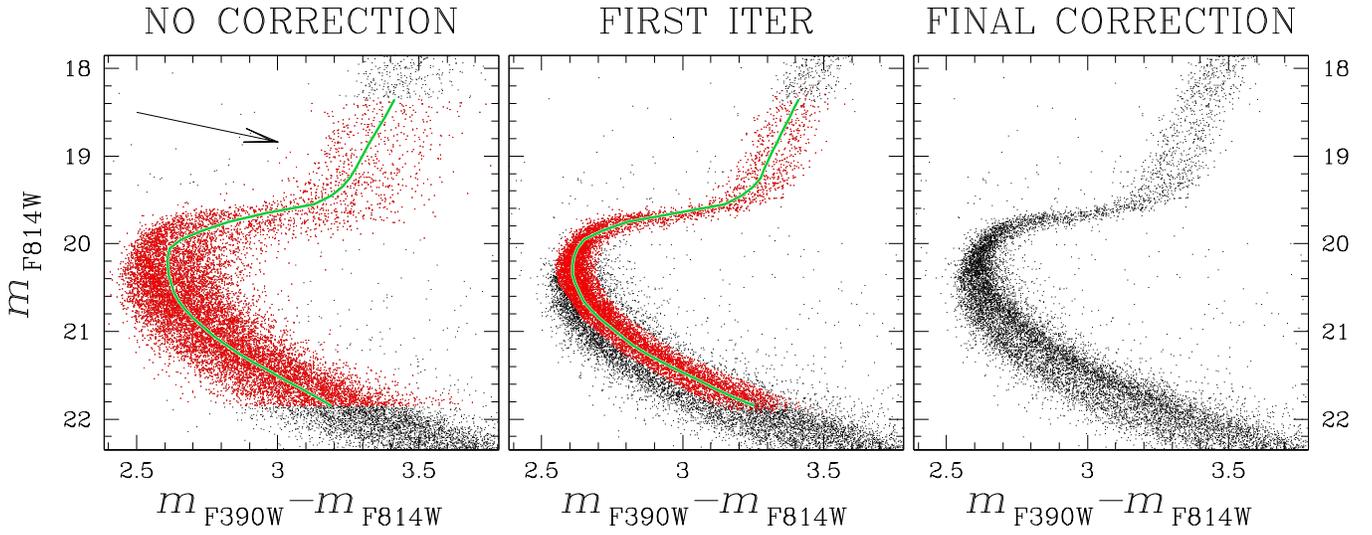}
\caption{Illustrating the correction for differential reddening.  The
  first panel shows the uncorrected $m_{\rm F814W}$ vs.\ $m_{\rm
    F390W}-m_{\rm F814W}$ CMD of NGC~6441 around the SGB region. Stars
  used as a reference for the first-guess DR correction are marked in
  red, and the fiducial sequence in green.  The arrow shows the
  direction of the reddening vector.  After a trial correction (second
  panel), the sequences in the CMD are significantly narrower, and we
  are able to refine our choice of reference stars and adjust the
  fiducial sequence a little.  In the final panel is the result of
  three iterations.  Note the sharpness of the SGB, and the splits in
  the MS and RGB.}
\label{f:redd_ba}
\end{figure*}

%F3
\begin{figure*}[t!]
\centering
\includegraphics[width=18cm]{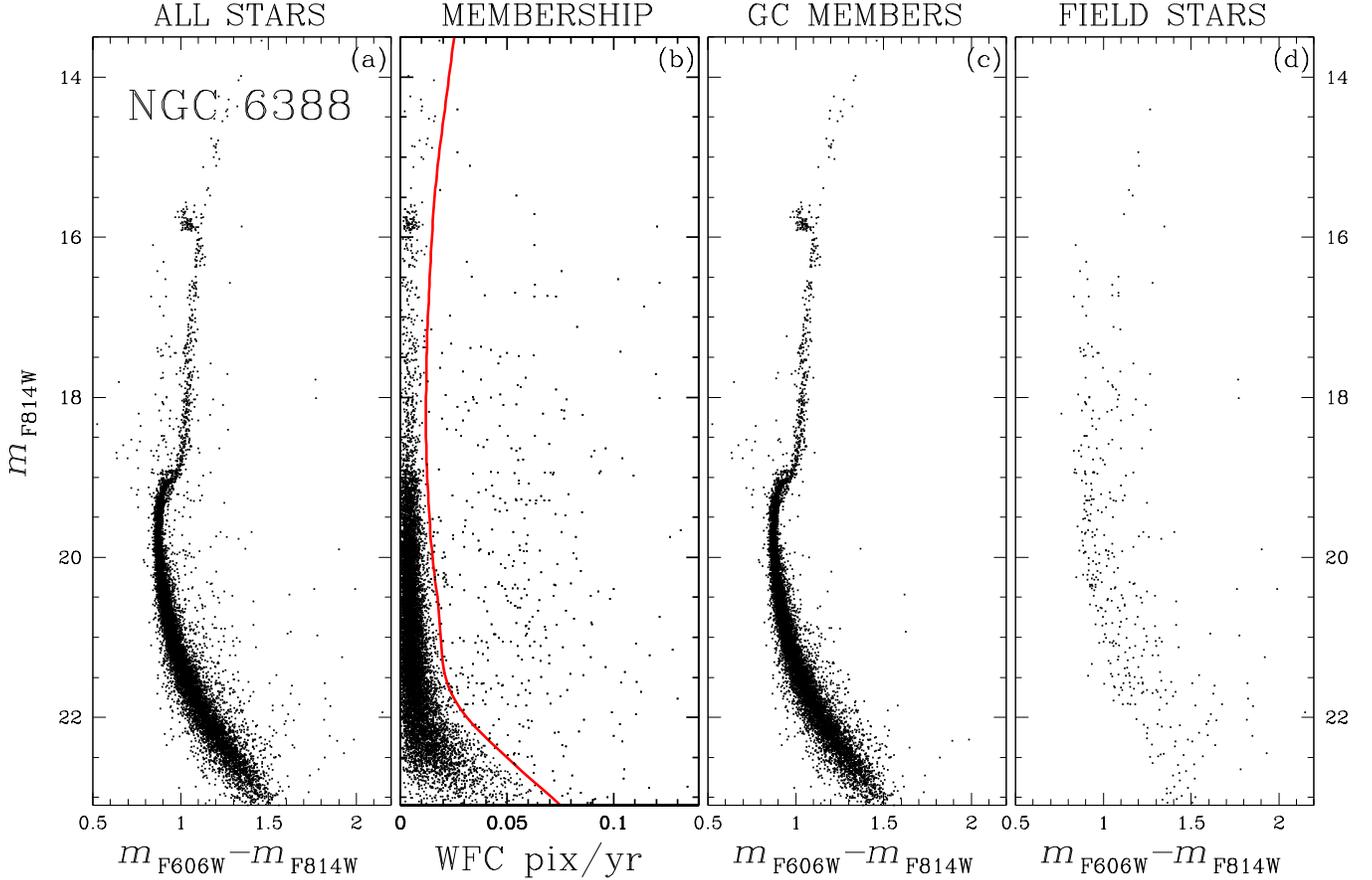}
\caption{Illustrating the proper-motion separation for NGC~6388.  (a)
  CMD for stars in common between GO-10775 and GO-11739.  (b) Sizes of
  the proper-motion displacements (in pixel/yr); the red line (drawn
  by hand) shows our membership criterion.  (c) CMD of
  proper-motion-selected cluster members; (d) stars rejected as field
  stars (even though this means the loss of a few members).}
\label{f:6388_pm}
\end{figure*}

%F4
\begin{figure*}[t!]
\centering
\includegraphics[width=18cm]{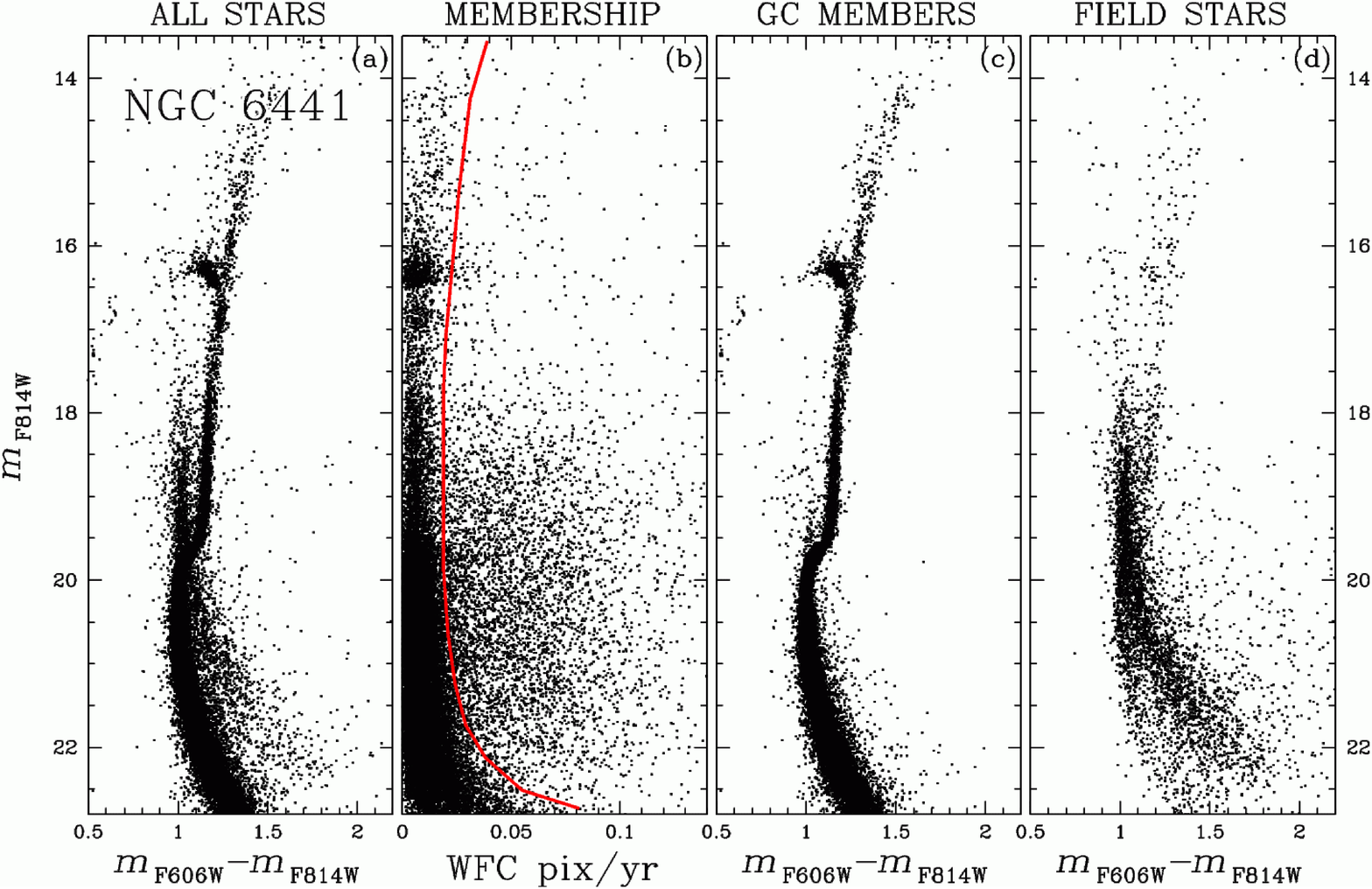}
\caption{Same as Fig.~\ref{f:6388_pm}, but for NGC~6441.  Although the
  field contamination is much stronger, we have nevertheless selected
  an almost clean sample of cluster members.}
\label{f:6441_pm}
\end{figure*}

\subsection{Sample selections}

Since our aim is fine photometric discrimination, it was a great
advantage to select the best-measured stars in our samples.  To
accomplish this, we closely followed the recipes and used the same
selection criteria as given in Milone et al.\ (2009), which makes use
of several diagnostics:\ photometric and astrometric rms, PSF-fit
residuals, and amount of scattered light from neighboring stars.
(This last diagnostic is available only for ACS/WFC catalogs because,
as we stated above, different codes where used for different cameras,
and only the one for the ACS/WFC provides the additional diagnostic.)

For more information about our selection procedures we refer the
reader to Milone et al.\ (2009).

\subsection{Differential reddening}

Both NGC~6388 and NGC~6441 are in the direction of the Galactic bulge,
where extinction is a serious problem ($E(B-V)$ 0.37 and 0.47
respectively, according to Harris 1996, 2010 edition).  Such large
reddenings can be expected to be accompanied by serious differential
reddening, which we treated with particular care.

As a first step, for each CMD we computed the direction of the
reddening vector.  For this we used the extinction coefficients in
Tables 14 and 15 of Sirianni et al.\ (2005) for G2 stars.  For WFC3
F390W, which is not in those tables, we interpolated linearly between
ACS/HRC F330W and F435W, getting $A_{\rm F390W}/E(B-V)=4.49$.  For the
F160W filter we adopted the value $A_{\rm F160W}/E(B-V)=0.57$ (from
YES, the York Extinction
Solver\footnote{\texttt{http://www3.cadc-ccda.hia-iha.nrc-cnrc.gc.ca/community/\\ YorkExtinctionSolver/}}),
and used the Fitzpatrick (1999) extinction law.  As we did for the
WFPC2, we took the extinction values from Schlegel, Finkbeiner \&
Davis (1998).

We dealt with differential reddening (DR) by means of a method that is
explained in detail in Section 3.1 of Milone et al.\ (2012a), to which
we refer the reader for details; here we only describe the method
briefly.

The overall method consists of iterating a core procedure which itself
has five steps: (1) We rigidly rotate the CMD to make the reddening
direction parallel to the $x$ axis, so that DR becomes a simple shift
in $x$, and the $x$ distance of a star from the now-vertical fiducial
becomes an estimate of its DR value.  (2) We next choose a set of
``reference stars'' --- stars whose DR values are best suited to serve
as standards.  Ideally these stars should lie in the parts of the CMD
where the sequences are almost perpendicular to the reddening vector,
i.e., the SGB and the faint parts of the RGB; to get enough reference
stars, however, we had to include the upper part of the MS.  (3) We
draw a fiducial line along the MS, SGB, and RGB using reference stars.
(4) Our actual DR correction makes use of the fact that DR varies
systematically with position in the field.  What we do is to correct
the color of each star by the median reddening of the nearest $n_\ast$
reference stars (see below).  (5) We rotate the corrected values
(i.e., corrected $x$ and unchanged $y$) back into the CMD frame.

A final step in each iteration was to redraw the fiducial sequence in
reaction to the changes in the choice of reference stars. (In
practice, changes after the first iteration are trivial, though.)

For any CMD that we need to correct, we choose $n_\ast$ by trial and
error, as a compromise between a large $n_\ast$ for robustness and a
small one for spatial resolution.  Typical $n_\ast$ values are in the
range 50--70.

The reason we need to iterate the process is to optimize the choice of
the set of reference stars.  As this choice improves, the DR
correction gets better, and the sequences in the corrected CMD get
sharper, allowing a more perceptive choice of reference stars --- and
so on, until the choice of reference stars no longer improves.  It is
important to keep in mind, however, that the DR correction is always
made to the original photometry, but with the new reference stars; it
is the optimization of the reference stars that makes the method work.

We illustrate this procedure with our DR correction of NGC~6441,
several steps of which are shown in Figure \ref{f:redd_ba}.  The first
panel shows the $m_{\rm F814W}$ vs.\ $m_{\rm F390W}-m_{\rm F814W}$ CMD
of the cluster around the SGB region, with our initial choice of
reference stars marked in red, and our initially-drawn fiducial
sequence in green.

The next panel shows the result of the first application of the
procedure; the DR correction was good enough to produce a significant
narrowing of all the sequences in the CMD; moreover, the RGB is now
bimodal, and the MS is showing suggestions of bimodality.  This
improvement allows us to choose a new set of reference stars (marked
in red), from which we have removed stars of the blue MS, and a number
of binaries and field stars.  The green line is the redrawn fiducial
sequence.

As indicated, we iterated this procedure.  Three iterations were
enough to reach our final choice of reference stars and the consequent
DR correction.  (A fourth iteration yielded negligible improvement.)
The final DR-corrected CMD is shown in the right panel of
Fig.~\ref{f:redd_ba}.

Our procedure also allows us to construct reddening maps, by averaging
the DR value used to correct the stars in each region.  (Note that we
will return to this subject in Sect.~\ref{ss:ms}, to present a
specific validation of the NGC~6441 reddening map.)

Finally, we note that from this point on, all the CMDs that we show
are corrected for differential reddening.

\subsection{Proper motions}
\label{ss:pm}

As noted in an earlier subsection, field-star contamination presents a
serious problem in our analysis of the fine structure of the CMDs.
There is no problem for the ACS/HRC and WFPC2 fields, which are at the
cluster centers, where field stars represent a negligible fraction.
The issue arises when we analyze ACS/WFC data (in particular GO-10775)
and the new WFC3 exposures (GO-11739), which were taken off-center to
avoid crowding. Since the GO-10775 and GO-11739 exposures overlap for
both clusters, we were able to measure proper motions, with a baseline
of at least 4 years.  In addition, some of GO-11739 exposures of
NGC~6441 were purposely taken a year apart, in order to measure proper
motions in the parallel ACS/WFC field, where it was clear that
field-star contamination was very serious.  This provides us with up
to 5 years of time baseline between GO-10775 and GO-11739 for NGC~6441
(and of course the 1-year baseline within GO-11739 --- needed for CMDs
that use only that program).

Proper motions were measured using ACS/WFC F606W and F814W exposures
as first epoch and WFC3/UVIS F390W exposures as second
epoch\footnote{The only exception is the proper-motion cleaning of the
  GO-11739 CMD of NGC~6441, using only F390W images, with a 1-year
  time baseline.}.  We did not use WFC3 IR exposures for proper
motions, because of the more than twice-as-large pixel size of the IR
detector and its severe undersampling.

We used the catalogs created by Anderson et
al.\ (2008)\footnote{\texttt{\allowbreak{http://www.astro.ufl.edu/$\sim$ata/public\_hstgc/databases.html}}}
to define a master coordinate frame.  As we did in (e.g.) Bellini et
al.\ (2009a), we first defined for each cluster a reference set of
probable members, on the basis of their position in the $m_{\rm
  F814W}$ vs.\ $m_{\rm F606W}-m_{F814W}$ CMD. These included only
bright, unsaturated stars, which have smaller position errors.

For each star we took each and every pairing of a position measurement
from the first epoch and another from the second epoch, and
transformed each of the two catalog positions into the master frame,
by means of a six-parameter linear transformation using only a local
set of 50 reference stars; we then differenced the two epochs to
derive a displacement for that pairing.  We took a 3$\sigma$-clipped
median of all these displacements of the star, and assigned it a sigma
that was the 68.27 percentile of the residuals around the median.
Since the reference stars are nearly all cluster members, the
displacements of stars are all relative to the mean motion of the
cluster.

After this first estimate of the proper motions, we excluded from our
selection of reference stars those whose motion differed from the mean
by more than 3 sigmas, and re-derived the motions.  After three
iterations the process converged.

Even though the mean motions of both NGC~6388 and NGC~6441 are similar
to the motions of the field stars, the dispersion of internal motion
in each cluster is about five times smaller than that of field stars
($\sim$18 km/s vs.\ 100 km/s). Thus by selecting stars whose motion is
close to the mean motion of the cluster, we exclude the majority of
field stars.

%F5
\begin{figure*}[t!]
\centering
\includegraphics[width=18cm]{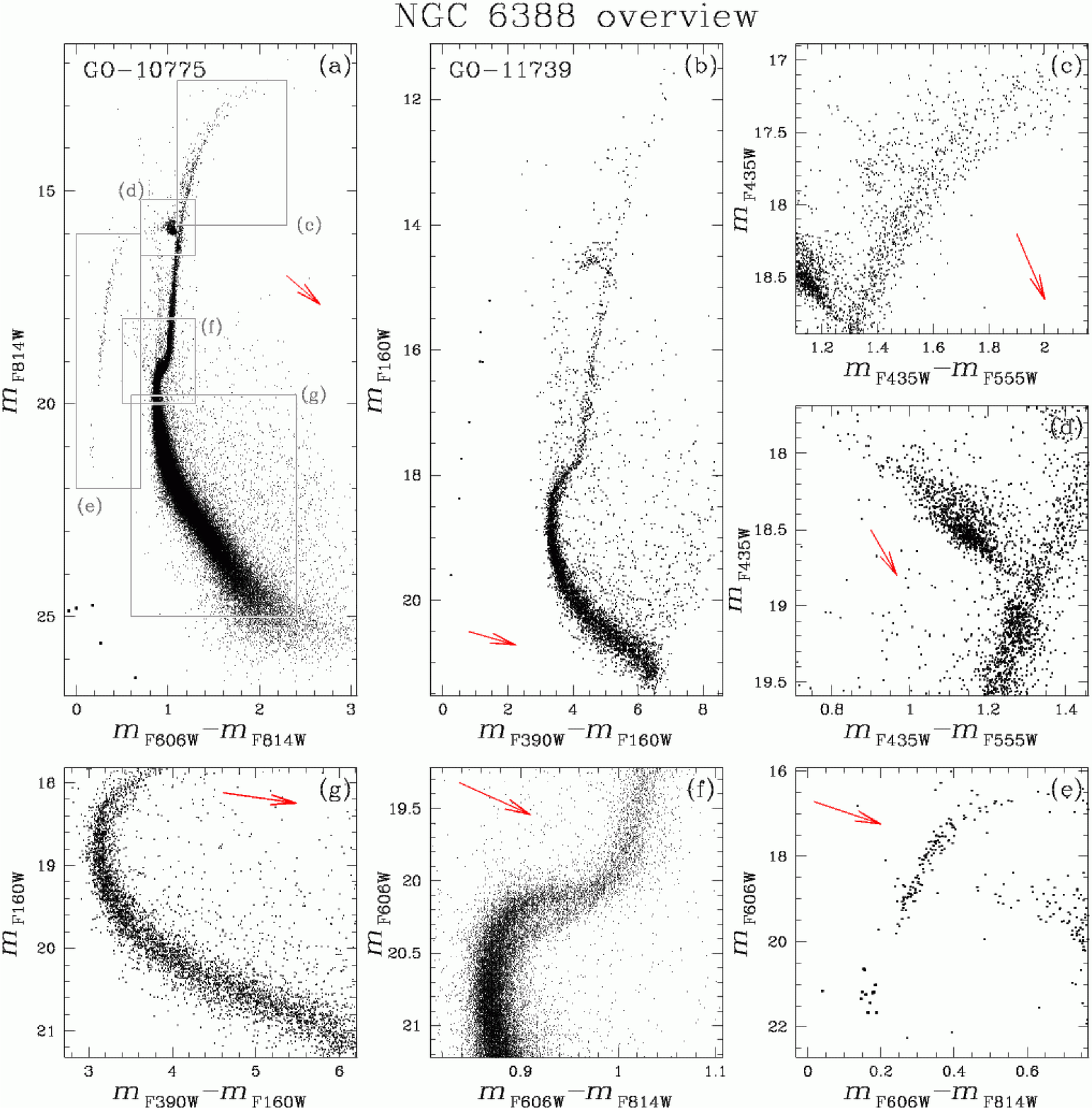}
\caption{Overview of the main features in the CMDs of NGC~6388, from
  different combinations of magnitude and color.  Panel (a) shows a
  CMD from GO-10775 (the most complete CMD), while in (b) is the CMD
  from GO-11739 WFC3 (the CMD with the widest color-baseline). In (a)
  we have highlighted five regions, labeled (c)--(g), around main
  cluster features. These regions are then shown in panels (c) to (g)
  using the filter pairs that best enhance each CMD region, e.g., a
  double RGB in panel (c), or a double SGB in panel (f).}
\label{f:6388_overview}
\end{figure*}

For NGC~6388, the overlap region can be seen in the left panel of
Figure 1, involving the bifurcated red and the bifurcated blue
footprints at the lower left.  In Figure~\ref{f:6388_pm} we illustrate
our proper-motion selection.  Successive panels show the CMD of all
the stars, the separation of cluster and field based on the sizes of
motions, the CMD of the cluster stars, and the CMD of the field stars.
From the last panel it is clear that in CMDs where we do not have
proper-motion separation, the risk of field-star contamination will be
greatest for the SGB and MS regions.  (Note: although we had three
filters available, we have chosen here to display the CMD using the
more familiar F606W--F814W pairing; F390W will come into play in our
ensuing discussion of the details of the sequences of the CMDs.)

Figure~\ref{f:6441_pm} is the analog of Fig.~\ref{f:6388_pm}, for
NGC~6441, in the overlap region of the same two programs (the same
colors and shapes of footprint, near the top of the right-hand panel
of Fig.~1).  In this case there are many more stars, both because of a
larger overlap and also because these exposures are near the cluster
center.  Also, field-star contamination is much more severe than for
NGC~6388, making cluster-field separation for NGC~6441 even more
important.  Nevertheless, our proper-motion separation yields an
almost clean sample of cluster members in panel (c), even though panel
(d) shows that we have lost a few cluster members.

Proper-motion-selected CMDs will be used extensively throughout the
paper.  Note also that when proper-motion measurements were available
we used them to select bona-fide cluster members as reference stars
for improved DR corrections.

\section{Evidence of multiple populations}

In this section we marshal the resources of our cameras and filters to
examine each of the clusters in detail, in order to uncover all the
evidence of multiple populations that we can find.  We begin with an
overview of a carefully chosen array of CMDs of each cluster; then we
take up the regions of the CMD in detail, one by one.

%F6
\begin{figure*}[t!]
\centering
\includegraphics[width=18cm]{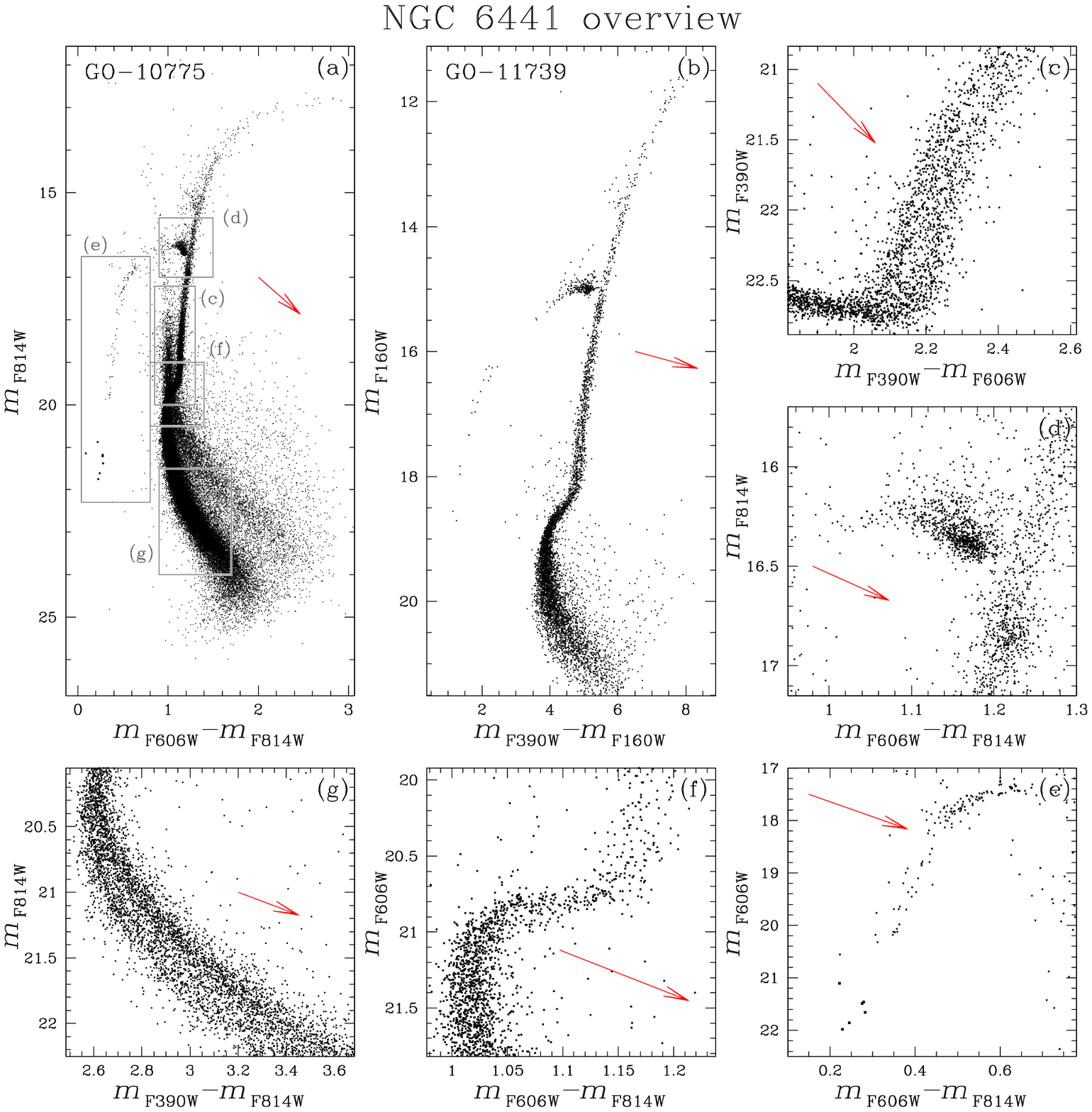}
\caption{Overview of the main features in the CMDs of NGC~6441, as in
  the previous figure for NGC~6388, except that the examples of
  enhancement are now the double RGB in panel (c) and the double MS in
  panel (g).}
\label{f:6441_overview}
\end{figure*}

\subsection{Detailed color-magnitude diagrams of NGC~6388}

Figure~\ref{f:6388_overview} begins with two overall CMDs of NGC~6388.
In panel (a) we show the most familiar form of CMD, using F606W and
F814W from the ACS/WFC images of GO-10775; it extends from the tip of
the RGB down to five magnitudes below the MS turn-off.  (These include
the stars that we showed in the nearly identical CMD of panel 4(a),
but now we include the fainter stars that did not figure in the
proper-motion discussion.)  Panel (b) shows a CMD from F390W and F160W
of the WFC3/UVIS and IR respectively, from GO-11739; there are fewer
stars here, because the smaller WFC3 images are off-center, but this
is our widest color baseline, and this CMD is new.

In panels (c)--(g) we show in detail the five regions that are
outlined in panel (a); for each detail panel, however, we use the
combination of filters that best enhances the features of that part of
the CMD.  We will return to these in later subsections, but even now
we readily see the splits of the RGB in (c) and the SGB in (f) as well
as the complex structure of the red clump of the HB in (d).  (It is
not obvious, but we shall see that panel (g) shows a greater MS width
than photometric errors can account for.  In any case, the
corresponding panel for NGC~6441 will show a clear split.)  Panel (e)
shows the extended blue HB, which has the anomalous blue tail of stars
(highlighted by larger dots).  In each panel the reddening vector is
shown as a red arrow.

In these panels we have plotted only those stars less affected by the
severe crowding in these central fields.  We omitted stars closer to
the center than a magnitude-dependent limiting radius, which we chose
by hand in a way that excludes larger regions for the fainter stars.

Proper-motion selection is not used in any of the panels of this
figure. The purpose of panels (a) and (b) is simply to show the whole
CMD and to define panels (c)--(g).  Panels (c), (d), and (e) show CMD
regions where field-star contamination is minimal, while in (f) and
(g) any gain in sharpness would be greatly outweighed by the loss in
star numbers from cutting back to the small overlap region between the
GO-10775 and GO-11739 images.

\subsection{Detailed color-magnitude diagrams of NGC~6441}

Figure~\ref{f:6441_overview} is an overview of the CMDs of NGC~6441,
similar to Fig.~\ref{f:6388_overview} for NGC~6388.

The CMD from GO-10775 data alone (panel (a)) shows strong
contamination from stars of the Galactic bulge, as already mentioned.
The 1-year time span of the GO-11739 data set allowed us to remove
most of the field stars from the largest-color-baseline CMD shown in
panel (b).  (This is why there is less field contamination here than
in Fig.~\ref{f:6388_overview}b.)  Panels (a) and (b) here have the
same scale as those of Fig.~\ref{f:6388_overview}, so as to allow a
direct visual comparison of the complete CMDs of these two clusters.
(Similarly, in the following subsections we will show a direct
comparison of the different evolutionary sequences of these clusters
using CMDs with the same filters and axis scale.)

Again, in the $m_{\rm F814W}$ vs.\ $m_{\rm F606W}-m_{\rm F814W}$ plane
we highlight five regions, labeled (c) to (g), where the main cluster
CMD features are present.  The CMDs zoomed around these regions are
shown in the other panels of the figure, using the combinations of
filters that best enhance these features, and taking advantage of
proper motions to remove field stars, whenever useful (panels (c),
(d), (f), and (g)).

We did not use proper-motion selections for panel (e).  First, the
blue HB is only marginally contaminated by field stars.  Second,
because proper motions are confined to the overlap area between
GO-10775 and GO-11739, we would have ended up with only half as many
stars, in a region of the CMD that is thinly populated.  We need also
to note that for panel (d), for the RC of the HB, we had to use here
the $m_{\rm F814W}$ vs.\ $m_{\rm F606W}-m_{\rm F814W}$ CMD instead of
the $m_{\rm F435W}$ vs.\ $m_{\rm F435W}-m_{\rm F555W}$ CMD that we
used for NGC~6388, because for NGC~6441 the equivalent WFPC2 F439W
images do not go deep enough.

Unlike NGC~6388, the RGB of NGC~6441 clearly splits at its base when
using the F390W filter (panel (c)).  Furthermore, for the first time
we can see that the MS of NGC~6441 is split into \emph{two distinct
  sequences} (panel (g), more details later).

In the following subsections we will compare and analyze each
evolutionary stage of the two clusters.  We want to emphasize once
again that all the CMDs presented in this work are corrected for
differential reddening effects.

%F7
\begin{figure*}[t!]
\centering
\includegraphics[width=16cm]{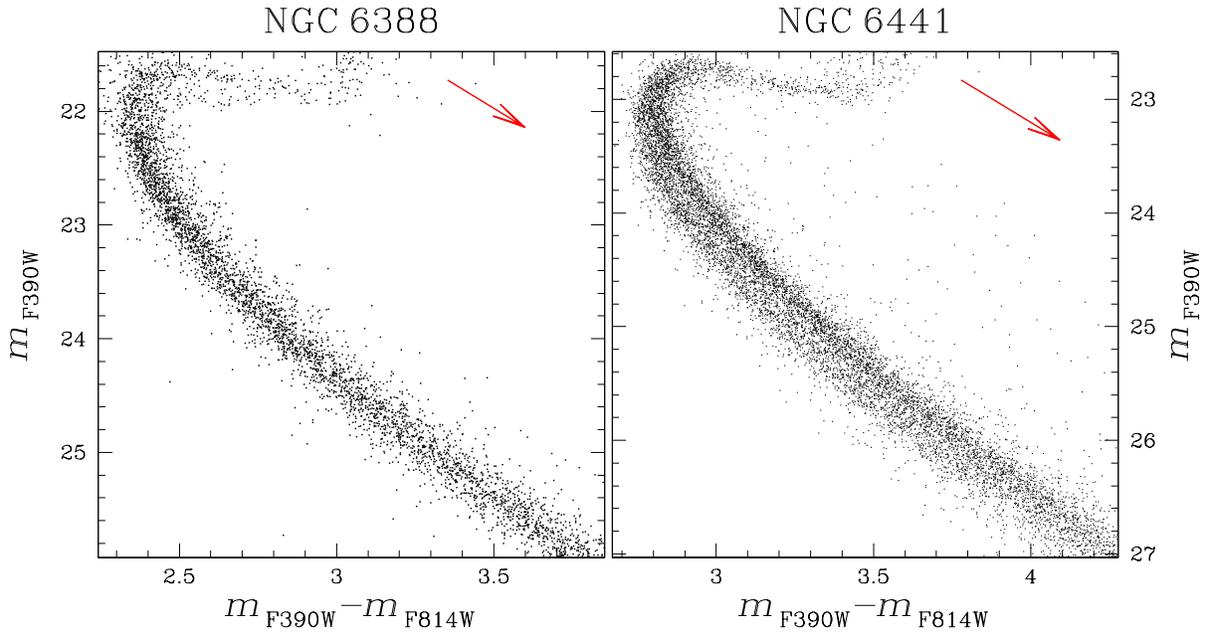}
\caption{$m_{\rm F390W}-m_{\rm F814W}$ CMD of the upper-MS and SGB
  regions of NGC~6388 (left) and NGC~6441 (right), on identical
  scales, contrasting their sequences.  As usual, reddening vectors
  are shown.}
\label{fig:msd}
\end{figure*}

%F8
\begin{figure*}[t!]
\centering
\includegraphics[width=16cm]{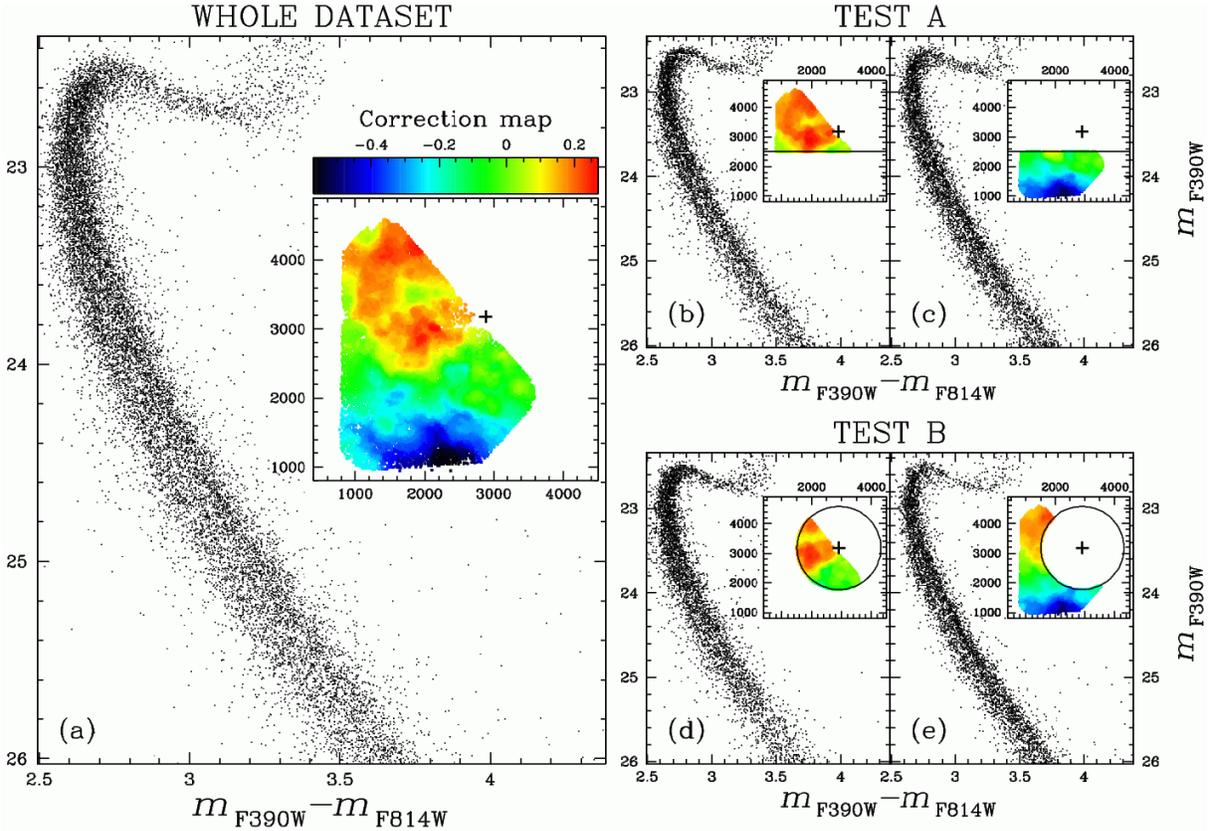}
\caption{(a) CMD of NGC~6441 around its MS and TO; inset shows our
  differential reddening map, with a ``+'' sign at the cluster center.
  Panels (b)--(e) show the results of the two tests described in the
  text, which demonstrate that the MS width cannot be explained by
  differential reddening.  The inset in each of those panels
  identifies the region whose CMD is shown.}
\label{f:redd_test}
\end{figure*}

\subsection{The Main Sequences}
\label{ss:ms}

Figure~\ref{fig:msd} shows the CMDs of NGC~6388 (on the left) and
NGC~6441 (on the right) around the MS region in the $m_{\rm F390W}$
vs.\ $m_{\rm F390W}-m_{\rm F814W}$ plane, adopting the same axis
scale. The reddening vectors are also displayed. It is clear from the
figure that the MS morphology of these ``twin'' clusters is quite
different. The MS of NGC~6388 is wider than expected from
observational errors alone (more in the following), but the CMDs made
using all combinations of filters at our disposal fail to provide any
hints of a MS split.  On the other hand, there seems to be no doubt
that the MS of NGC~6441 is bimodal.

Because of the relatively high reddening values of the two clusters,
one might suspect that the spread/split in the MS of the two clusters
is the result of badly corrected differential reddening.  We note,
however, that the reddening vector runs parallel to the MS, so that
reddening cannot affect the width of the MS, nor should DR correction
affect it either.  This near-parallelism of the reddening vector to
the MS should be kept in mind, as it applies to nearly all CMDs.  It
will be seen, moreover, that a similar parallelism holds for the
reddening vector and two-color diagrams of MS stars for all color
combinations.  Indeed, if we plot an uncorrected CMD of a region
  of NGC~6441 small enough that differential reddening cannot be an
  issue, the MS split is visible.

To further verify that we are not being deceived by serious
systematics in our DR corrected photometry, we performed two specific
tests.  The left panel of Fig.~\ref{f:redd_test} shows the $m_{\rm
  F390W}$ vs.\ $m_{\rm F390W}-m_{\rm F814W}$ CMD of NGC~6441 around
its MS and TO regions, for all stars in common between GO-11739 and
GO-10775 FoVs.  (We chose these two filters because they are the only
ones that clearly split the MS; we are confined to the stars in the
overlap region because the two filters come from those two different
programs.)

The inset shows the color-coded DR map for all the stars, in the
coordinate reference system of the GO-10775 data set, with map
coordinates in ACS/WFC pixels.  The cluster center is marked by a
``$+$''.  Each point corresponds to a star, and is colored according
to the DR correction found for it.  The maximum extent of the DR
corrections is 0.86 mag along the direction of the reddening vector.

Since the DR correction map shows a vertical gradient, for the first
test (TEST A) we split the FoV into upper and lower subsets containing
the same numbers of stars, and performed the DR correction procedure
on each of the two independent subsets, as a test of whether the
apparent breadth of the MS could be just high/low reddening.  The
resulting CMDs, in panels (b) and (c), each show the MS split of panel
(a) --- somewhat more clearly in (c), because the stars of panel (b)
suffer photometrically from the crowding in the cluster center.  In
the less perturbed photometry of panel (c) one can even discern a
dominant red component and a bluer tail in the colors.  But the main
point is that Test A totally refutes the idea that the bluer and
redder parts of the MS spread could simply be stars with greater and
less reddening.

To investigate whether the MS spread/split might in fact result
directly from crowding in the cluster center, we performed TEST B, in
which the two halves of the sample were an inner one and an outer one,
as shown in panels (c) and (d).  The MS width is still evident in both
samples, and the outer sample, in panel (d), if anything shows the
split even more strongly than panel (c), above it.

These tests show that the MS spread of NGC~6441 is real. Moreover,
when only less-crowded, well-measured stars are selected (e.g., right
panel of Fig.~\ref{fig:msd}, but see also panel (a) of
Fig.~\ref{f:redd_test}), the MS splits into two branches, the redder
one being the more populated.

No MS split was visible in the case of NGC~6388, but the tests just
performed can be taken as evidence that the MS broadening observed
there is also real, rather than an artifact of differential reddening.

%F9
\begin{figure*}[t!]
\centering
\includegraphics[height=7.65cm]{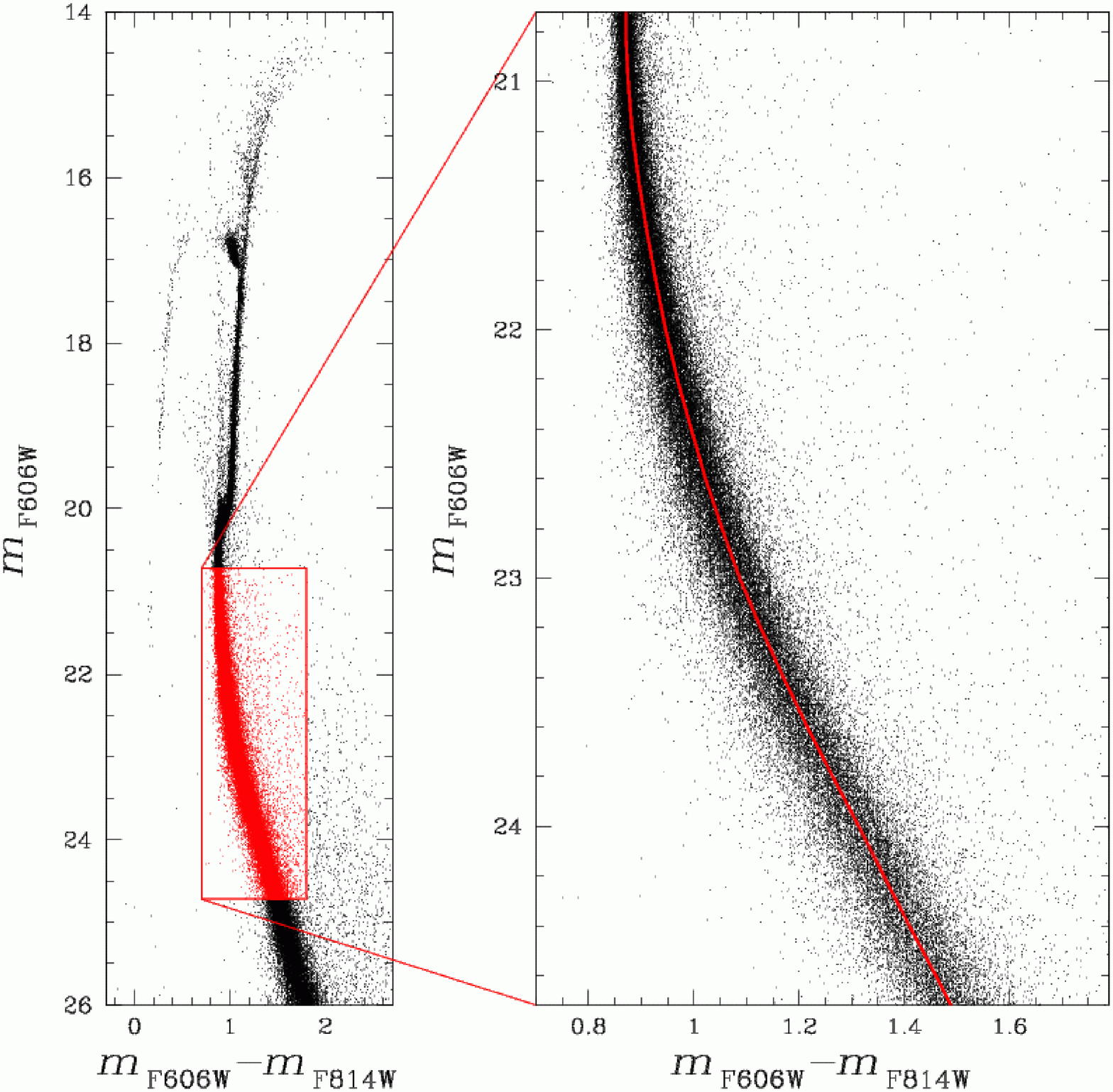}
\includegraphics[height=7.65cm]{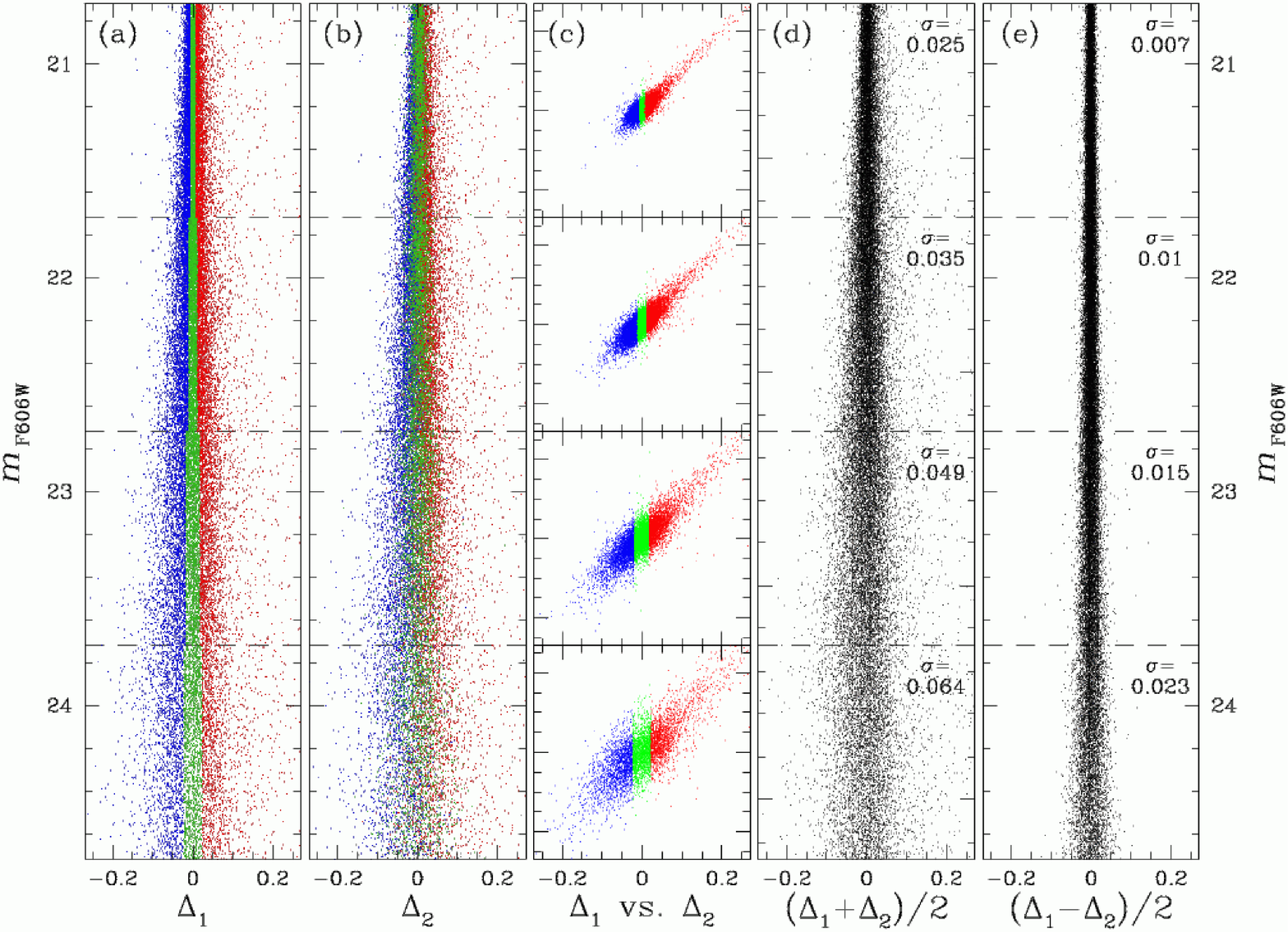}
\caption{The two leftmost panels show the GO-10775 $m_{\rm F606W}$
  vs.\ $m_{\rm F606W}-m_{\rm F814W}$ CMD of NGC~6388 and a zoom-in of
  the upper MS.  The fiducial line is in red in the second panel.  The
  remainder of the figure shows the complicated procedure that is
  described in the text, in which we divide the images into two sets
  and show that the stars show similar color residuals in the two
  independent data sets.}
\label{f:6388_ms_ata}
\end{figure*}

%F10
\begin{figure*}[t!]
\centering 
\includegraphics[height=7.65cm]{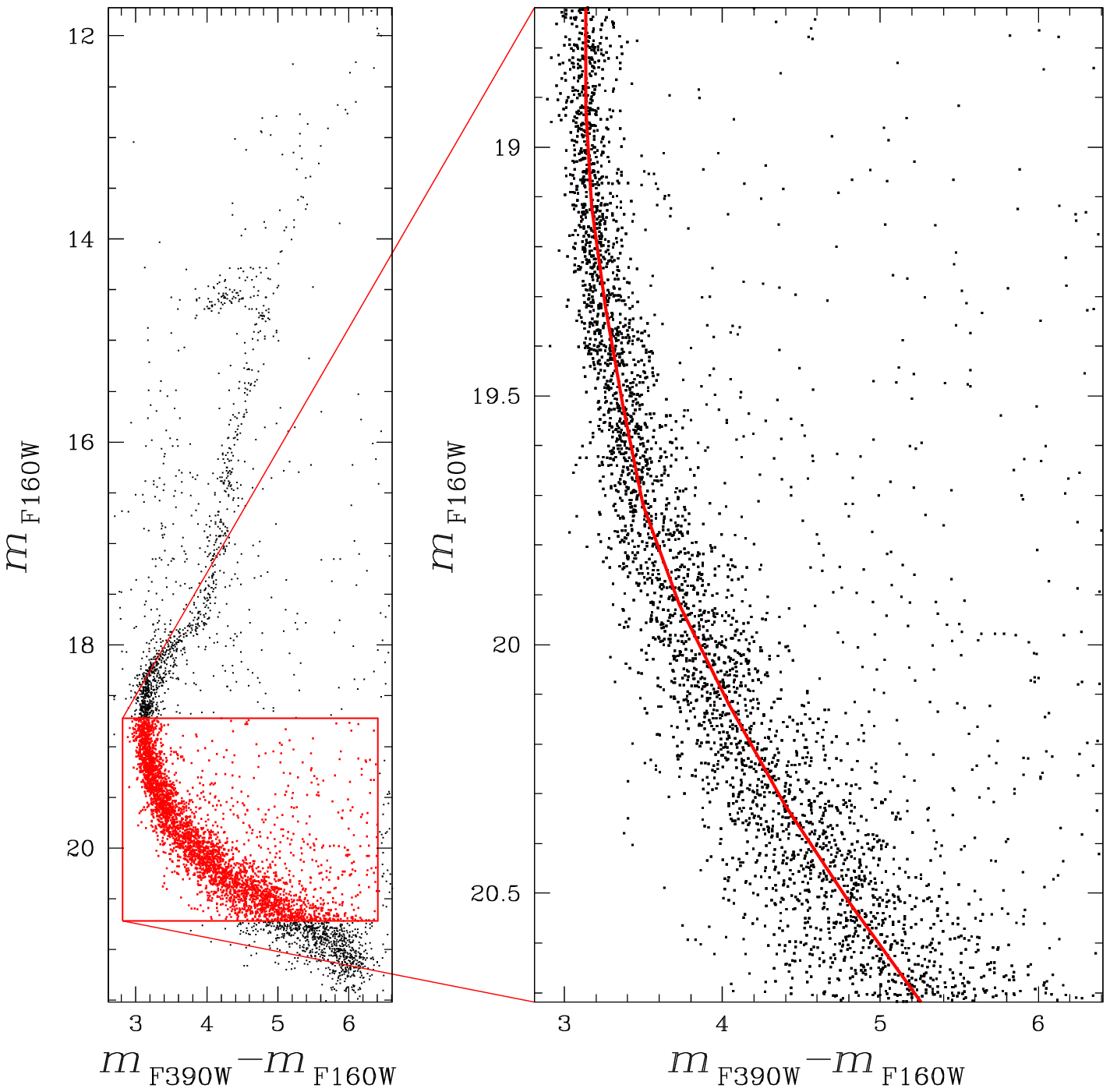}
\includegraphics[height=7.65cm]{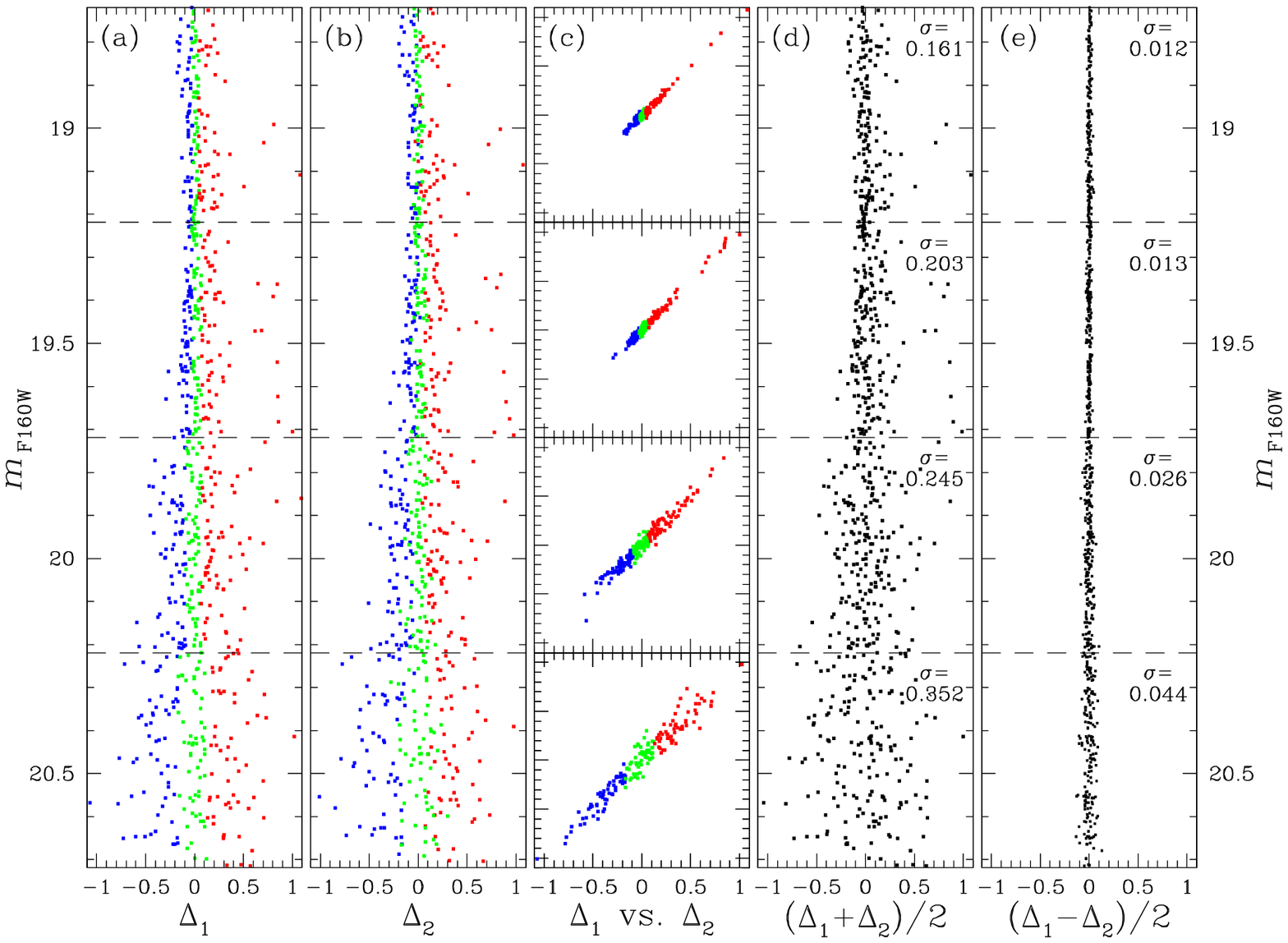}
\caption{Same as Fig.~\ref{f:6388_ms_ata}, but using $m_{\rm F160W}$
  vs.\ $m_{\rm F390W}-m_{\rm F160W}$.}
\label{f:6388_ms_p8}
\end{figure*}

%F11
\begin{figure*}[t!]
\centering
\includegraphics[width=18cm,height=8.5cm]{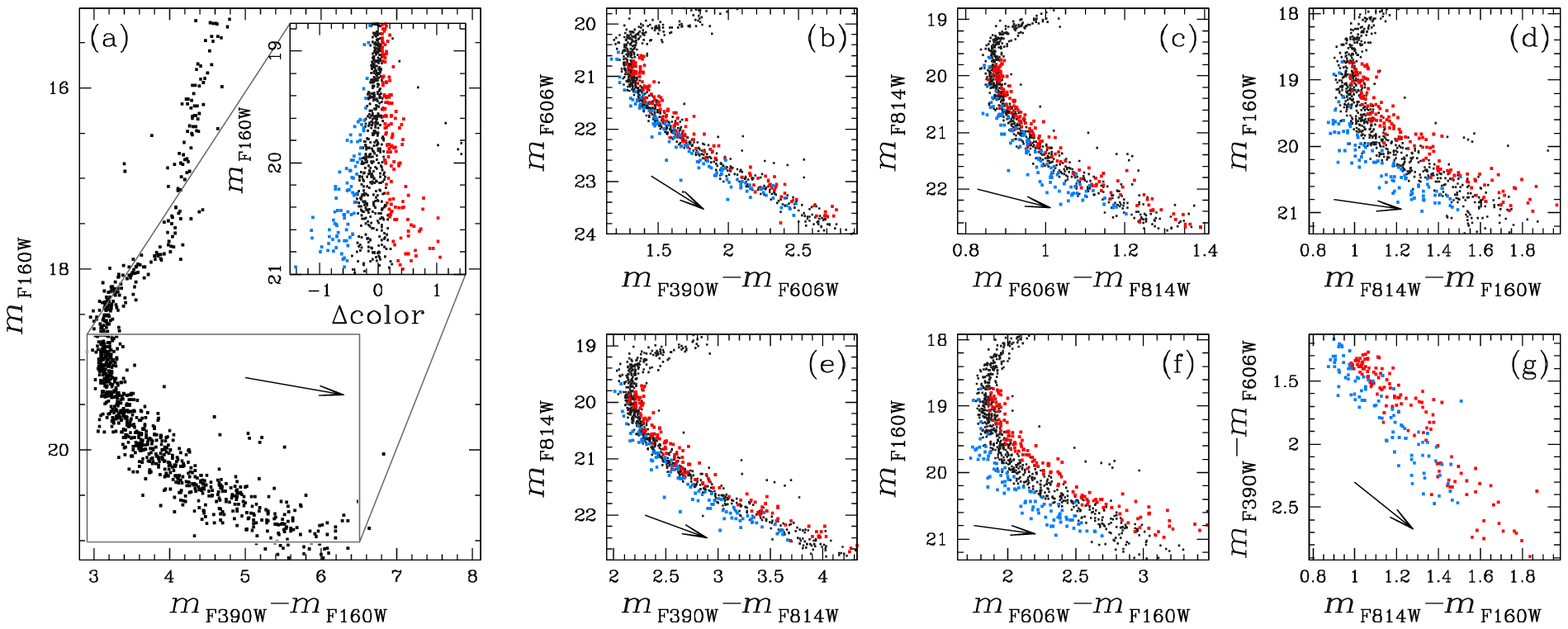}
\caption{Multicolor analysis for the MS of NGC~6388. Panel (a):\ the
  $m_{\rm F160W}$ vs.\ $m_{\rm F390W}-m_{\rm F160W}$ CMD. The inset
  shows the rectified CMD, with blue, black, and red stars plotted in
  those colors.  Panels (b) through (f) show CMDs using different
  filter combinations; the stars keep the same color coding; panel (g)
  is the $m_{\rm F390W}-m_{\rm F606W}$ vs.\ $m_{\rm F814W}-m_{\rm
    F160W}$ two-color diagram, showing only the blue- and red-coded
  stars. The two extreme groups remain distinct in all panels.  Each
  panel includes a reddening vector.}
\label{f:6388_mca}
\end{figure*}

\subsubsection{The MS of NGC~6388}

As shown in the study by Anderson et al.\ (2009) of the MS broadening
in 47~Tuc, and by Milone et al.\ (2010b) for NGC~6752, an effective
way of testing for a true CMD-sequence broadening is to divide the
images of the field of stars into two independent halves.  If the
broadening is intrinsic, the stars that have redder (or bluer) colors
in a CMD made from one half of the images will have redder (or bluer)
colors in the CMD from the other half. But if the broadening is due
only to measuring errors, a star that is redder in the first half of
the images will have an equal chance of being redder or bluer in the
second half, and the bluer stars of the first subset are equally
likely to be red or blue in the second half.  Again, we note that the
reddening vector is almost parallel to the MS (see, e.g., panel (a) of
Fig.~\ref{f:6388_overview}), and therefore DR cannot change the
locations of the stars with respect to the MS fiducial line.  In the
following, we will apply this technique, basically following the
procedure described in Anderson et al.\ (2009).

The leftmost panel of Fig.~\ref{f:6388_ms_ata} shows the $m_{\rm
  F606W}$ vs. $m_{\rm F606W}-m_{\rm F814W}$ CMD of NGC~6388.  The
second panel shows a zoom-in of the upper part of the MS (colored in
red in the first panel). The continuous curve in red marks the adopted
MS ridge line.  We subtracted from the color of each star that of the
MS ridge line at the same magnitude level.  The rectified MS obtained
in this way, for the two independent set of images, is shown in panels
(a) and (b).  In panel (a) the stars were divided into three
equally-populated groups, color-coded in red, green and blue.  In
panel (b) we have kept for each star the same color that it had in
panel (a).  (We decided to divide the MS into three groups rather than
two, as we had done in our previous papers, to better understand the
color behavior of the stars in the two subset of images.)

It is clear that the color of stars is maintained very well (with only
a small scatter due to photometric errors) from panel (a) to panel
(b). To emphasize this fact, we show in panel (c) the correlation
between the color of each star in the two halves of the data, divided
in four different magnitude bins (since measuring errors increase at
fainter magnitudes).  A star redder than the MS ridge line in the
first half of the data is also measured as redder in the great
majority of cases, in the second half of the data, and
vice-versa. This is the mark of a true spread in color.

In panel (d) we show the color distribution of the rectified MS as
measured from the whole set of images, $(\Delta_1+\Delta_2)/2$.  The
distribution of half the difference between the colors measured
separately from each half of the images, $(\Delta_1-\Delta_2)/2$, is
presented in panel (e).  This distribution is a statistical estimator
of the error in the color distribution in panel (d).

At each magnitude level of panels (d) and (e), an estimate of the
$\sigma$ of the spread of the distribution is given (calculated here
as the 68.27 percentile of the absolute deviation from the median).
As one can see, the intrinsic width of the MS is about three times as
large as what would be expected from photometric errors alone.  
  (As a further test, artificial stars within the magnitude interval
  shown in Fig.\ 9, and chosen to have $m_{\rm F606W}-m_{\rm F814W}$
  colors along the MS of NGC~6388, gave MS widths ranging from 0.005
  to 0.027 --- only a small fraction of the width observed in
  Fig.~9.)

A similar procedure applied to the $m_{\rm F390W}-m_{\rm F814W}$ color
yields a 1$\sigma$ width of the MS of 0.09 mag, at 1.1 mag below the
turnoff (after subtracting in quadrature the photometric errors of
$\sim 0.03$ mag).

The presence of an intrinsic breadth in the MS in NGC~6388 is further
confirmed when we apply the technique of Anderson et al.\ (2009) to
the GO-11739 data set (Fig.~\ref{f:6388_ms_p8}).  In this case, the
color baseline is six times larger, and the MS appears to be about
\textit{ten times wider} than what is expected from photometric errors
alone.

The GO-10775 and GO-11739 fields overlap marginally (only about a
thousand stars with a measure in all the 4 available filters), but
this is enough to show, in Fig.~\ref{f:6388_mca}, a multicolor
analysis, similar to what was done for $\omega$~Centauri by Bellini et
al.\ (2010), 47~Tuc by Milone et al.\ (2012b), and NGC~6397 by Milone
et al.\ (2012c).  Moreover, here we can use proper motions to
eliminate field stars --- much more important here than in the sparse
fields of Milone et al.

We rectified the MS in the $m_{\rm F160W}$ vs. $m_{\rm F390W}-m_{\rm
  F160W}$ plane, in the magnitude interval $18.75<m_{\rm F160W}<21$
(see panel (a) and the inset of Fig.~\ref{f:6388_mca}). Note that at
$m_{\rm F160W}>20.5$ the MS is already as wide as about 1 magnitude in
the proper-motion-selected CMD.  We defined three groups of stars
(marked with blue, black, and red colors), with the property of being
bluer, of average color, and redder, respectively, than an arbitrary
line drawn through the middle color of the MS (not shown in the
Figure).  We then kept the same colors for the stars of these two
groups in plotting different CMDs (panels (b) through (f)), made by
using all other available color combinations.  A final panel (g) shows
the two-color diagram $m_{\rm F390W}-m_{\rm F606W}$ vs.\ $m_{\rm
  F814W}-m_{\rm F160W}$ for the two extreme groups of stars.  It is
interesting to note that blue and red stars always remain on the blue
and on the red side of the MS, respectively.  The reddening vectors
are drawn in order to emphasize that they are approximately, or even
closely, parallel to the faint MS in the CMDs and, most importantly,
in the two-color diagram.  Thus differential reddening has little or
no effect on the MS width or on the separation of the red and blue
points.
  
We have demonstrated that the MS of NGC~6388 shows a significant color
spread. This can be seen both by using optical colors and, more
importantly, by using the wide color baseline offered by the new
GO-11739 F390W and F160W images.  Even so, our careful investigation
was unable to reveal any hint of MS splitting with the data sets at
our disposal.

%F12
\begin{figure*}[t!]
\centering \includegraphics[width=18cm]{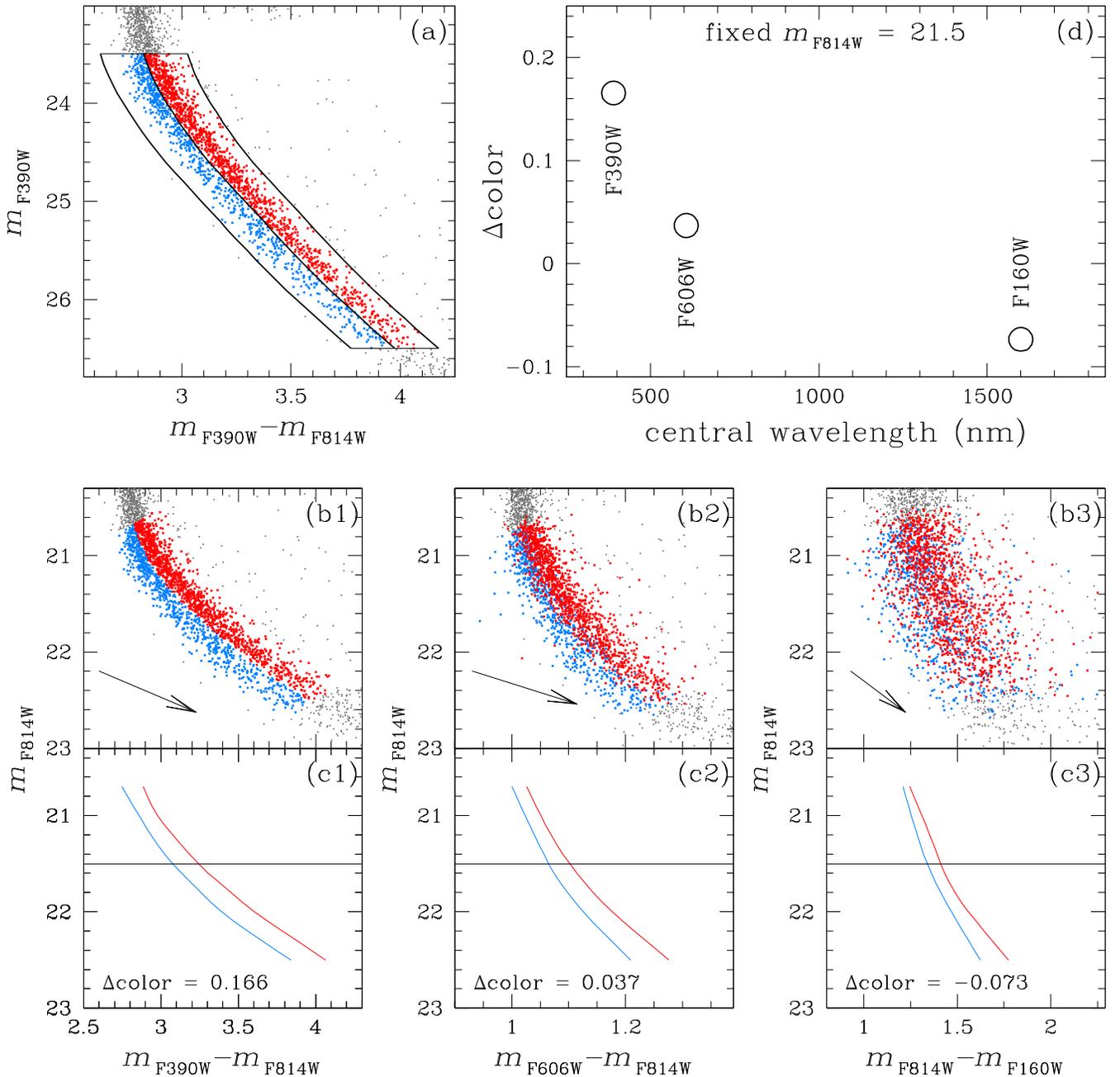}
\caption{Multicolor analysis of the MS of NGC~6441. The two MS
  populations are selected in panel (a).  Panels (b) and (c) show the
  behavior of the sequences and their fiducials in the planes $m_{\rm
    F814W}$ vs.\ $m_{\rm X}-m_{\rm F814W}$, where X is one of the
  other filters.  The color differences between the two sequences at
  fixed $m_{\rm F814W}=21.5$ are plotted in panel (d).}
\label{f:6441_mca}
\end{figure*}

\subsubsection{The MS of NGC~6441}
\label{sss:ms_6441}

In NGC~6388 we were able only to demonstrate that the MS is broad,
without managing to reveal possible MS subpopulations.  In NGC~6441,
by contrast, the MS components are visibly split, and we can therefore
study their behavior in a much more straightforward way.  What is
more, GO-10775 and GO-11739 overlap by about half the ACS/WFC field
--- much more than in the case of NGC~6388 --- and this makes for much
better statistics.  What we do for NGC~6441, then, is to perform a
multicolor analysis in order to model the difference in the chemical
content of the two MS populations, as we did in 47~Tuc (Milone et
al.\ 2012b) and NGC~6397 (Milone et al.\ 2012c).

We drew a fiducial line by hand passing in between the two MSs, and
defined as rMS stars those within a 0.2-mag interval of color redward
from this fiducial, and as bMS stars those within 0.2 mag blueward
from it, in the magnitude interval $23.5<m_{\rm F390W}<26.5$ (see
panel (a)) of Fig.~\ref{f:6441_mca}).

Panels (b) show the $m_{\rm F814W}$ vs.\ $m_{\rm X}-m_{\rm F814W}$
CMDs of the two MSs, where X is one of the other three filters (F390W,
F606W, and F160W). In panels (c) we plotted the location of the ridge
lines of the two MSs in the same planes as panels (b).  We measured
the color difference $\Delta$color between the blue and the red ridge
lines at a fixed magnitude level ($m_{\rm F814W}=21.5$, black
horizontal lines).  Panel (d) collects these color differences as a
function of the central wavelength $\lambda$ of the $m_{\rm X}$ band
(open circles).

The color difference between the two sequences depends strongly on the
adopted color baseline.  In particular, it is highest in the $m_{\rm
  F390W}-m_{\rm F814W}$ color, where it amounts to $\sim 0.166$ mag
(1.1 mag below the turnoff). This compares to 0.09 mag as the width of
the MS of NGC~6388 and we conclude that the MSs of the two clusters
are intrinsically different, with the MS of NGC~6388 being narrower
than the separation of the two MSs of NGC~6441.  We will demonstrate
in Section~\ref{s:discussion} that:
 \begin{itemize}
\item The color spread of the MSs of the two clusters is similar in
  $m_{\rm F606W}-m_{\rm F814W}$, implying that the difference in He
  content among the different stellar populations in the two clusters
  must be similar (as strongly suggested by the morphology of their
  HBs);
\item Other chemical differences (which would be C, N, and O) must be
  invoked to explain the sharp difference of the two MSs in the
  $m_{\rm F814W}$ vs.  $m_{\rm F390W}-m_{\rm F814W}$ CMD.
\end{itemize}

%F13
\begin{figure*}[t!]
\centering
\includegraphics[width=17.0cm]{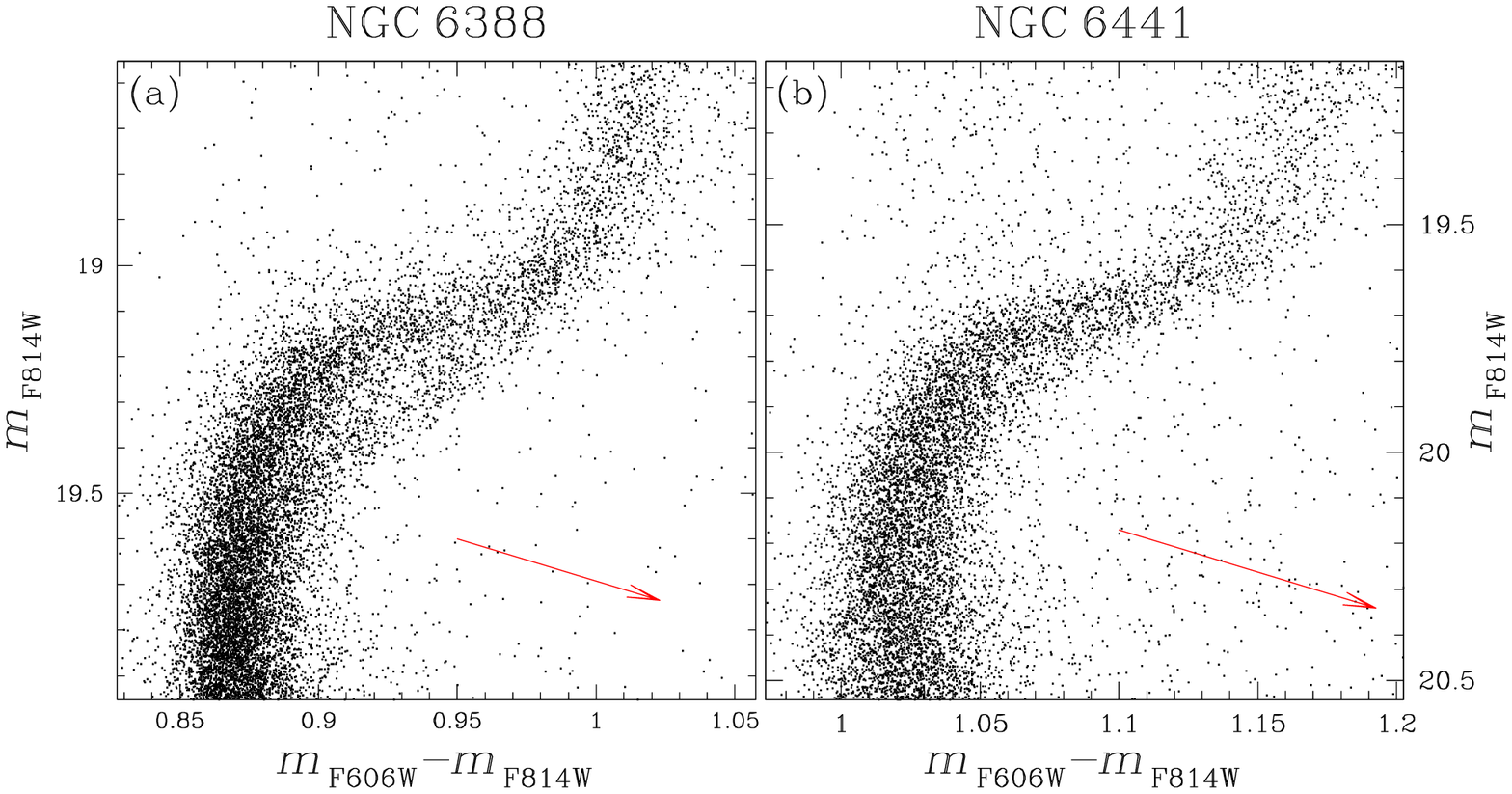}\\
\includegraphics[width=17.0cm]{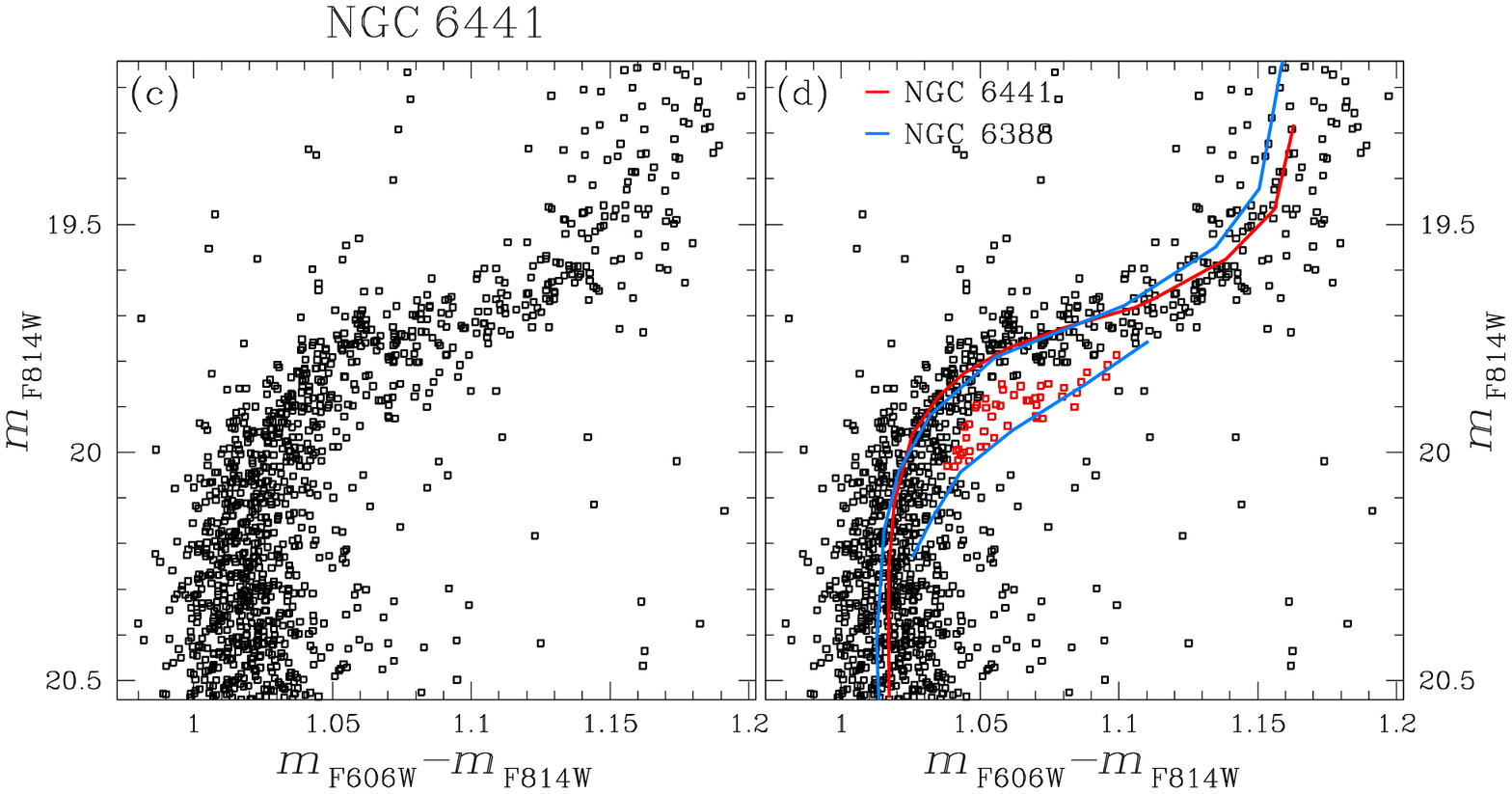}
\caption{(a) and (b):\ The SGB region of the CMDs of NGC~6388 (left)
  and NGC~6441 (right), showing all stars from GO-10775.  (c):
  Proper-motion-selected members of NGC~6441, from the overlap region
  of GO-10775 and GO-11739.  Panel (d) is a replica of panel (c), but
  with the ridge lines of the MS, the bright part of the SGB, and the
  lower RGB of NGC~6441 marked in red, and those of NGC~6388
  over-plotted in blue---the latter shifted to fit the stars of
  NGC~6441.  The stars that are colored red are the fSGB that is
  discussed in the text.}
\label{f:sgb_diff}
\end{figure*}

%F14
\begin{figure*}[t!]
\centering
\includegraphics[width=18cm]{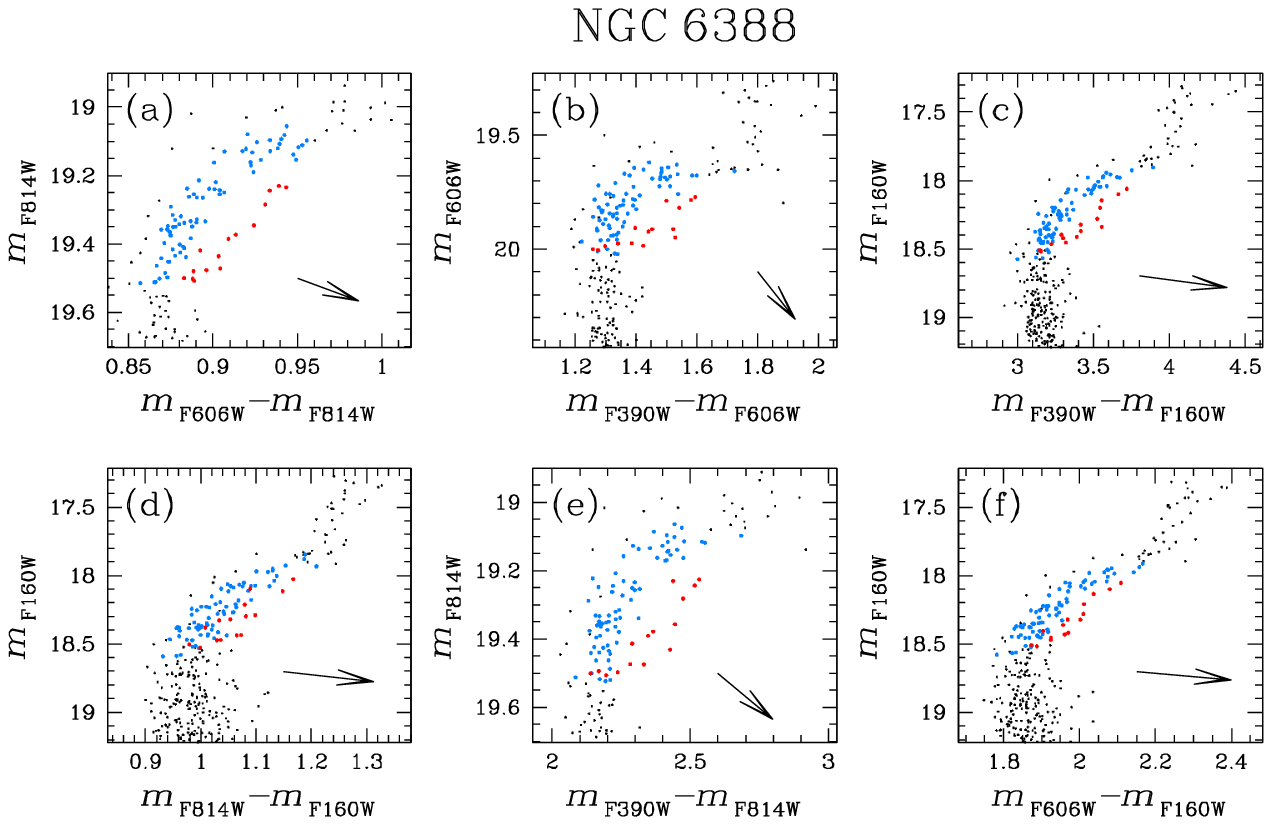}
\caption{(a): $m_{\rm F814W}$ vs.\ $m_{\rm F606W}-m_{\rm F814W}$ CMD
  of the SGB region of NGC~6388 for proper-motion-selected stars in
  the overlap between GO-10775 and GO-11739, with bSGB and fSGB stars
  colored azure and red, respectively. The remaining panels show the
  other color combinations that are available, with each star
  retaining the color that it was assigned in panel (a).  The bSGB and
  fSGB stars remain well separated in all panels. Reddening vectors
  are also shown.}
\label{f:sgb_6388_mca}
\end{figure*}

%F15
\begin{figure*}[t!]
\centering
\includegraphics[width=18cm]{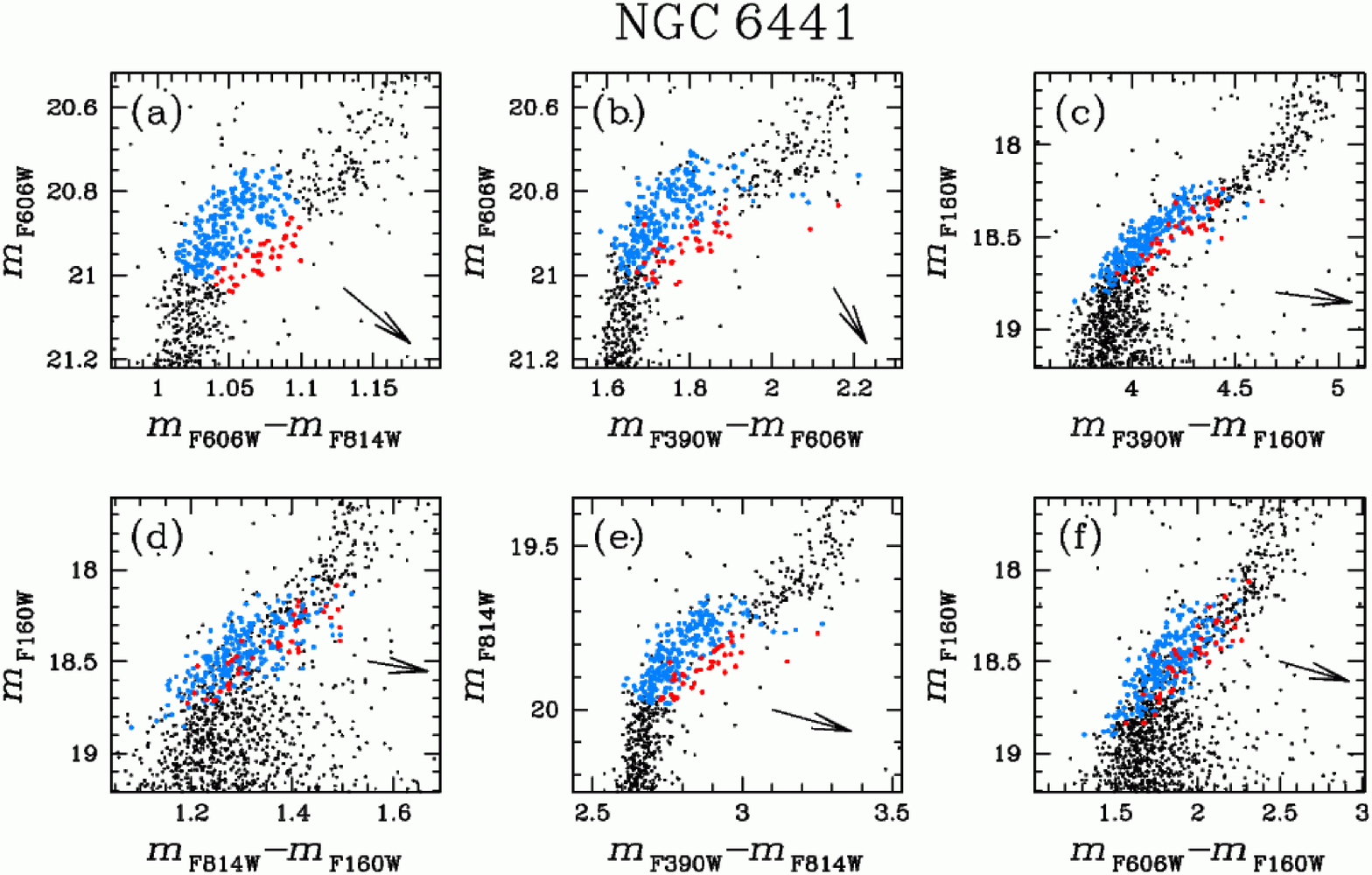}
\caption{Same as Fig.~\ref{f:sgb_6388_mca}, but for the SGB of
  NGC~6441.  (There are more stars because the overlap between
  GO-10775 and in GO-11739 is larger for NGC~6441.)}
\label{f:sgb_6441_mca}
\end{figure*}

\subsection{The Sub-Giant Branches}

The top panels of Fig.~\ref{f:sgb_diff} show the CMDs of NGC~6388 and
NGC~6441 around the SGB region in the $m_{\rm F814W}$ vs.\ $m_{\rm
  F606W}-m_{\rm F814W}$ plane, both with the same axis scale. All the
stars plotted come only from GO-10775.  While the SGB split in
NGC~6388 is clear, field-star contamination prevents us from analyzing
the SGB morphology of NGC~6441 in detail.  Nevertheless, it is clear
that the overall width of the SGB of NGC~6388 is decidedly larger than
that of NGC~6441.

Thanks to the overlap between the fields of view of GO-10775 and
GO-11739, we can take advantage of the $\sim$5-year time baseline
between the two and use proper motions to isolate probable members of
NGC~6441; panel (c) shows the resulting CMD.  Now that field-star
contamination is no longer an issue, there seems to be a second,
fainter and less populated SGB even in NGC~6441 (as could already be
suspected in panel (f) of Fig.~\ref{f:6441_overview}).

Panel (d) is a replica of panel (c) on which we have drawn (in red)
the ridge line of the MS, SGB, and RGB as follows. We divided the
cluster stars in the figure into 0.15-mag bins (0.4 mag for RGB stars)
and derived 3$\sigma$-clipped median values of magnitude and color for
each bin.  We selected only the brighter SGB for this purpose, while
stars that seemed to belong to a fainter SGB have been colored in red.
Using the same technique, we derived in a similar way for NGC~6388 the
ridge lines for the MS, SGB (both components), and RGB, and drew them
in blue. We manually shifted the main-component ridge line of NGC~6388
to best overlap the MS and turn-off regions of NGC~6441.

The SGB split of NGC~6441 is much less than that of NGC~6388.  This
means that, if a difference in C+N+O is the cause for the SGB split in
NGC~6441 and NGC~6388 (as suggested for the case of NGC~1851 by, e.g.,
Cassisi et al.\ 2008), that difference must be smaller for NGC~6441
than for NGC~6388.

We performed a multicolor analysis on the SGB region of the clusters.
A close-up view of the SGB of NGC~6388 is shown in panel (a) of
Fig.~\ref{f:sgb_6388_mca}, for proper-motion-selected stars in common
between GO-10775 and GO-11739.  The blue and red colors are used to
define the bright and the faint SGB components, hereafter bSGB and
fSGB, respectively (i.e., in the remaining panels each star will keep
the color that it is assigned here).  Panels (b) through (f) show the
other five CMDs made with the remaining color combinations at our
disposal (six total combinations with four filters). The reddening
vector is displayed in all panels. Bright and faint SGB stars are
color-coded as in panel (a). In all panels the bSGB is brighter than
the fSGB, and the two components remain well separated.

Similarly, Fig.~\ref{f:sgb_6441_mca} shows the multicolor analysis for
the SGB of NGC~6441.  The bright and faint SGB components for NGC~6441
are selected in the $m_{\rm F606W}$ vs.\ $m_{\rm F606W}-m_{\rm F814W}$
CMD (panel (a)), and again we plot only proper-motion-selected stars.

%F16
\begin{figure*}[t!]
\centering
\includegraphics[width=17cm]{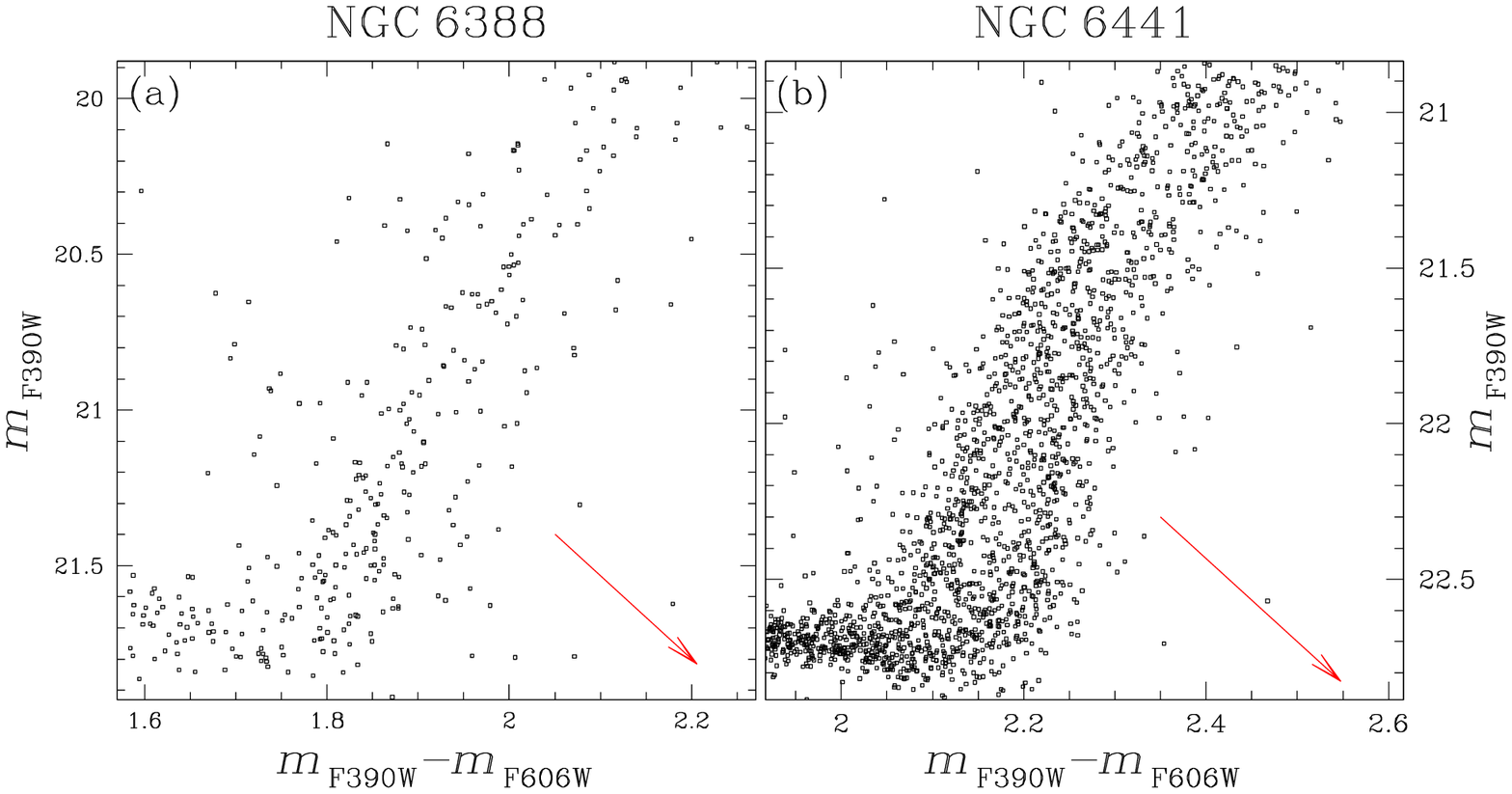}
\includegraphics[width=17cm]{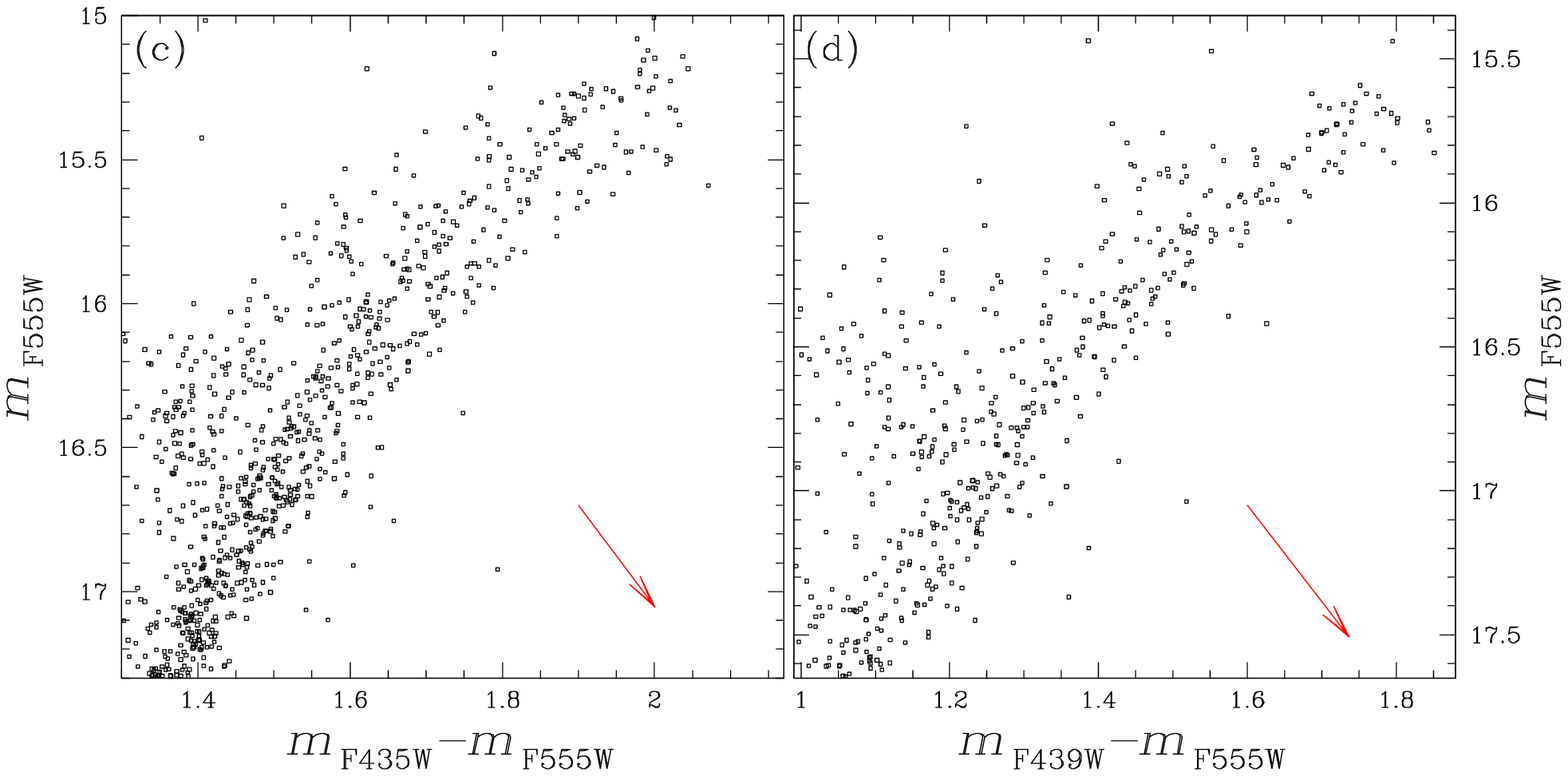}
\caption{RGBs, plotted on the same scale to facilitate comparison.
  NGC~6388 is on the left and NGC~6441 on the right.  Top and bottom
  show different color baselines. (F435W and F439W are equivalent
  filters in two different cameras.)  Because the CMDs come from
  overlap fields of different size, NGC~6388 has fewer stars than
  NGC~6441 in the top row and more in the bottom row.}
\label{f:rgb_diff}
\end{figure*}

Unlike the case of NGC~6388, the two SGB components of NGC~6441 do not
remain distinct in all CMDs, but the bSGB stars again stay, on
average, brighter than the fSGB stars.  The separation of the two
groups, however, is much less obvious than for NGC~6388.

A deeper analysis of the SGBs of NGC~6388 and other globular clusters
is presented in a separate paper (Piotto et al.\ 2012).

\subsection{The Red-Giant Branches}
\label{ss:rgb_diff}

%F17
\begin{figure*}[t!]
\centering
\includegraphics[width=18cm]{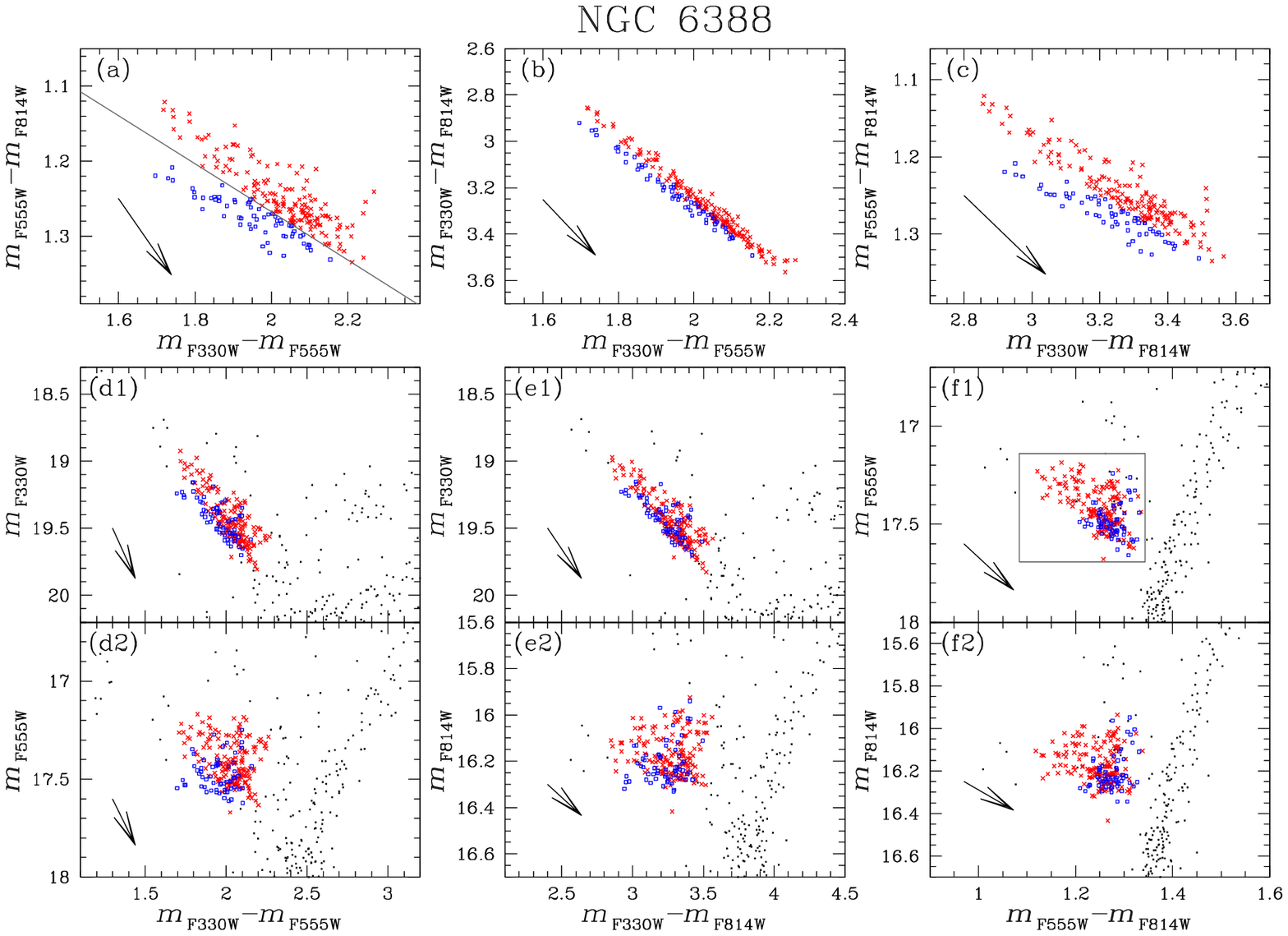}
\caption{Two-color diagrams (top panels) and CMDs (middle and bottom
  panels) of the red clump of NGC~6388. The RC sample is defined in
  panel (f1).  The selection of the main and secondary RC components
  (mRC and sRC, respectively in red and blue) is shown in panel (a).}
\label{f:hb:6388}
\end{figure*}

%F18
\begin{figure*}[t!]
\centering
\includegraphics[width=18cm]{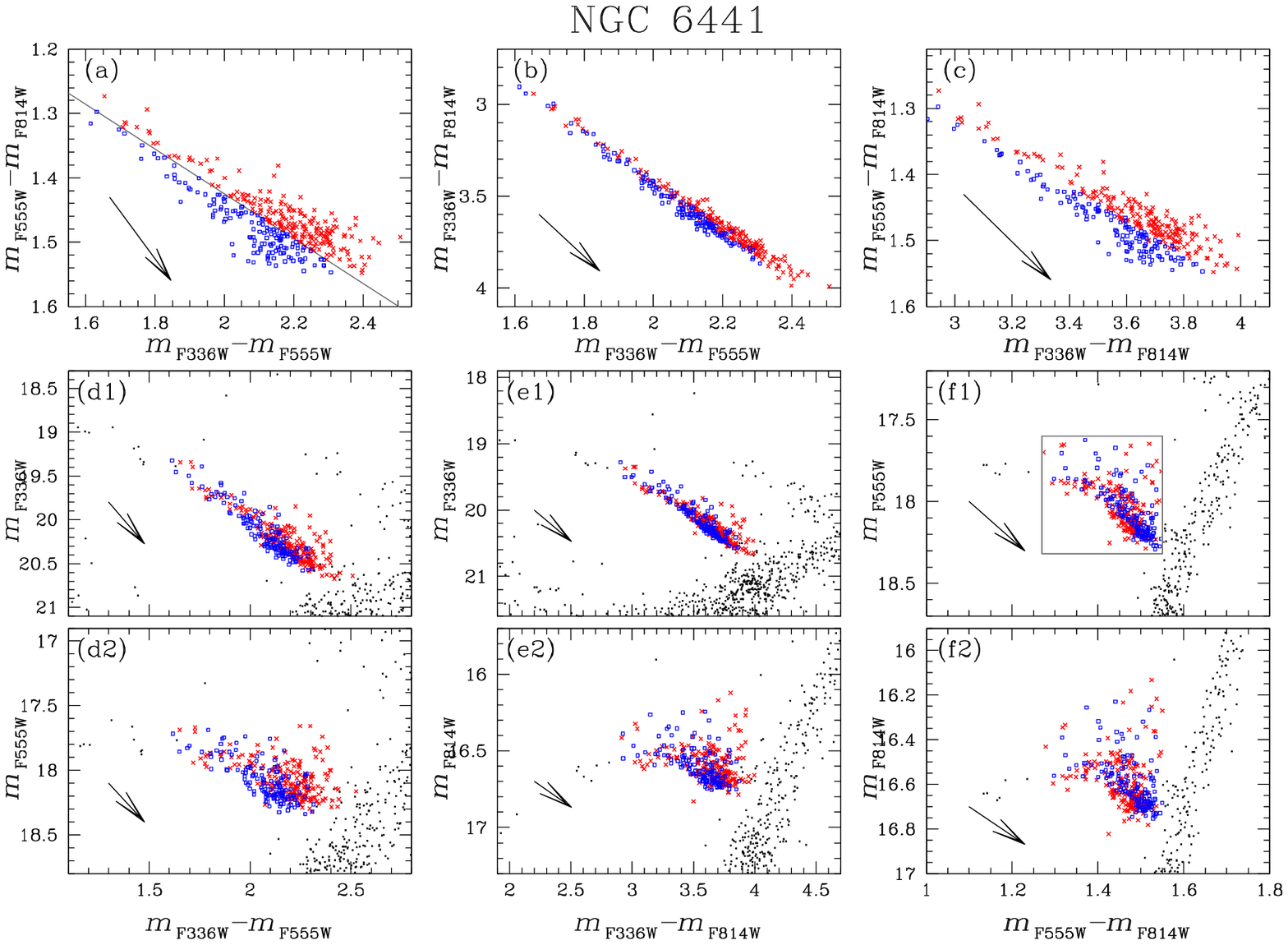}
\caption{Similar to Fig.~\ref{f:hb:6388}, but for NGC~6441.}
\label{f:hb:6441}
\end{figure*}

Both NGC~6388 and NGC~6441 have a split RGB, but the splits are
different, as shown in Figure~\ref{f:rgb_diff}, in which we see CMDs
of the two clusters side by side, with two different color baselines,
top and bottom.  Unfortunately, in CMDs drawn from quite disparate
sources the numbers of stars are not comparable; moreover, the
presence of the asymptotic giant branch (AGB) next to the RGB confuses
the picture further.  Nevertheless, we shall show how to use these
CMDs to trace where the brighter and fainter RGB sequences (bRGB and
fRGB) run in each cluster.

In panel (c) of Fig.~\ref{f:rgb_diff} a split in the RGB of NGC~6388
is evident for stars brighter than $m_{\rm F555W}=16.2$.  We can
exclude the possibility that the brighter RGB sequence might be
composed of AGB stars, since the latter are bluer and well separated
from the RGB.  Panel (a) seems to be suggesting a separation for the
entire length of the RGB, but it is hard to draw definite conclusions
from so few stars.

For NGC~6441 panel (b) shows a clear split in the faint part of the
RGB, but the split becomes less convincing at brighter magnitudes.
Panel (d) offers weak confirmation of a split at the bright end, but
the star numbers are just too small to be at all sure.

Interestingly, RGB splits are usually found only in ultraviolet
photometry (or else in Str\"omgren filters that are sensitive to
molecular bands see, e.g., Gratton et al.\ 2012, and
  references therein, for a review); panels (a) and (b) are in
agreement with this.  By contrast, the cluster M~4, for example, has a
very narrow RGB in the $B-I$ color (Marino et al.\ 2008), but is
broad, or even split, in $U-B$.  Thus the apparent split of the RGB of
NGC~6388 in an optical color (panel (c)) comes as a surprise,
  and may be suggesting an additional population phenomenon in
NGC~6388 and NGC~6441, different from that at work in M~4.
Unfortunately, all the stars for which an abundance analysis
  have been performed (e.g., Carretta et al.\ (2007), lie at radial
  distances outside our FoV, and therefore we cannot link a given RGB
  branch to a particular Na-O family.

In conclusion, the RGBs of these two clusters show evidence of splits,
but firmer conclusions await richer samples of stars, or a wider
choice of filters, or both.

\subsection{The Horizontal Branches}

We will focus our HB analysis only on the red clump (RC) region, and
not on the extended HB (which is discussed in Section \ref{ss:chem}
instead, in conjunction with the helium abundance).

Milone et al.\ (2012b) showed that UV two-color diagrams allow a clear
separation of the two main populations of 47~Tuc in all evolutionary
sequences, and we can profitably use a similar technique on the RC of
NGC~6388.  Here we chose the small but high-resolution ACS/HRC
exposures in F330W, F555W, and F814W.  This data set was the best
choice we could make.  We tried adding F435W from WFPC2 SNAP-9821 for
NGC~6388, but it turned out not to add any useful information.  We
could not combine ACS/HRC with ACS/WFC from GO-10775, because of the
severe crowding in the latter.  Had we tried instead to use WFC3 F390W
exposures, combined with GO-10775, the small spatial overlap would
have given too few stars for adequate statistics.

The top row of panels of Fig~\ref{f:hb:6388} shows the three possible
two-color diagrams from the HRC data set. Panels (d1), (d2), (e1),
(e2), and (f1), (f2) show the CMDs of the RC for different
color-magnitude combinations.

The choice of red-clump stars that we made is shown by the box in
panel (f1).  In panel (a) we drew by hand a line that separates the
stars into a main and a secondary group (mRC and sRC).  The groups are
colored red and blue, respectively, and each star retains its color
from this panel throughout the rest of the figure.

Panels (a) and (c) clearly show a double-horned RC, a feature that is
less obvious in panel (b).  We found the sRC to be fainter overall
than the mRC (with perhaps the exception of panel (e1)). Moreover, sRC
stars are bluer than mRC stars in panels (d1) and (e1), but redder in
(f1) and (f2).

Anomalies in the RC morphology have been reported also in the
  CMDs of NGC~6440 and NGC~6569 by Mauro et al.\ (2012) and, on a
  different scale, in the CMD of Terzan~5 by Ferraro et al.\ (2009),
  both using IR filters. Since here we instead see the RC split in a
  two-color diagrams employing UV filters, we think that the cause of
  the RC split in NGC~6388 is more similar to that of, e.g., 47~Tuc
  (Milone et al.\ 2012b), rather than those seen in IR CMDs.

%F19
\begin{figure*}[t!]
\centering
\includegraphics[width=18cm]{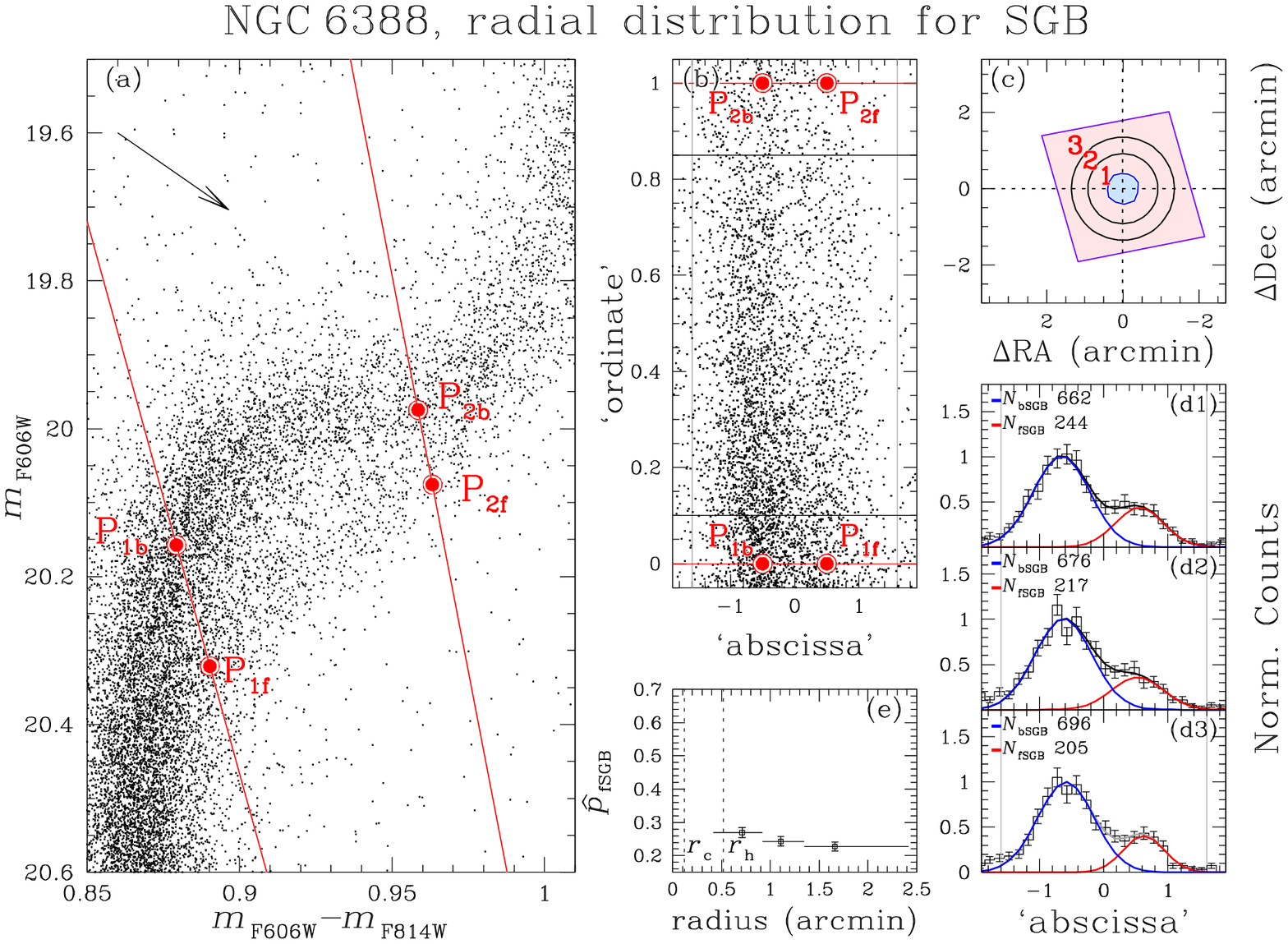}
\caption{(a) Blow-up of a region of the SGB of NGC~6388, with the four
  points and two lines marked that are used in the transformation to
  the coordinate system of the next panel.  (See text for details.)
  (b) Rectification of the SGB region that was shown in (a).  The
  horizontal black lines delimit the 'ordinate' range used in the
  analysis, and the vertical gray lines the 'abscissa' range.  (c) The
  parallelogram is our field.  The azure circle is the excluded region
  in the cluster center; the pink area is the working region that we
  divided into the three subregions that are marked.  (d)
  Dual-Gaussian fits, in the `abscissa' interval from $-$1.6 to 1.6,
  of the color distribution of stars in each radial subregion;
  `abscissa' is the same scale as in (b).  Vertical gray lines show
  the range that was fitted.  Black lines are the fits, red and blue
  lines the individual Gaussians.  (e) Relative proportions of the two
  populations, expressed as the fraction of the total that is
  contributed by fSGB.  Horizontal lines show radial extent of each
  subregion; error bars come from the sample size and the binomial
  distribution.  Dotted lines mark the core radius and half-light
  radius.}
\label{f:6388:sgb_rapp}
\end{figure*}

We performed the same multi-color analysis also for the RC of NGC
6441. To better compare the results between the two clusters, we made
use of observations taken through filters as similar as possible to
those used for NGC~6388. This choice implied the use of the WFPC2
observations from GO-5667 and GO-8718, which include data collected
with the F336W filter instead of the ACS/HRC F330W one. Because of
this, results on the RC analysis of NGC~6441 are not one-to-one
comparable with those of NGC~6388, but they can still tell us
something about similarities or differences between the RC of the two
clusters.

Figure~\ref{f:hb:6441} is the equivalent of Fig.~\ref{f:hb:6388} for
NGC~6441. Red-clump stars are again selected in the gray-line box in
panel (f1), and stars of the mRC and sRC are again delineated in panel
(a) and carry the same colors throughout.

The first difference from the RC of NGC~6388 is that here none of the
two-color diagrams has a double-horned shape.  In all the CMDs, stars
of the mRC and sRC are better mixed than in NGC~6388. In panels (d1)
and (d2) sRC stars appear to be bluer (and fainter) than mRC stars.
This mRC/sRC dichotomy seems to carry over, although in a less evident
way, into panels (e1) and (e2) as well.

The HB regions of these two clusters are at first glance so similar
that many investigations (e.g., Rich et al.\ 1997) lumped the two
clusters together; nevertheless, when we look closely the two HBs (in
particular their red clumps) are rather different.  [Note,
    however, that the double-horned red clump in NGC~6388 is visible
    only when the ACS/HRC F330W filter is used. Although the WFPC2
    filter F336W throughput is quite similar to that of the ACS/HRC
    F330W, small differences in filter response may produce quite
    large effects on the CMD (see, e.g., Momany et al.\ 2003).]  We
will resume the discussion of the HBs in Section~\ref{s:discussion}.

\section{Radial distributions}
\label{sect:rdistr}

NGC~6388 and NGC~6441 have both been revealed to host multiple stellar
populations.  In this section we will study the radial distributions
of these subpopulations, for different CMD regions.

\subsection{NGC~6388}

%F20
\begin{figure*}[t!]
\centering
\includegraphics[width=18cm]{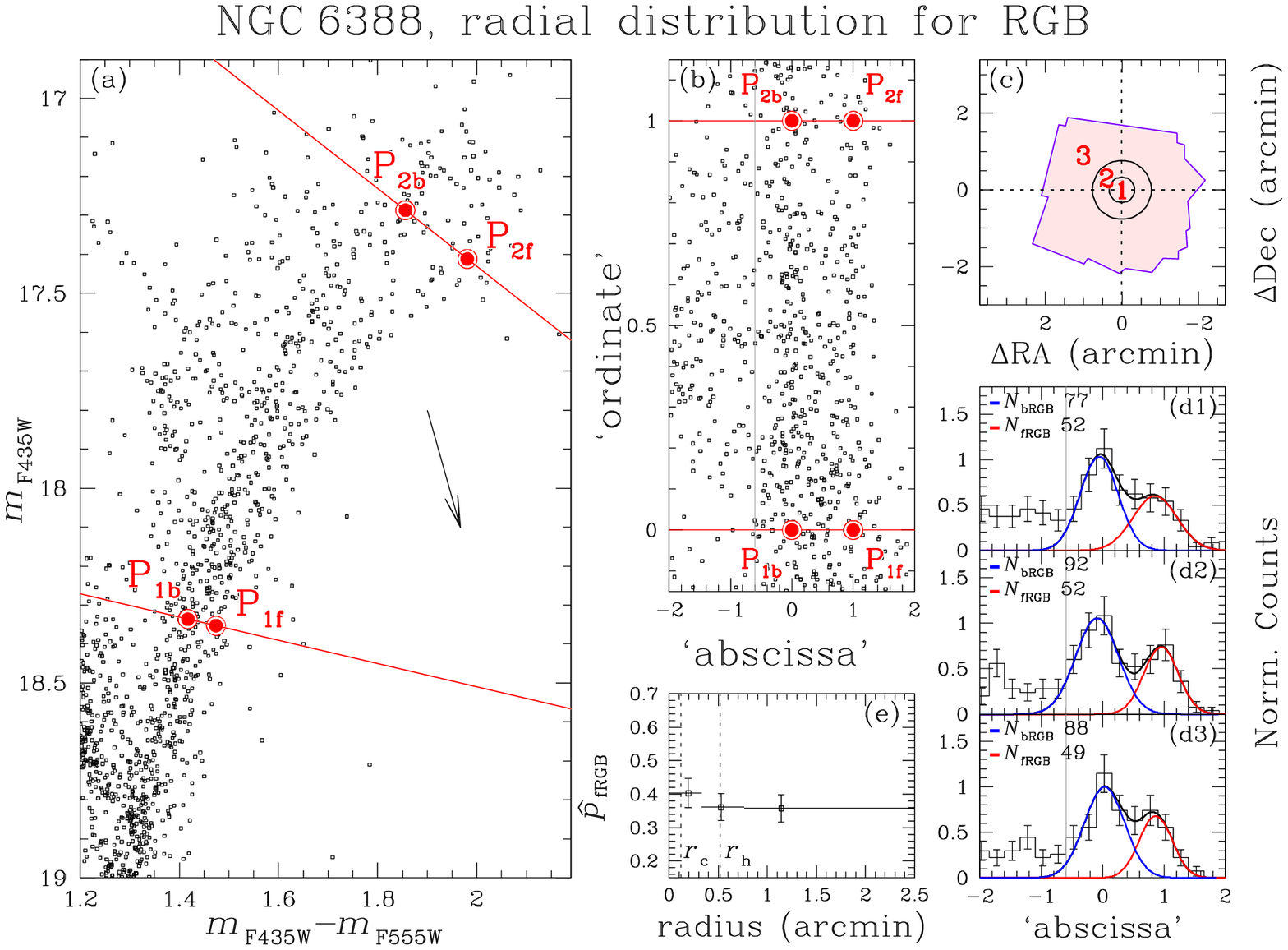}
\caption{(a) Blow-up of a region of the RGB of NGC~6388, with the four
  points and two lines marked that are used in the transformation to
  the coordinate system of panel (b).  The analysis makes use only of
  stars that are to the right of the gray vertical line and between
  the two red horizontal lines in panel (b).  (c) The irregular
  outline is the total field of view of SNAP-9821, which we used in
  its entirety, because of the paucity of stars.  The three radial
  zones contain equal numbers of stars.  (d) Dual-Gaussian fits of
  color distribution of stars in each radial subregion; `abscissa' is
  the same scale as in (b). Black lines are the fits, red and blue
  lines the individual Gaussians.  (e) Relative proportions of the two
  populations, expressed as the fraction of the total that is
  contributed by fRGB.  Horizontal lines show radial extent of each
  subregion; error bars come from the sample size and the binomial
  distribution.  Dotted lines mark the core radius and half-light
  radius.}
\label{f:6388:rgb_rapp}
\end{figure*}

%F21
\begin{figure*}[t!]
\centering
\includegraphics[width=18cm]{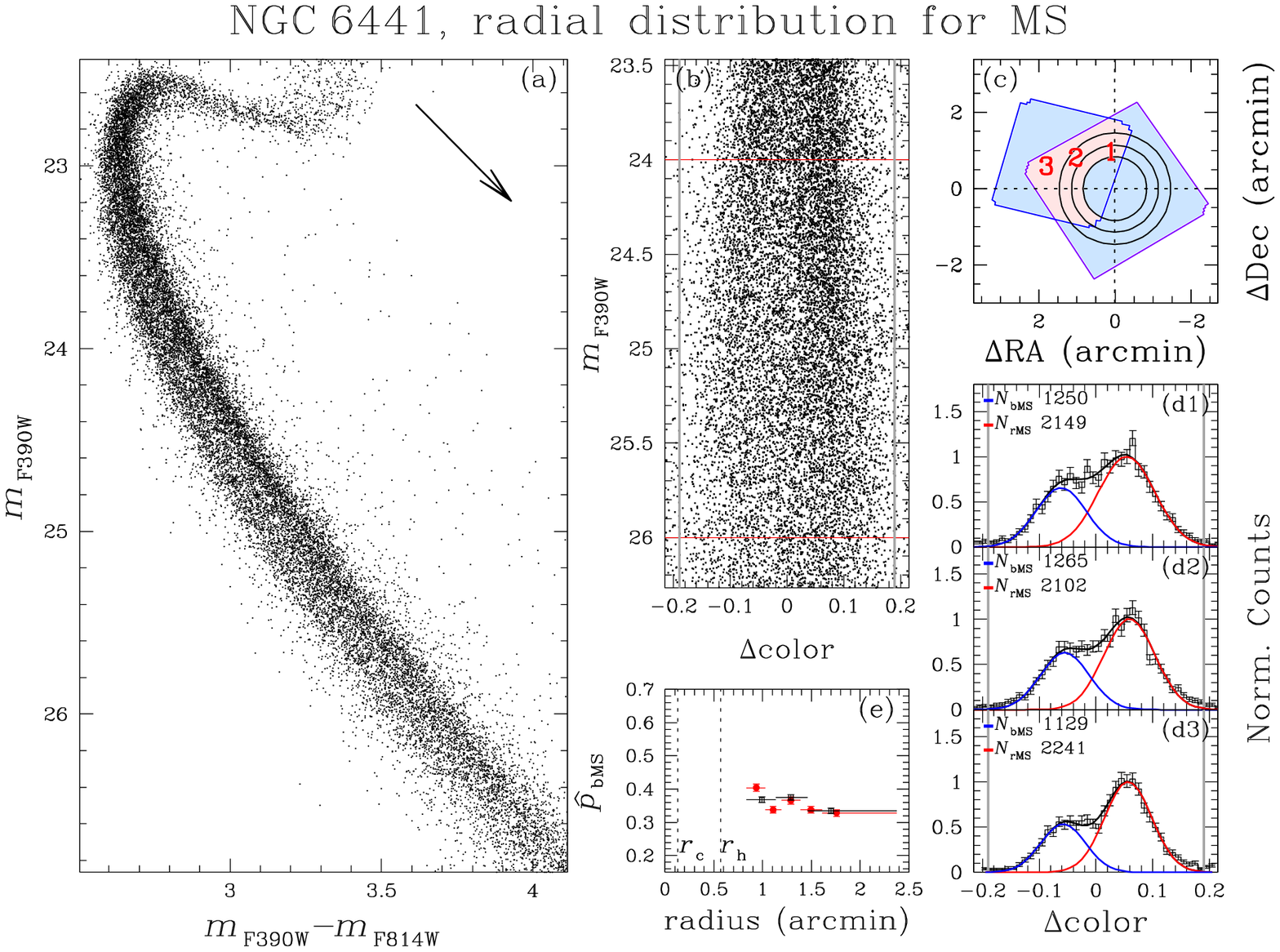}
\caption{(a) Blow-up of a region of the MS of NGC~6441.  For
  rectification we used the average MS ridge line; the result is shown
  in panel (b).  The stars selected for the analysis are in the region
  bounded by the red horizontal lines and the gray vertical lines.
  (c) The footprints of the GO-10775 ACS/WFC images (large
  parallelogram) and the GO-11739 WFC3/UVIS images (small
  parallelogram); the pink area is our working region, with our three
  radial zones marked.  (d) Dual-Gaussian fits of color distribution
  of stars in each radial zone; $\Delta$color has the same scale as
  (b).  Black lines are fits, red and blue lines the individual
  Gaussians.  (e) In black, the fraction of the MS total that is bMS,
  for the three radial intervals shown in panel (c).  Horizontal lines
  show radial extent of each subregion; error bars come from the
  sample size and the binomial distribution.  The large number of
  stars allowed a finer radial sampling as an alternative, using five
  subregions instead of three. Those results are shown in red.  Dotted
  lines mark the core radius and half-light radius.}
\label{f:6441:ms_rapp}
\end{figure*}

\subsubsection {The Sub-Giant Branch}
\label{sss:6388_sgb_rad}

We started with the double SGB of NGC~6388. We chose to analyze the
$m_{\rm F606W}$ vs.\ $m_{\rm F606W}-m_{\rm F814W}$ CMD of GO-10775
because of the wide separation between the two SGBs in this CMD, and
the large number of available stars.

Figure \ref{f:6388:sgb_rapp} illustrates our analysis of the radial
distributions of the SGB components.  To rectify the SGB sequence we
used a procedure developed by Milone et al.\ (2009), which we describe
here in some detail, as it is rather complicated.  First we selected
by hand two points ($\rm P_{\rm 1f},P_{\rm 2f}$) on the faint SGB
(fSGB) and two points ($\rm P_{\rm 1b},P_{\rm 2b}$) on the bright SGB
(bSGB).  These points define the two lines in panel (a) of
Fig.~\ref{f:6388:sgb_rapp}.  We then linearly transformed the CMD into
a reference frame (which we will refer to simply as `abscissa',
`ordinate') in which the origin corresponds to $\rm P_{1b}$ and the
coordinates of $\rm P_{1f}$, $\rm P_{2b}$ and $\rm P_{2f}$ are (1,0),
(0,1) and (1,1), respectively.  (The transformation equation is given
in the Appendix of Milone et al.\ 2009.)  This transformation
equalizes the separation of the sequences at the two ends of the
interval, but they are curved in between the ends.  (We do not show
this intermediate frame.)

We then used the following procedure to effect a straightening that
was a compromise for the two sequences.  First we fitted a fiducial
line to each of the two sequences, as follows: We divided the bSGB
into `ordinate' intervals of 0.025 and the fSGB into intervals of
0.05.  For each sequence, within each of these intervals we computed
2.5-$\sigma$-clipped median `abscissa' and `ordinate' values, and then
put a smooth spline curve through those points.  Then as the line to
use for rectification we took a line midway between the smooth spline
curves of the two sequences.

The rectification itself consisted of subtracting from the `abscissa'
of each star the `abscissa' of this line at the same `ordinate' level.
Panel (b) shows the two SGBs rectified in this way. For simplicity, we
will refer to the newly obtained $\Delta$`abscissa' as just
`abscissa'.  We also show in panel (b) the transformed locations of
$\rm P_{\rm 1f}$, $\rm P_{\rm 2f}$, $\rm P_{\rm 1b}$, and $\rm P_{\rm
  2b}$.

Stars with `abscissa' between $-$1.6 and 1.6 (gray vertical lines) and
`ordinate' between 0.1 and 0.85 (in red) were selected for the
radial-distribution analysis.

We excluded from the analysis the innermost 25 arcsec because of high
crowding, and divided the remainder of the GO-10775 field into three
radial intervals (marked 1, 2 and 3 in panel (c)) in such a way as to
have the same number of stars in each interval.  The excluded
innermost 25 arcsec are colored in azure.

For each radial interval (panels (d1), (d2), and (d3)) we extracted
the `abscissa' distribution of the stars within the limits, and did a
simultaneous fit of a pair of Gaussians to it, as done in Bellini et
al.\ (2009b).  The fit was limited to the `abscissa' values between
$-$1.6 and 1.6. The fit is marked with a black curve, while the
individual Gaussians for the bright and the faint SGB are colored in
blue and red, respectively. In the top-left corner of each of these
panels we reported the areas of each of the two Gaussians ($N_{\rm
  bSGB}$ and $N_{\rm fSGB}$, for the bright and the faint component,
respectively).

After this somewhat complicated fitting procedure, it is not
immediately obvious how to assign error bars to the numerical ratios
of the two components, but the answer becomes clear when we step back
and consider what we have done.  Our basic aim has been to determine,
at each distance from the cluster center, what fraction of the
population belongs to each component;\ the fitting of the dual
Gaussians was only a device for estimating how many stars belong to
each population component.  The statistics of the problem are
therefore binomial:\ if a fraction $p_{\rm fSGB}$ of the stars
actually belongs to the fainter component, random samplings from this
population will have a binomial distribution, with $\sigma_{N_{\rm
    fSGB}}=\sqrt{Np(1-p)}$, where $N=N_{\rm fSGB}+N_{\rm bSGB}$ is the
sample size.  For each observed case, we use as the estimator of the
fraction $p$ the quantity $\hat p_{\rm fSGB}=N_{\rm fSGB}/N$. Then
$\sigma_{\hat p_{\rm fSGB}}=\sigma_{N_{\rm fSGB}}/N=\sqrt{\hat p_{\rm
    fSGB}(1-\hat p_{\rm fSGB})/N}$; this is our best estimate of the
error bar of $\hat p$.

That is what we have done.  One should bear in mind, however, that in
so far as our estimation methods might be somewhat imperfect, the true
uncertainty could be somewhat larger than this error bar.  (We prefer
to make this simple statement, rather than attempt to doctor the
statistics.)

In panel (e) we summarize the results of the analysis: values of $\hat
p_{\rm fSGB}$ are plotted as a function of the median radial distance
of the stars in each radial interval, with errors. The horizontal
lines show the radial range to which each point applies.  Vertical
dashed lines mark the core radius $r_{\rm c}$ and the half-light
radius $r_{\rm h}$.  We ran artificial-star experiments to check
  for possible influence of incompleteness on the radial gradients. We
  found that in the radial interval covered by our images the
  completeness is always higher than 84\%, with marginal differences
  (at most 2\%) between bSGB and fSGB stars. Therefore incompleteness
  cannot significantly change the results displayed in Fig.~19.

We notice some marginal evidence of a gradient, with fSGB stars more
concentrated than bSGB ones.

%F22
\begin{figure*}[t!]
\includegraphics[width=18cm]{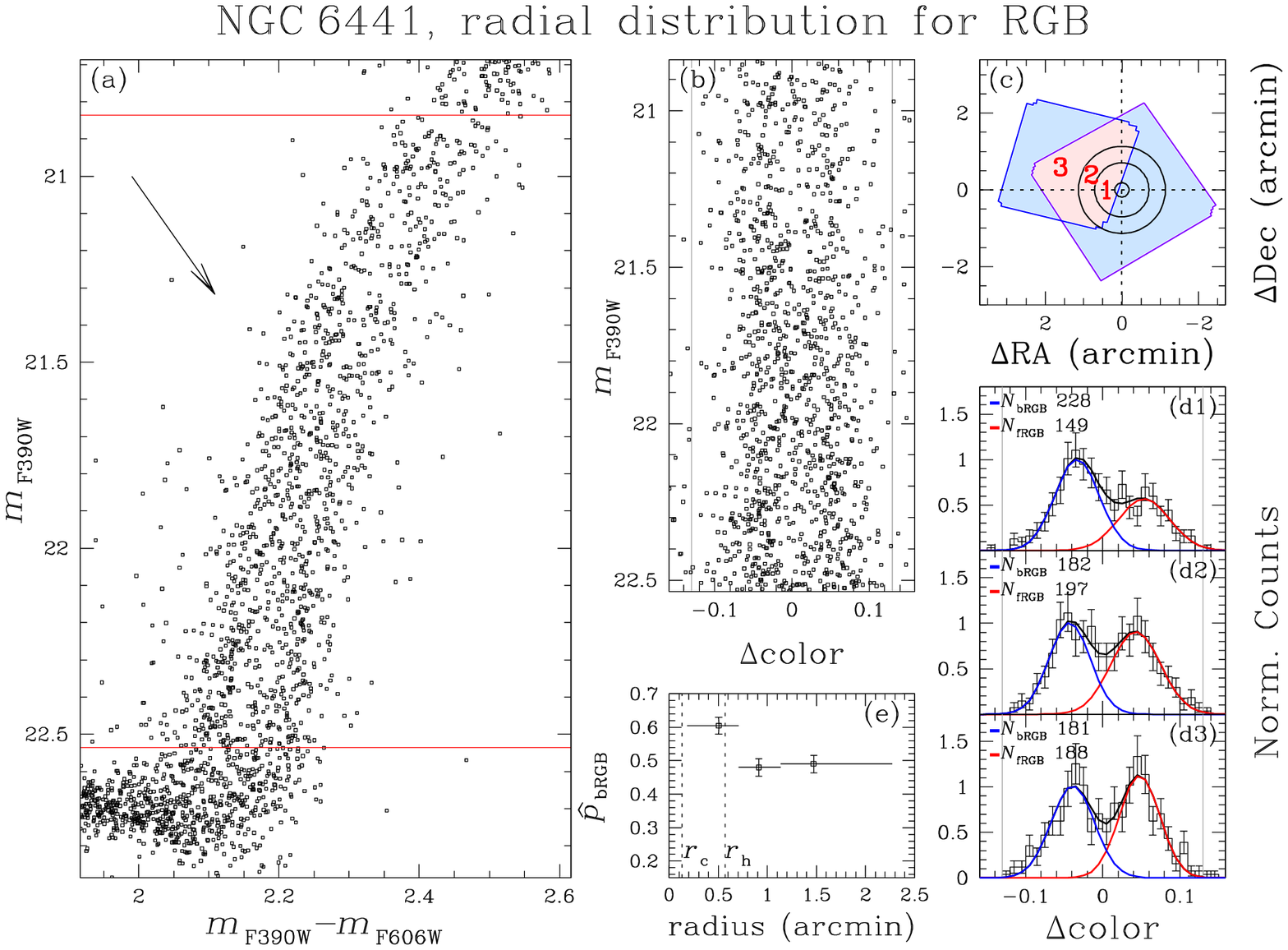}
\caption{As in Fig.~\ref{f:6441:ms_rapp}, but for the RGB
  sub-populations of NGC~6441.}
\label{f:6441:rgb_rapp}
\end{figure*}

\subsubsection {The Red-Giant Branch}

To study the radial distributions of the RGB of NGC~6388 we chose the
ACS/WFC F435W--F555W CMD of SNAP-9821, the data set that was able to
split the RGB in Fig.~\ref{f:rgb_diff}.

In panel (a) of Figure~\ref{f:6388:rgb_rapp} we see the bRGB and fRGB
(as we again call the two components) well separated at the brighter
end but badly blended at fainter magnitudes --- a situation for which
our dual-Gaussian technique is well suited.

We began with a transformation and rectification procedure like the
one that we described for the SGB.  This case was a little easier,
because the two population branches turned out to be parallel after
the transformation, so that fewer steps were required for the
rectification.

A glance at panel (b) reveals a serious problem, however: the
encroachment of the AGB on the bRGB.  (We purposely show a wide range
of `abscissa' so as to expose this issue.)  Because of the AGB, the
bluest range of `abscissa' will have to be ``out of bounds'' for our
fitting of the Gaussians, with a cut-off not far to the left of the
peak of the Gaussian that fits the bSGB.  The placement of the cutoff
is then a delicate question, to answer which we invent a special
procedure.  Without any question of radial binning yet, we analyze
{\it all} the stars that are between `ordinate' values 0 and 1 in
panel (b).  We make use of the fact that the number of AGB stars is a
very different function of `ordinate' from the number of RGB stars.
As a criterion for absence of AGB interference, we divide the
`ordinate' range from 0 to 1 into three separate parts, and require
that the ratio $N_{\rm fRGB}$/$N_{\rm bRGB}$, as derived from our
dual-Gaussian fitting with the lowest values of `abscissa' excluded,
come out the same for all three ranges of `ordinate'.  We carry out
this test (which is not illustrated here) for various placements of
the cut-off, and choose the cut-off that is farthest to the left but
still satisfies the criterion.  That value is the vertical gray line
in panel (b).

(A thoughtful reader may ask how we can be sure that the two
populations represented by the bRGB and the fRGB should have the same
distribution of stars with `ordinate' value.  The answer is that this
distribution is set by gross rules of stellar structure, as a star
moves up the RGB; these rules are quite insensitive to the small
differences that separate these two populations.  AGB evolution is
governed by quite different physics, however, and leads to a very
different distribution of `ordinate' values.)

With this problem disposed of, we can proceed to the selection of
radial bins.  Crowding is not an issue for these bright stars; we can
study them even in the very core of the cluster.  Farther from the
center, however, the star numbers become too small.  In panel (c) the
azure outline is our SNAP-9821 field, and the region that we studied
is in pink.

We had found the best `abcissa' interval to have its left limit at
$-0.6$, which is marked with a gray vertical line in panels (d1),
(d2), and (d3).  (We show the full `abscissa' range of panel (b),
however.)  The best-fit dual Gaussians are shown in each of the panels
(d), in blue and red for the bright and the faint RGB component,
respectively. We also show the areas of the Gaussians (i.e., star
numbers) in the top-left corner.

Finally, in panel (e) we report $\hat p_{\rm fRGB} = N_{\rm
  fRGB}/(N_{\rm fRGB}+N_{\rm bRGB})$ as a function of the median
distance of the stars from the cluster center in each radial interval,
with error bars calculated as explained in the previous subsection.

There is no evidence for a radial gradient, though it must be noted
that the error bars are much larger than in the previous case, because
of the smaller number statistics.

\subsubsection{A possible relation between the radial distributions of 
the SGB and the RGB}

The radial distribution of $\hat p_{\rm fSGB}$ extends from the
half-light radius, at $0\farcm5$, out to $\sim2\farcm4$; our results
give marginal evidence of a decrease (at $\sim$2$\sigma$).  Our RGB
analysis is extended to larger radii $r<2\farcm65$, but with a much
smaller number of stars and, for what it its worth, no evidence for a
radial gradient in the population fractions.  In both cases one
component (bSGB or bRGB) is significantly more numerous than the
other.  It is tempting to consider the bRGB as the progeny of the
bSGB, although $\hat p_{\rm fRGB}$ is larger than the inward
extrapolation of $\hat p_{\rm fSGB}$.  We will discuss possible
implications of these results in the final section.

\subsection{NGC~6441}

\subsubsection{The Main Sequence}

We have seen that NGC~6441 hosts two distinct stellar populations,
clearly detectable both on the MS and on the RGB (whereas for NGC~6388
the separations were on the SGB and RGB).

To measure the population fraction of the two MSs we again used the
dual-Gaussian fitting technique.  Panel (a) of
Fig.~\ref{f:6441:ms_rapp} shows a close-up of the MS of NGC~6441 in
the $m_{\rm F390W}$ vs.\ $m_{\rm F390W}-m_{\rm F814W}$ CMD, chosen
because this combination of filters enhances the split of the MSs.  We
computed the MS ridge line as done in
subsection~\ref{sss:6388_sgb_rad} for the SGBs of NGC~6388, then we
subtracted from the color of each star the color of the ridge line at
the same magnitude as the star. The rectified MS ($\Delta$color) is
plotted in panel (b), in the magnitude interval $23.05<m_{\rm
  F390W}<27.05$.  Stars with $24<m_{\rm F390W}<26$ (red horizontal
lines) and $\Delta$color between $-$0.19 and 0.19 (gray vertical
lines) were used for the radial-distribution analysis.

Panel (c) shows the footprints of GO-10775 and GO-11739 (larger and
smaller parallelograms, respectively). The overlap region used for the
analysis is colored pink.  We excluded the centermost 50 arcsec
because of crowding problems.  As done for the other
radial-distribution analyses, we defined three radial bins, each
containing the same number of selected stars.

Panels (d1), (d2), and (d3) show the histograms of the $\Delta$color
distribution within each radial interval. The best-fit dual Gaussians
are shown in blue and red for the bMS and the rMS, respectively.  The
areas of the Gaussians are also shown, as done in
Figs.~\ref{f:6388:sgb_rapp} and \ref{f:6388:rgb_rapp}, in the top-left
of each (d) panel.  The `abscissa' interval used to fit the dual
Gaussian is marked with gray vertical lines.

For each radial interval we derived the values of $\hat p_{\rm
  bMS}=N_{\rm bMS}/(N_{\rm bMS}+N_{\rm rMS})$, with errors, which are
in black in panel (e).  Motivated by the fact that there are more than
3000 stars in each radial interval, we decided also to derive a finer
sampling of the radial distribution of $\hat p_{\rm bMS}$. Therefore,
we divided the pink area in panel (c) into five equally populated
radial intervals.  For each of them we fitted a dual Gaussian (not
shown) and computed $\hat p_{\rm bMS}$ values, which are reported, in
red with errorbars, in panel (e).  There is strong evidence of a
radial gradient, with the bMS more concentrated than the rMS.

\subsubsection{The Red Giant Branch}

Finally, we analyzed the radial distribution of the two RGB components
of NGC~6441.  A close-up view of the RGBs in the $m_{\rm F390W}$
vs.\ $m_{\rm F390W}-m_{\rm F606W}$ CMD is shown in panel (a) of
Fig.~\ref{f:6441:rgb_rapp}. To rectify the sequences, we derived a
fiducial line for the whole RGB, and subtracted from the color of each
star the color of the fiducial line at the same magnitude level. The
rectified RGB is shown in panel (b). For the radial-distribution
analysis we selected stars in the magnitude range $22.52\le m_{\rm
  F390W}\le 20.84$ (red horizontal lines in panel (a)), and between
$\Delta$color $-$0.13 and 0.13 (gray vertical lines in panel (b)).

Since we are using the same data set as for the MS of NGC~6441, panel
(c) shows the same footprints as Fig.~\ref{f:6441:ms_rapp}c.  We
excluded the centermost 11.1 arcsec because of crowding, and defined
three radial intervals, each containing the same number of stars.

In panels (d1), (d2), and (d3) we show the histograms of the
$\Delta$color distributions for the three radial intervals, together
with the best-fit Gaussians and their areas. In blue we mark the bRGB
component, in red the fRGB one, and the combined curve is in black.

For each radial interval, we computed $\hat p_{\rm bRGB}=N_{\rm
  bRGB}/(N_{\rm bRGB}+N_{\rm rRGB})$ along with its error, and plotted
these as a function of median radial distance, in panel (e).

As for the MS, there is some evidence of a radial gradient, with the
bRGB more concentrated than the rRGB.  Analogously with the SGB and
RGB of NGC~6388, the similarity in radial gradients here is consistent
with the populations of the RGB of NGC~6441 being the progeny of those
of the MS -- although here again there is a disagreement in the mean
numbers.

\subsubsection{A possible relation between the radial 
distributions of the SGB and the RGB}

In summary, our values of $\hat p_{\rm bMS}$ extend from
$\sim$$0\farcm8$ (inside of which the crowding is too great) out to
$\sim$$2\farcm4$; in this range, the number of bMS stars relative to
the rMS shows a significant outward decrease.  Similarly, our study of
the RGB showed an outward decrease of the number of bRGB stars
relative to fRGB.  However, as in the case of SGB and RGB stars in
NGC~6388, here too there is an inconsistency between the overall
levels of $\hat p_{\rm bMS}$ and $\hat p_{\rm bRGB}$.  We will discuss
the implications of these results in the final section.

\begin{table}[th!]
\label{tab:3}
\small{
\begin{center}
\begin{tabular}{cccc|ccccc}
\multicolumn{8}{c}{\textsc{Table 3}}\\
\multicolumn{8}{c}{\textsc{Results of Radial Distribution Analyses
    for NGC~6388}}\\
\hline\hline
\multicolumn{4}{c|}{SGB}&\multicolumn{4}{|c}{RGB}\\
\hline
$\!\!$$r_{\rm med}$ $(^{\prime})$ $\!\!\!$& $N$ & $\hat p_{\rm fSGB}$ & $\!\!$$\sigma_{\hat p_{\rm fSGB}}$$\!\!$ &
$\!\!$$r_{\rm med}$ $(^{\prime})$ $\!\!\!$& $N$ & $\hat p_{\rm fRGB}$ & $\!\!$$\sigma_{\hat p_{\rm fRGB}}$$\!\!$\\
\hline
$\!\!$0.711& $\!\!\!$906$\!\!\!$ & 0.269& 0.015& 0.195&$\!\!\!$ 129$\!\!\!$&$\!\!$ 0.403&0.043\\
$\!\!$1.108& $\!\!\!$893$\!\!\!$ & 0.243& 0.014& 0.529&$\!\!\!$ 144$\!\!\!$&$\!\!$ 0.361&0.040\\
$\!\!$1.661& $\!\!\!$901$\!\!\!$ & 0.228& 0.014& 1.141&$\!\!\!$ 137$\!\!\!$&$\!\!$ 0.358&0.041\\
\hline
\end{tabular}
\end{center}}
\end{table}

\subsection{Population ratios and their radial gradients}
\label{ss:rad_grad}

In this section we summarize the results illustrated in Figures
19--22, concerning the number ratios between the two populations, and
the radial gradients of the ratios. In the outer regions of NGC~6441
(outside $\sim$$0\farcm8$), the blue MS represents $\sim$1/3 of the
total population ($\hat p_{\rm bMS}$$\sim 0.33$), with a small radial
gradient; closer to the center the crowding prevents us from
distinguishing the two MSs.  For the RGB we can reach the cluster
center, where the fRGB actually predominates by a little, with a
radial gradient bringing $\hat p_{\rm fRGB}$ below 0.5 farther out.
Unfortunately, radial gradients make it difficult to compare
population ratios for different evolutionary sequences, but it does
look, at first sight, as if: (i) the CMD locus of the bMS population
crosses through the locus of the rMS population; (ii) the two SGBs
overlap, and it is difficult to say whether the bMS population joins
the bRGB or the fRGB.  For NGC~6388 we could not split the MS, so the
population ratio and its radial trend were derived only for the SGB
and the RGB --- complicating comparisons, by leaving us with MS and
RGB for NGC~6441, but SGB and RGB for NGC~6388.

The results for the two clusters are summarized in Tables 3 and 4,
respectively, which give the median radius of each zone, the number of
stars in the zone, the fraction that belong to the fainter component
(or in the case of the MS, to the bluer branch), and the statistical
uncertainty of the fraction.  Alternatively, the same results can be
viewed graphically in the (e) panels of Figures 19--22.

The overall tendency is to have one sequence more concentrated towards
the cluster center, although this is not always statistically
significant. In particular, for NGC~6441 both bMS and bRGB are more
centrally concentrated, and the latter can be considered the progeny
of the former. For NGC~6388, instead, while the fSGB population is
more centrally concentrated, the distribution of the two populations
in the RGB region is found to be flat. Based solely on star counts, we
can tentatively consider bRGB as the progeny of bSGB, each being the
more populated branch.

\begin{table}[th!]
\label{tab:4}
\small{
\begin{center}
\begin{tabular}{cccc|cccc}
\multicolumn{8}{c}{\textsc{Table 4}}\\
\multicolumn{8}{c}{\textsc{Results of Radial Distribution Analyses
    for NGC~6441}}\\
\hline\hline
\multicolumn{4}{c|}{MS}&\multicolumn{4}{|c}{RGB}\\
\hline
$\!\!$$r_{\rm med}$ $(^{\prime})$$\!\!\!$&$N$&$\hat p_{\rm bMS}$&$\!\!$$\sigma_{\hat p_{\rm bMS}}$$\!\!$&
$\!\!$$r_{\rm med}$ $(^{\prime})$$\!\!\!$&$N$&$\hat p_{\rm bRGB}$&$\!\!$$\sigma_{\hat p_{\rm bRGB}}$$\!\!$\\
\hline
$\!\!$0.993& $\!\!\!$3399$\!\!\!$& 0.368& 0.008& 0.505& 377&0.605&$\!\!$ 0.025\\
$\!\!$1.290& $\!\!\!$3367$\!\!\!$& 0.376& 0.008& 0.914& 379&0.480&$\!\!$ 0.026\\
$\!\!$1.703& $\!\!\!$3370$\!\!\!$& 0.335& 0.008& 1.472& 369&0.491&$\!\!$ 0.026\\
\cline{1-4}
$\!\!$0.936&  $\!\!\!$2089$\!\!\!$& 0.404& 0.011&&&&\\
$\!\!$1.108&  $\!\!\!$2087$\!\!\!$& 0.338& 0.010&&&&\\
$\!\!$1.289&  $\!\!\!$2088$\!\!\!$& 0.366& 0.011&&&&\\
$\!\!$1.496&  $\!\!\!$2088$\!\!\!$& 0.338& 0.010&&&&\\
$\!\!$1.757&  $\!\!\!$2089$\!\!\!$& 0.328& 0.010&&&&\\
\hline 
\end{tabular}
\end{center}}
\end{table}

\begin{table}[ht!]
\label{tab:5}
\small{
\begin{center}
\begin{tabular}{ccccrcc}
\multicolumn{7}{c}{\textsc{Table 5}}\\
\multicolumn{7}{c}{\textsc{Statistical Analysis of the Radial Gradients}}\\
\hline\hline
Fig. &  NGC  & component & $N$ &   $A$~~~   &  $\sigma$  & signif \\
\hline
19  &  6388  &    fSGB   &  3  & $-$0.042  &  0.021  &  1.95 \\
20  &  6388  &    fRGB   &  3  & $-$0.041  &  0.062  &  0.67 \\
21  &  6441  &     bMS   &  3  & $-$0.050  &  0.016  &  3.14 \\
21  &  6441  &     bMS   &  5  & $-$0.069  &  0.016  &  4.31 \\
22  &  6441  &    bRGB   &  3  & $-$0.112  &  0.037  &  3.01 \\
\hline
\end{tabular}
\end{center}}
\end{table}

\begin{table}[ht!]
\label{tab:6}
\small{
\begin{center}
\begin{tabular}{lcc}
\multicolumn{3}{c}{\textsc{Table 6}}\\
\multicolumn{3}{c}{\textsc{Multipopulation Summary}}\\
\hline\hline
sequence       &       NGC~6388      &     NGC~6441\\
\hline
MS     &    ~~~~broadened~~~~    &     split\\
SGB    &       split         &  hints of a split\\
RGB   &  split  &   split \\
HB (red clump)  & split & broadened \\
\hline
\end{tabular}
\end{center}}
\end{table}

%F23
\begin{figure*}[t!]
\centering
\includegraphics[width=8cm]{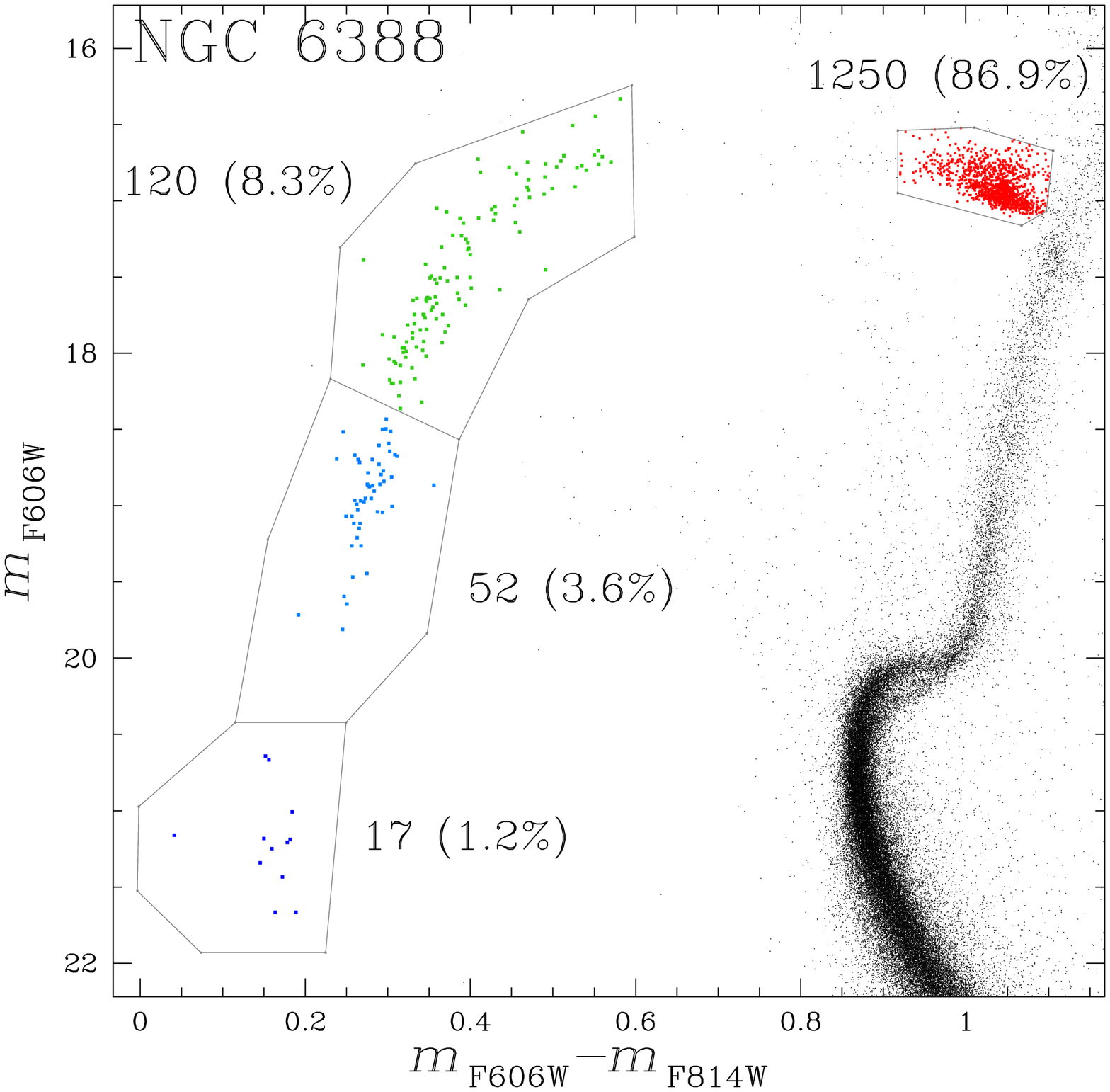}\phantom{ppp}
\includegraphics[width=8cm]{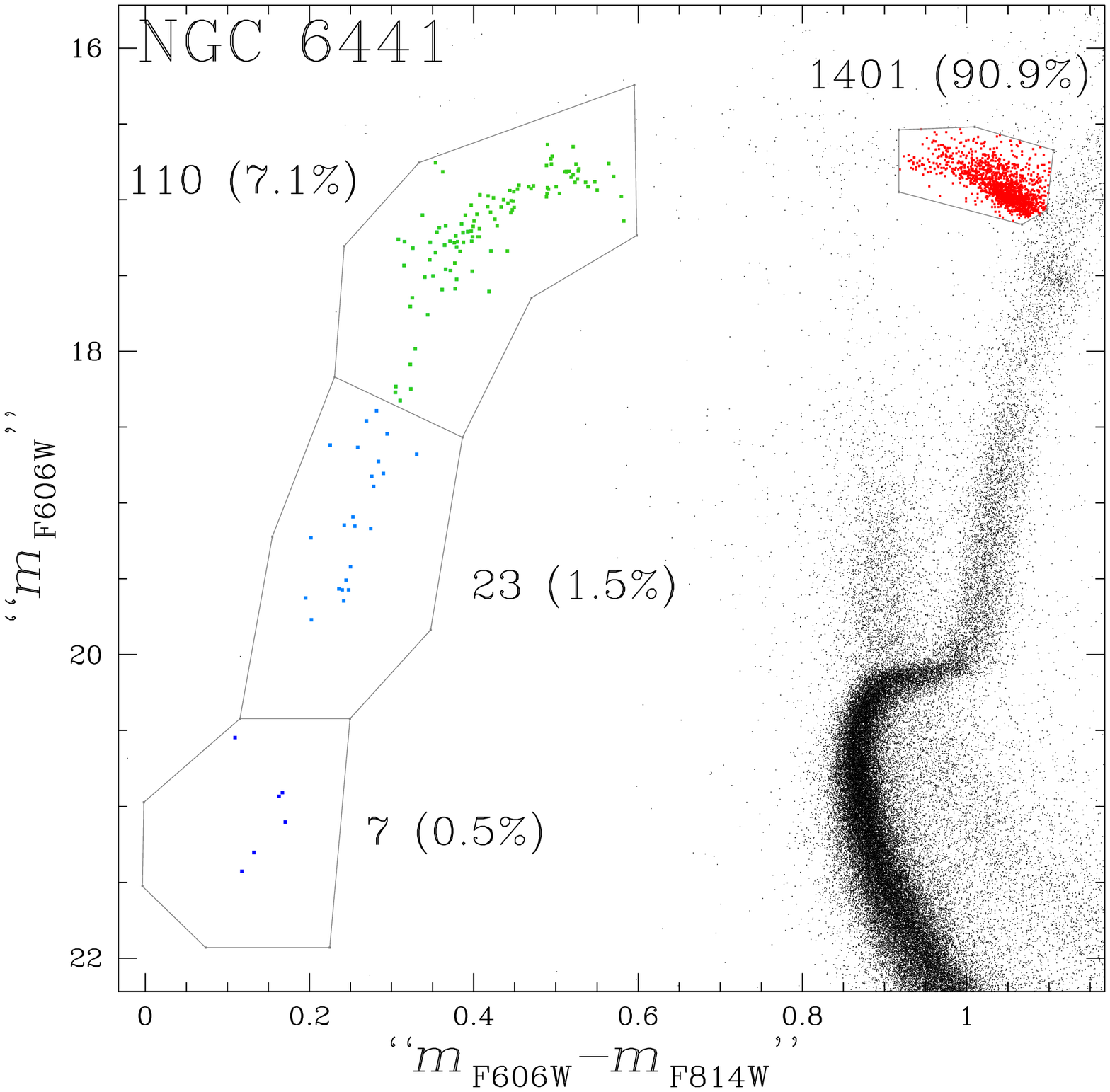}
\caption{The horizontal-branches of NGC~6388 and NGC~6441 in the
  $m_{\rm F606W}$ vs.\ $m_{\rm F606W}-m_{\rm F814W}$ CMD. We
  identified by hand four regions, colored in red for the RC, and
  green, azure, and blue for three distinct segments of the blue HB,
  from brightest to faintest. Internal labels give the number of stars
  in each of these regions, and the fraction of the total HB
  population that it constitutes.}
\label{f:hb_con}
\end{figure*}

In the study of the population ratios listed in Tables 3 and 4, it is
important to judge whether apparent radial gradients are real or not.
To test this, we calculated a least-squares line through each set of
points.  With at most five observed points, any formal error for the
slope of the line would be meaningless, so we instead made a Monte
Carlo simulation.  In place of each observed point we took a Gaussian
with mean equal to the value of the point, and with the sigma of the
point, and we calculated the slope $A$ for each of a million random
samplings from these Gaussians.  The million values of $A$ had a
distribution that was closely Gaussian; Table 5 gives the resulting
slopes (per arcmin) and their sigmas.  In the last column of the table
is the ratio of $\vert A\vert$ to $\sigma$, which is an index of how
reliably a slope differs from zero; a $2\sigma$ result is of course
only marginal, while a $3\sigma$ result is quite significant.  Thus
the radial gradient in Fig.\ 19 is marginal, that in Fig.\ 20 dubious,
and those in Figs.\ 21 and 22 highly significant.

\section{Discussion}
\label{s:discussion}

\subsection{Summary of the observational evidence of multiple stellar 
populations}

We have analyzed a large sample of proprietary (GO-11739) and archival
images of the two massive, metal-rich Galactic-bulge GCs NGC~6388 and
NGC~6441, have used proper motions to remove most of the field stars,
have corrected all our CMDs for differential reddening, and have made
a thorough multicolor analysis of the stellar populations of the two
clusters.

In both clusters we found significant evidence for the presence of at
least two stellar generations, which manifest themselves as broadened
or multiple sequences in all the main evolutionary branches of the
CMDs.  Table~6 collects the main results on the multiplicity of the
various sequences.  These results are distilled from a quite complex
data set, whose most important aspects are summarized next.

The MS of NGC~6388 is analyzed with various filter combinations in
Figs.~9--11. Most of the CMDs show a broadening of the MS, but none of
them succeeds in splitting it into distinct sequences as instead
happens for NGC~6441 when using the $m_{\rm F390W}-m_{\rm F814W}$
color, as shown in Figures ~6(g) and 21.  The separation of the two
MSs of this latter cluster as a function of color baseline is
illustrated in Fig.~\ref{f:6441_mca} Figure~\ref{f:redd_test} further
confirms the reality of these results for both clusters, showing that
neither of them could result from differential reddening (for which
all CMDs have been corrected). Note, moreover, that the average
reddening of NGC~6388 is lower than that of NGC~6441 (see
subsection~2.5), hence it should have been easier to recognize two MSs
(if present) compared to the case of NGC~6441. We conclude that the MS
structures of the two clusters are different.

The SGBs of the two clusters are examined in Fig.~\ref{f:sgb_diff}
which shows a clear split for the SGB of NGC~6388 in panel (a) but at
best a doubtful one for NGC~6441 in panels (c) and (d), where the SGB
components apparently overlap; Figures~14 and 15, respectively,
explore the SGB for each cluster in multicolor domains. Again, we note
this more subtle difference between the two clusters, with two
distinct SGBs in NGC~6388 and just a broadened SGB in NGC~6441 (the
opposite situation compared to the MS).

The structure of the RGB of the two clusters is shown in Fig.~16. In
both cases there is a hint of a double RGB, more evident in the upper
part of the RGB of NGC~6388 and in the lower part for NGC~6441.  The
evolutionary connection between the MS and the RGB is somewhat
ambiguous in the case of NGC~6441: the population ratio seems to
indicate that bMS and fRGB stars belong to the same population, but
the radial gradient (which complicates the calculation of the
population ratios) suggests instead to connect the bMS to the
bRGB. For NGC~6388 it appears that bSGB and bRGB stars belong to the
same population.

For the HB we have thus far examined only the red clump; we take up
the extended blue part of the HB in the following. The split of the RC
of NGC~6388 can be seen in the two-color diagrams in panels (a) and
(c) of Fig.~17, while for NGC~6441 the broadening is seen in most of
the panels of Fig.~18.  Therefore there is some evidence that the RC
is also a mixture of the first and second generations.  However, what
makes these two clusters unique, among the 150 globular clusters of
the Milky Way, is their blue HB extension in spite of their high
metallicity. Furthermore, their blue HBs are remarkably similar, as
illustrated in Fig.~\ref{f:hb_con}. Note also the similarity of the
relative number of stars in the distinct sections into which we have
divided the HB.  This similarity is particularly surprising given that
--- as is well known --- the distribution of stars on the HB is
extremely sensitive to many different parameters (age, composition,
RGB mass loss). Thus, the differences in the structure of the MS, SGB
and RGB of the two clusters demonstrate that there must be a sizable
difference in some of those parameters, and yet the HBs turn out to be
very similar. This {\it conundrum} is further discussed in the last
subsection.

%F24
\begin{figure}[t!]
\centering
\includegraphics[width=8cm]{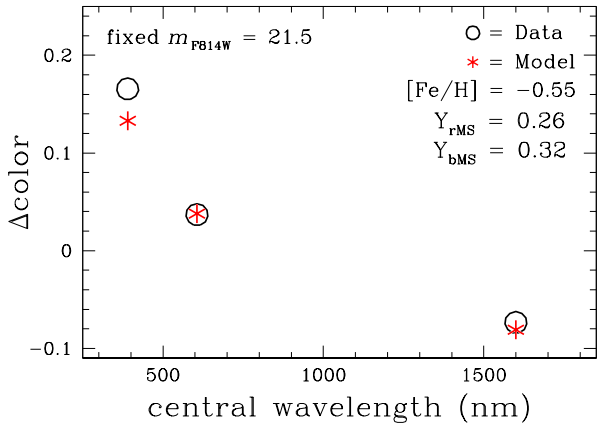}
\caption{Points from panel (d) of Fig.~\ref{f:6441_mca}, on which we
  have superimposed values of $\Delta$color that follow from synthetic
  colors calculated with our best-fit value of the helium abundance of
  the bMS of NGC~6441.}
\label{f:ms_chem}
\end{figure}

\subsection{The helium abundance of the secondary populations}
\label{ss:chem}

Photometric differences in the CMD between multiple stellar
populations in GCs have been attributed to differences in helium
(e.g., Bedin et al. 2004; Norris 2004; D'Antona et al.\ 2005; Piotto
et al.\ 2005, 2007) or in light-element abundances (e.g., Sbordone et
al.\ 2011). Different photometric bands can enhance different effects
of the chemical pattern of the population.  As noted in the
introduction, the presence of helium-enriched stars in these two
clusters was first proposed based on the exceptional morphology of
their HBs and on the properties of their RR Lyrae variables. Now, the
discovery of a double main sequence in NGC~6441 lends support to this
interpretation, and favors a scenario in which the helium-rich stars
belong to a second stellar generation, as opposed to the possibility
that they would owe high helium to deep-mixing on the upper RGB, as
proposed by Sweigart \& Catelan (1998).

Our studies of 47~Tuc (Milone et al.\ 2012b), NGC~6397 (Milone et al.\
2012c), and NGC~6752 (Milone et al.\ 2010) demonstrate that
light-element variations should not significantly affect the $m_{\rm
  F606W}-m_{\rm F814W}$ or $m_{\rm F814W}-m_{\rm F160W}$ colors of MS
stars, so we explore the possibility of helium being the chief cause
of the observed MS split in NGC~6441 (or of MS broadening, depending
on the specific bands).  To this end, we calculated synthetic colors
for the two MSs, adopting a primordial helium $Y$=0.26 for the rMS,
while trying for the bMS $Y$ values between 0.27 and 0.36, in steps of
0.01.  We assumed for each sub-population the same iron abundance
([Fe/H]=$-$0.55 Harris 1996, 2010 edition, see also Carretta et
al. 2009b, Origlia et al. 2008).  Note that a difference of 0.1 dex in
[Fe/H] (which represents the present uncertainty in the metallicity
difference between the two clusters) in the calculated models does not
change our conclusions.

In previous papers we have used a similar approach to estimate helium
differences between GC sub-populations, but there we knew the
light-element abundance of the two MSs and here we do not.  Hence, we
assume the abundance of C, N and O to be the same, in spite of likely
differences in at least their relative proportions between the first
and second generations.  In practice, we assumed [C/Fe]=$-$0.45,
  [O/Fe]=0.27 (Origlia et al.\ 2008), and solar N abundance.  In both
  cases we used [$\alpha$/Fe]=0.3 (Origlia et al. 2008).   Neglecting
possible differences of CNO proportions should not affect our
conclusion about helium because, as noted above, the F606W, F814W and
F160W filter bands are almost completely insensitive to C, N, and O,
while only F390W is expected to be somewhat affected by abundance
variations in these light-elements.

We adopted the synthetic colors from the BaSTI isochrones
(Pietrinferni et al.\ 2004) in order to compare observations and
synthetic colors, and determined $T_{\rm eff}$ and $\log g$ for MS
stars at $m_{\rm F814W}$=21.5. The ATLAS12 code (Kurucz 2005; Castelli
2005; Sbordone et al.\ 2007) was then used, which allows us to work
with arbitrary chemical compositions. We performed spectral syntheses
from 3,400 \AA\ to 20,000 \AA\ by using the SYNTHE code (Kurucz 2005),
and integrated the synthetic spectra over the transmission curves of
the four filters.  From the resulting fluxes we formed, for each value
of $Y$, the differences $\Delta{\rm color}=m_{\rm X}-m_{\rm F814W}$.

Our result is shown in Fig.~\ref{f:ms_chem}.  We found that the value
$Y_{\rm bMS}$=0.32 optimizes the agreement between models and
observations. The overall good agreement suggests that helium can
indeed be the main cause of the observed split, with bMS stars being
He-enhanced by $\Delta Y$$\sim$0.06 relative to rMS stars.  Best-fit
model and data are less in agreement for the F390W point, as expected
because this band-pass includes CN and CH molecular bands, hence the
flux through this band is sensitive to CNO variations (e.g., Milone et
al.\ 2012b).

%F25
\begin{figure*}[t!]
\centering
\includegraphics[width=9.0cm]{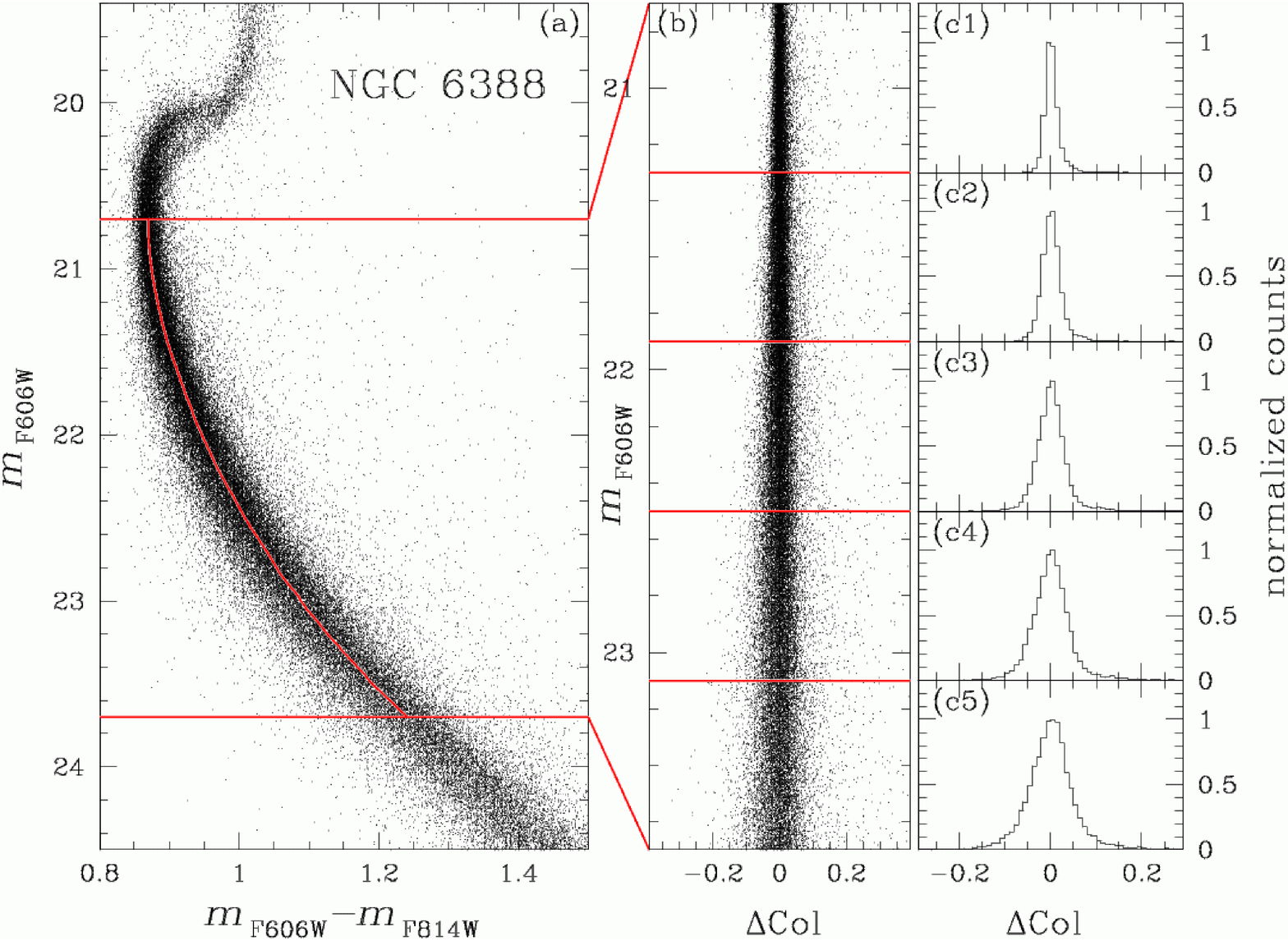}
\includegraphics[width=9.0cm]{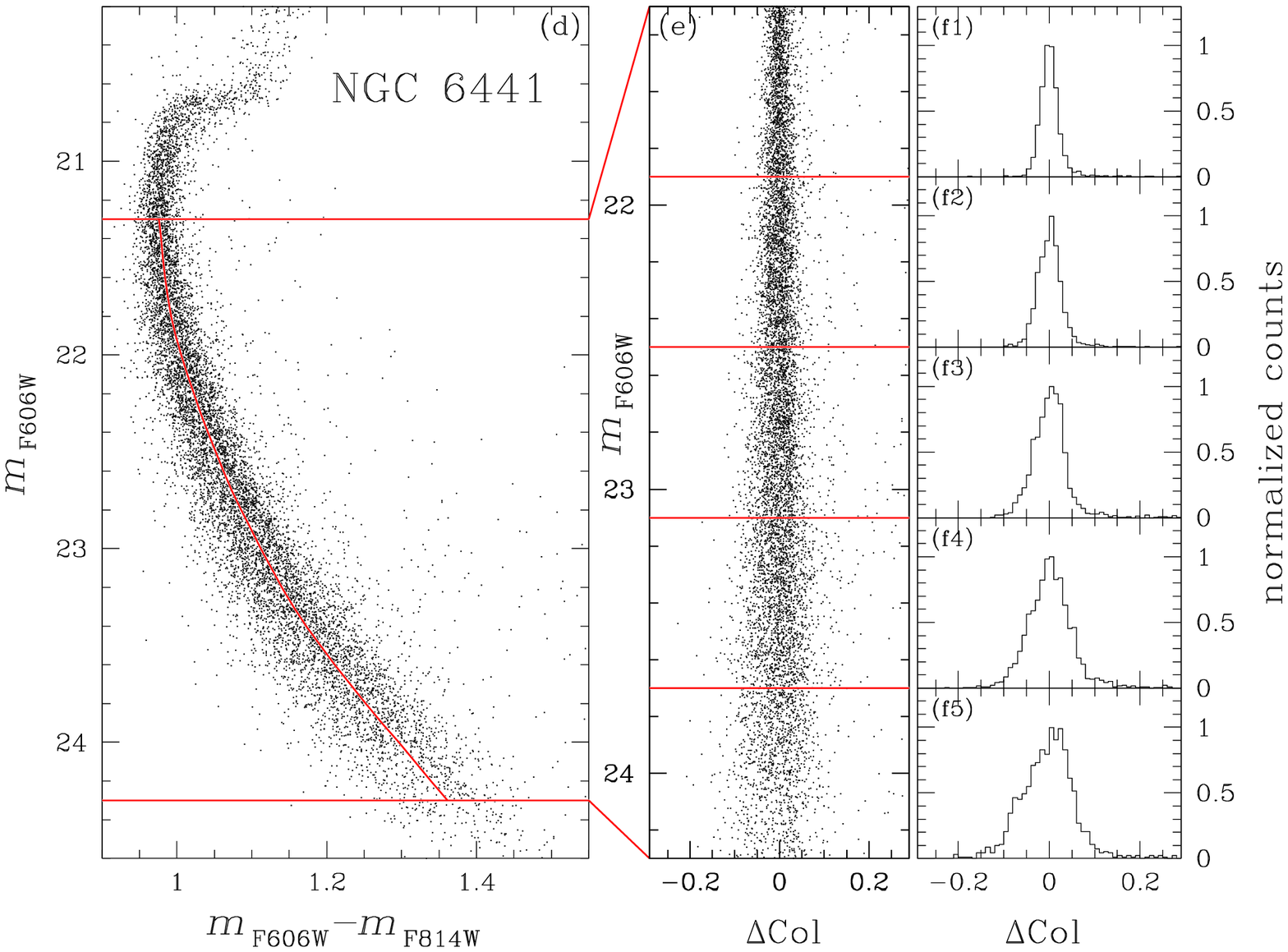}
\caption{Illustration of the procedures used to determine the MS
  intrinsic breadth of the two clusters in $m_{\rm F606W}- m_{\rm
    F814W}$. Panels (a), (b), and (c) refer to NGC~6388, Panels (d),
  (e), and (f) to NGC~6441. See text for details.}
\label{t1_2}
\end{figure*}

In the case of NGC~6388 there is no evidence of a split in its MS,
which is clearly broader than expected from photometric errors, but
the color spread (in $m_{\rm F390W}-m_{\rm F814W}$) is smaller than
that of NGC~6441.  However, as previously emphasized, the $m_{\rm
  F606W}- m_{\rm F814W}$ color baseline is more sensitive to helium
variation, but is insensitive to variations of CNO.  In Figure
\ref{t1_2} we compare the $m_{\rm F606W}$ vs.\ $m_{\rm F606W}- m_{\rm
  F814W}$ CMDs of NGC~6388 and NGC~6441.  Panels (a) and (d) are the
CMDs themselves (from the overlapping regions of GO-11739 and
GO-10775, proper-motion-selected members only for NGC~6441).  Panels
(b) and (e) show the rectified MSs, extending $\sim$3 magnitudes below
the TO. The rectified $\Delta$color distribution, in five magnitude
intervals, is shown in panels (c) and (f).

\begin{figure}[t!]
\centering
\includegraphics[width=8cm]{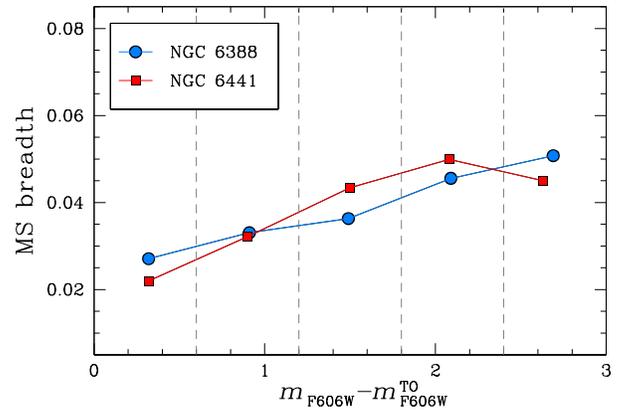}
\caption{MS breadth in $m_{\rm F606W}-m_{\rm F814W}$ as a function of
  the magnitude $m_{\rm F606W}$ below the TO. Blue for NGC~6388, red
  for NGC~6441.}
\label{t3}
\end{figure}

The color distributions appear to be symmetric in NGC~6388 but
slightly skewed bluewards in NGC~6441.  Table~7 gives the MS breadths,
from fitting Gaussians to the distribution of $\Delta$color.  The
first column identifies the panel in the figure, and the final column
gives the intrinsic breadths, calculated as the quadrature difference
between the preceding two columns.  These breadths are also plotted in
Fig.~\ref{t3} as a function of the $m_{\rm F606W}-m_{\rm F606W}^{\rm
  TO}$ magnitude\footnote{To choose the TO value in $m_{\rm F606W}$ we
  derived a fiducial line along the MS and took it's bluest point.}.
We conclude that the intrinsic $m_{\rm F606W}- m_{\rm F814W}$ color
dispersions of the two MSs are quite similar, so that despite the
different manifestation of the two populations, at least their helium
distributions must likewise be quite similar.

Though a complete and thorough analysis and interpretation of the
numerous CMDs presented here is deferred to a future paper (Cassisi et
al. in preparation), we wish to present here some further remarks on
the estimates of the helium abundance in the stellar populations of
NGC~6388 and NGC~6441.

The $Y$$\sim$0.32 that we found for the bMS stars in NGC~6441 from the
MS separations in the various color baselines can be compared with the
He content estimated by Caloi and D'Antona (2007) and by Busso et
al.\ (2007) from their analysis of the HB morphology and of the
pulsational properties of the RR Lyrae stars in this cluster. Busso et
al.  were able to reproduce the blue and tilted HB morphology of
NGC~6441 by assuming a He abundance $Y$$\approx$0.35 for the
second-generation stars, only $\sim$0.03 higher than our estimate from
the MS split.  At the same time Caloi \& D'Antona suggested that a
fraction of the NGC~6441 stars would have a He content as high as
$Y$=0.38--0.40 in order to account for the extreme blue part of the
HB. Part of this discrepancy with our present result and that of Busso
et al.\ could be due to the use of optical bands by Caloi \& D'Antona,
which surely are not optimal for studying the very hot HB extension
(see Dalessandro et al.\ 2011).  Moreover, in order to match synthetic
to observed HBs, both Busso et al.\ and Caloi \& D'Antona appeal to a
substantial spread of helium from $Y$$\simeq$0.25 in the first
generation up to the their maximum estimated values (0.35 and 0.40,
respectively). Our result, with two distinct MSs in NGC 6441 suggests
instead that (at least in this cluster) there are basically two
distinct values of the helium abundance, with a clear gap in between.

Furthermore, there is yet another puzzling inconsistency between the
results of those two investigations based on the HB and the results of
the present paper.  According to Caloi \& D'Antona only $\sim$38\% of
the stars in NGC~6441 would have primordial He content ($Y\sim$0.25),
whereas according to Busso et al.\ only 15--16\% of the stars should
belong to the second generation.  Though in different ways, both
results are at variance with what we obtained from the two MSs of this
cluster, i.e., the majority of stars ($\sim 65$\%) are on the red
(presumably first-generation) MS and only $\sim 1/3$ appear to belong
to the second generation.  We shall come back to this issue in the
next subsection.

%F26

\begin{table}[ht!]
\label{tab:7}
\small{
\begin{center}
\begin{tabular}{ccccccc}
\multicolumn{6}{c}{\textsc{Table 7}}\\
\multicolumn{6}{c}{\textsc{Intrinsic MS Breadth in $m_{\rm F606W}-m_{\rm F814W}$  Colors}}\\
\hline\hline
Ref. & $N_{\rm stars}$& $\langle m_{\rm F606W}\rangle$ & $\Delta$Col rms &
$\sigma$ & Breadth\\
\hline
\multicolumn{6}{c}{NGC 6388}\\
(c1)&   10148&  21.02&   0.015&  0.031&    0.027\\
(c2)&   12109&  21.61&   0.018&  0.038&    0.033\\
(c3)&   12119&  22.19&   0.023&  0.043&    0.036\\
(c4)&   10801&  22.79&   0.030&  0.054&    0.046\\
(c5)&   8987 &  23.39&   0.041&  0.065&    0.051\\
\hline
\multicolumn{6}{c}{NGC 6441}\\
(f1)&   1626&  21.62&   0.016&  0.027& 0.022\\
(f2)&   1935&  22.20&   0.021&  0.038& 0.032\\
(f3)&   2048&  22.80&   0.028&  0.051& 0.043\\
(f4)&   1947&  23.38&   0.037&  0.062& 0.050\\
(f5)&   1134&  23.93&   0.052&  0.069& 0.045\\
\hline
\end{tabular}
\end{center}}
\end{table}

At least part of these discrepancies can be traced to the assumption
made by Busso et al.\ that the whole red clump consists of
first-generation stars, with second-generation stars confined to the
blue HB extension; indeed, the RC represents $\sim$12--15\% of the
total HB population. We have found instead that the RC is a mixture of
first- and second-generation stars. It is then possible that all
second-generation stars arrive on the HB at the RC, spend there
roughly 50\% of the HB lifetime, and then evolve into an extended blue
loop for the remaining $\sim$50\% of the time. In fact, as emphasized
by Caloi \& D'Antona, the evolution of the helium-rich and metal-rich
HB models may be confined to the RC region, or, after spending some
time on the RC, the stars may experience an extended excursion
reaching very blue colors for differences in total mass of order of
just $\sim 0.01\, M_\odot$ (Sweigart \& Gross 1978). For this reason,
it is difficult to precisely pinpoint the helium abundance from the
mere extension of the HB.

Finally, we note that the MS breadth in the $m_{\rm F606W}-m_{\rm
  F814W}$ color increases with decreasing luminosity (see Fig.~26 and
Table~7) in a fashion that apparently cannot be accounted for by the
increasing photometric errors.  Here, we simply observe that an
increase of the MS spread in the 2--3 magnitude interval below the TO
is also observed in the MSs of $\omega$~Cen (King et al.\ 2012) and
NGC 2808 (Piotto et al.\ 2007).  We shall return on this observational
evidence in a future paper.

\subsection{An NGC~6388--NGC~6441 conundrum?}
\label{conundrum}

As already emphasized, these two clusters are unique in having a major
blue horizontal branch extension in spite of their high metallicity
and in the properties of their RR Lyrae variables. Furthermore, their
HBs are remarkably similar in shape and extension and even in the
relative star numbers along the branches (see Fig.~\ref{f:hb_con}).
The two clusters also have in common the presence of two population
components whose distinctness is evident in various parts of their
CMDs, with the population that is (slightly) more concentrated to the
center likely being the second generation. The outstanding anomaly,
however, is that whereas the MS of NGC~6441 is clearly split into two
branches (in $m_{\rm F390W}-m_{\rm F814W}$ color), that of NGC~6388 is
single, though broadened.  Given the extremely high sensitivity of the
HB morphology to the helium abundance, we conclude that the helium
difference between the two stellar generations must be similar, in the
two clusters. The results presented in Fig.~\ref{t1_2} and \ref{t3},
showing the close similarity of the MS of the two clusters in the
$m_{\rm F606W}-m_{\rm F814W}$ color, strengthen this conclusion
further.  How can these clusters exhibit almost identical HBs while
having very different MS properties in $m_{\rm F390W}-m_{\rm F814W}$
color?  In other words, if the second generation populations in the
two clusters have nearly identical helium abundances, why is the MS
split in NGC~6441 but not in NGC~6388?  Clearly, something else must
differ in order to reduce below detection the MS split of NGC~6388, or
to amplify that of NGC 6441. This is the {\it conundrum} of what is
otherwise a pair of twin clusters.

 One possibility is metallicity [Fe/H], such that in NGC~6388 the
 helium-rich population is also somewhat metal-enriched, while it is
 not in NGC~6441.  This possibility appears to be excluded by the
 high-resolution spectroscopy of Carretta et al.\ (2007), according to
 which RGB stars in NGC~6388 have the same [Fe/H] within $\sim$0.01
 dex. Their sample consists of just seven stars, however, so that the
 chance of their having missed the minority population is not entirely
 negligible. However, among these seven stars there appears to be a
 spread in oxygen, sodium, and aluminum abundances, which should
 indicate that, though small, the sample actually includes stars from
 both the first and the second generation. This is confirmed also by a
 lower-resolution study of a larger sample of stars (Carretta et
 al.\ 2009a).  The option that the seven stars of Carretta et al.\ all
 belonged to a single population does not appear to be supported by
 observations.

 As emphasized above, the main difference between the two clusters
 arises when the F390W filter comes into play, and within its passband
 strong CN and CH bands are included (see Fig.~32 in Milone et
 al.\ 2012b).  A more attractive option is therefore that the
 secondary population in NGC~6388 is CNO-enhanced compared with the
 first generation, though this requires a chemical enrichment scenario
 that is quite ad hoc. In this way excess CN and/or CH molecular
 blanketing in the UV would make redder the MS of the secondary
 population in this cluster.  A similar effect could also be produced
 by the C:N:O proportions in the secondary population of NGC~6388
 being different from those in the secondary population of NGC~6441.

 We defer to a future paper a more quantitative exploration of these
 alternatives (Cassisi et al., in preparation), but we note that
 further progress will require an extensive spectroscopic exploration
 of these clusters at high spectral resolution, in particular having
 photometrically identified the targets as belonging to one or the
 other sub-population. More extensive \textit{HST} observations,
 especially at ultraviolet wavelengths, would also help in solving
 this conundrum.

 \acknowledgments \noindent \textbf{Acknowledgments.} A.B., S.C.,
 G.P., and A.R.\ acknowledge partial financial support by PRIN INAF
 ``Formation and early evolution of massive star clusters''. A.B.\ and
 G.P.\ acknowledge support by ASI under grants ASI-INAF I/016/07/0 and
 I/009/10/0, and by the Progetto di Ateneo of the Universit\'a di
 Padova (ref.\ num.\ CPDA103591). J.A.\ and I.R.K.\ acknoledge support
 from STScI grant GO-11739.

\small{
}

\end{document}